\documentclass[12pt]{article}%
\pdfoutput=1
\usepackage[nosort]{cite}
\usepackage{graphicx}
\usepackage{multicol}
\usepackage{amsfonts}
\usepackage{amssymb}
\usepackage{amsmath}
\usepackage{heck}
\usepackage{afterpage}
\usepackage{setspace}
\usepackage{verbatim}
\usepackage{color}
\usepackage{slashed}
\usepackage{longtable}
\usepackage{float}
\usepackage{subcaption}
\usepackage{epsfig}
\usepackage{enumerate}
\usepackage{epstopdf}
\usepackage[enableskew, vcentermath]{youngtab}
\usepackage{adjustbox}
\newsavebox{\mysavebox}

\usepackage{youngtab}
\usepackage{tikz}
\usepackage[margin=1in]{geometry}
\usepackage{titletoc}%
\usepackage{hyperref}
\hypersetup{colorlinks,%
citecolor=black,%
filecolor=black,%
linkcolor=black,%
urlcolor=black}
\providecommand{\U}[1]{\protect\rule{.1in}{.1in}}
\usetikzlibrary{decorations.markings}

\numberwithin{equation}{section}

\usepackage{mathtools}
\usepackage{pgfplots}
\usetikzlibrary{chains, external}
\usepgfplotslibrary{external}
\tikzset{node distance=2em, ch/.style={circle,draw,on chain,inner sep=2pt},chj/.style={ch,join},every path/.style={shorten >=4pt,shorten <=4pt},line width=1pt,baseline=-1ex}

\newcommand{\ba}{\begin{eqnarray}}
\newcommand{\ea}{\end{eqnarray}}

\newcommand{\mf}{\mathfrak}
\DeclareMathOperator{\SU}{\mathit{SU}}

\newcommand{\Tr}{\, {\rm Tr}}

\newcommand{\ov}{\overset }

\newcommand{\be}{\begin{equation}}
\newcommand{\ee}{\end{equation}}

\tikzstyle{startstop} = [rectangle, rounded corners, minimum width=3cm, minimum height=1cm,text centered, draw=black, fill=blue!10]
\tikzstyle{startstop} = [rectangle, rounded corners, minimum width=3cm, minimum height=1cm,text centered, draw=black, fill=blue!10]

\tikzstyle{io} = [trapezium, trapezium left angle=70, trapezium right angle=110, minimum width=3cm, minimum height=1cm, text centered, draw=black, fill=blue!30]

\tikzstyle{process} = [rectangle, minimum width=3cm, minimum height=1cm, text centered, draw=black, fill=orange!30]
\tikzstyle{decision} = [diamond, minimum width=3cm, minimum height=1cm, text centered, draw=black, fill=green!30]

\tikzstyle{arrow} = [thick,->,>=stealth]

\tikzset{->-/.style={decoration={
  markings,
  mark=at position #1 with {\arrow[scale=2.4]{>}}},postaction={decorate}}}

\makeatletter \@addtoreset{equation}{section} \makeatother

\begin{document}

\date{May 2018}

\title{Top Down Approach to 6D\ SCFTs}

\institution{UPENN}{\centerline{${}^{1}$Department of Physics and Astronomy, University of Pennsylvania, Philadelphia, PA 19104, USA}}

\institution{IAS}{\centerline{${}^{2}$School of Natural Sciences, Institute for Advanced Study, Princeton, NJ 08540, USA}}

\authors{Jonathan J. Heckman\worksat{\UPENN}\footnote{e-mail: {\tt jheckman@sas.upenn.edu}}
and Tom Rudelius\worksat{\IAS}\footnote{e-mail: {\tt rudelius@ias.edu}}}

\abstract{Six-dimensional superconformal field theories (6D SCFTs) occupy a central
place in the study of quantum field theories encountered in high energy theory.
This article reviews the top down construction and study of this rich class of quantum field theories, in particular,
how they are realized by suitable backgrounds in string / M- / F-theory.
We review the recent F-theoretic classification of 6D SCFTs, explain how to calculate
physical quantities of interest such as the anomaly polynomial of 6D SCFTs, and also explain recent progress in understanding
renormalization group flows for deformations of such theories. Additional topics covered by this review include
some discussion on the (weighted and signed) counting of states in these theories via superconformal indices. We also include
several previously unpublished results as well as a new variant on the swampland conjecture for general quantum field theories
decoupled from gravity. The aim of the article is to provide a point of entry into this growing literature rather than an exhaustive overview.}

\maketitle

\tableofcontents

\enlargethispage{\baselineskip}

\setcounter{tocdepth}{2}

\newpage

\section{Introduction \label{sec:INTRO}}

Quantum field theory (QFT) is the basic language for understanding a huge
swath of physical phenomena. It undergirds our understanding of the Standard
Model of particle physics, inflationary cosmology, as well as many condensed
matter systems. From this perspective, it is clearly important to understand
the structure of this formalism and all its possible manifestations.

Conformal field theories (CFTs) are a particularly important subclass of QFTs.
They arise in limits where all mass scales have disappeared. Many
quantum field theories can be viewed as a flow from one fixed point of the
renormalization group (RG)\ equations to another, so it is clear that
understanding the beginning and end of such flows can provide important
insights into general QFTs.

Given their central importance, it is perhaps surprising that so little is known
about the general structure of quantum field theory. For example, until quite
recently, it was not known whether interacting conformal fixed points existed
in more than four dimensions. The situation changed dramatically with the
advent of new methods from string theory. In the 1990's, a set of apparently
mysterious six-dimensional theories with \textquotedblleft
tensionless strings\textquotedblright\ \cite{Witten:1995zh,
Strominger:1995ac} were discovered. At the time of their discovery,
it was unclear whether such theories described
an ordinary quantum field theory with a spectrum of local operators, or
instead involved a more exotic non-local structure.

Reference \cite{Seiberg:1996qx} convincingly argued that these seemingly exotic theories
are actually strongly coupled conformal field theories in disguise. Crucial
to this analysis is the presence of a moduli space of vacua, as appears in
theories with sufficient supersymmetry. All known 6D\ CFTs have either eight
or sixteen real supercharges, so we focus our discussion on superconformal field theories
(SCFTs). Even with the aid of supersymmetry, as of this writing,
no Lagrangian description is known for any interacting 6D SCFT.

In spite of this fact, the mere existence of such theories leads to a number
of important conceptual and \textquotedblleft practical\textquotedblright%
\ uses. Conceptually, there is the feature that although these
theories seem to involve strings with vanishing tension, they are nevertheless
described by a local quantum field theory. Additionally, the absence of a
Lagrangian description challenges some of the conventional approaches to
understanding quantum field theory typically espoused in textbooks.

From a practical standpoint, these 6D\ SCFTs also serve as the
\textquotedblleft master theories\textquotedblright\ for understanding a wide
variety of lower-dimensional strongly coupled phenomena. Perhaps the best
known example of this kind is flat $T^{2}$ compactification of a
6D\ SCFT with sixteen real supercharges. This yields a 4D theory with sixteen
supercharges, namely $\mathcal{N}=4$ super Yang-Mills. The complex structure
of the $T^{2}$ translates to the holomorphic gauge coupling $\tau$ of the 4D
theory, and the celebrated Montonen-Olive duality \cite{Montonen:1977sn, Goddard:1976qe, Osborn:1979tq}
is interpreted as the redundancy in specifying the shape of the torus under transformations
$\tau\rightarrow\tau+1$ and $\tau\rightarrow-1/\tau$ (see reference \cite{Vafa:1997mh}).

More recent examples include the study of such 6D theories on Riemann
surfaces \cite{Witten:1997sc, Gaiotto:2009we}. Here again, changes in the shape of the Riemann surface translate to
highly non-trivial duality transformations in the 4D effective field theory.
Similar insights have followed for compactifications on other spaces, leading
to a beautiful correspondence between the structure of higher-dimensional
theories and their lower-dimensional counterparts.

Given their central role in a number of theoretical investigations, it
therefore seems important to provide a more systematic starting point for the
construction and study of 6D\ SCFTs. The aim of this review is to provide a
point of entry to this fast growing area of investigation.

Now in spite of the fact that these are quantum field theories, it turns out
that the only known methods for explicitly constructing these theories
inevitably involve taking a suitable singular limit of a string theory
construction, namely a \textquotedblleft top down\textquotedblright\ approach.
There are, of course very important consistency conditions which
\textquotedblleft bottom up\textquotedblright\ considerations impose, and we
shall explain how these considerations naturally mesh with the string theory
picture. In this vein, there is accumulating evidence that
the most flexible option for realizing a broad class of stringy vacua is based
on F-theory, a strongly coupled phase of type IIB\ string theory. Indeed, at
present, all known 6D\ SCFTs can be accommodated in this framework, and there
is even a conjectural classification of 6D\ SCFTs based on this approach.

The plan of this review article is as follows. First, in Section \ref{sec:WHATIS} we
discuss in general terms what is meant by a 6D\ SCFT, as well as ways
one might attempt to realize such a conformal fixed point. We follow this with
a short explanation of why, prior to the use of stringy methods, such theories
were long thought not to exist, and we provide some canonical examples. Section \ref{sec:BOTTOM}
reviews some of the known bottom up constraints on such theories, in particular the tight
structure of anomalies in chiral 6D supersymmetric theories. Section \ref{sec:FTHEORY} introduces
some preliminary aspects of F-theory, and Section \ref{sec:FAGAIN} explains the conditions
necessary to realize a 6D SCFT in this framework. In Section \ref{sec:CLASSIFY}
we review the central elements in the classification of F-theory backgrounds that yield a 6D SCFT, and we introduce a novel swampland conjecture for quantum field theories.
We then turn to the calculation of various properties of such theories, including the anomaly polynomial
of a 6D SCFT in Section \ref{sec:ANOMPOLY}, and the structure of RG flows in Section \ref{sec:RGFLOWS}.
Progress on the counting of microscopic states in these theories is reviewed in Section \ref{sec:INDICES}.
Section \ref{sec:CONC} summarizes the main elements reviewed in this article and briefly discusses particularly pressing
areas for future investigation. A number of mathematical details used in the study of these theories
are reviewed in a set of Appendices, including (previously unpublished) expressions for the anomaly polynomials of a number of 6D SCFTs.

\textbf{Omissions:} Due to space constraints and the fact that some areas of 6D\ SCFTs
are still undergoing rapid investigation, we have chosen to omit some topics
from our discussion. These include a detailed discussion of compactifications
of 6D\ SCFTs on various lower-dimensional spacetimes \cite{Gaiotto:2015usa,
Ohmori:2015pua, DelZotto:2015rca, Franco:2015jna, Ohmori:2015pia, Coman:2015bqq,
Morrison:2016nrt, Apruzzi:2016nfr, Razamat:2016dpl, Bah:2017gph,
DelZotto:2017pti, Kim:2017toz, Hassler:2017arf, Bourton:2017pee, Kim:2018bpg, Apruzzi:2018oge}
as well as the application of the conformal bootstrap to such theories \cite{Beem:2014kka,
Beem:2015aoa, Chang:2017xmr, Bobev:2017jhk}. We have also chosen to omit
the recent classification of $AdS_7$ supergravity backgrounds, as the methods are somewhat orthogonal
to the main elements of this review article \cite{Apruzzi:2013yva, Gaiotto:2014lca,
Apruzzi:2015wna, Cremonesi:2015bld, Passias:2015gya, Passias:2016fkm, Apruzzi:2017nck}.
Each of these areas is currently very active, and merit their own review articles.
Finally, we will aim to emphasize the conceptual elements
which are important from the modern perspective, and not try to reconstruct
a chronological account. Nevertheless, for a partial list of
references to what are now recognized as top down constructions of 6D SCFTs
from the 1990's, see e.g. \cite{Witten:1995zh, Strominger:1995ac,
Ganor:1996mu, Seiberg:1996vs, Witten:1996qb, Morrison:1996pp, Witten:1996qz,
Bershadsky:1996nu, Brunner:1997gf, Blum:1997mm, Blum:1997fw, Aspinwall:1997ye, Intriligator:1997dh, Hanany:1997gh}.

\section{What is a 6D\ SCFT? \label{sec:WHATIS}}

In this section we discuss in general terms what a 6D superconformal field
theory (SCFT)\ is, why they are difficult to study, and as a related point,
why they are interesting to study.

To begin, let us recall the definition of a conformal field theory in $D>2$
spacetime dimensions.\footnote{In $D=2$ SCFTs, it is best to couch the discussion in terms of
representations of the Virasoro algebra.} Our review of the algebra follows
the discussion of reference \cite{Minwalla:1997ka},
to which we refer the interested reader for additional details.

A CFT is a quantum field theory which enjoys, in addition to the usual Lorentz
and translation symmetries (known as the Poincar\'{e} symmetries) an
enhancement to the symmetry algebra $\mf{so}(D,2)$. For it to be a sensible theory,
we require the existence of a stress energy tensor, and some collection of
local operators which satisfy non-trivial correlation functions. These
correlators are subject to a number of constraints, as stems from the presence
of conformal symmetry. One can supplement this bosonic symmetry by a
supersymmetry. The appearance of such fermionic conserved charges imposes
additional constraints on the structure of a quantum field theory, and
accordingly, allows one to make more precise statements.

In a general supersymmetric theory, we have fermionic symmetry generators $Q$
which transform as a spinor of the Lorentz algebra. Schematically, these are
the \textquotedblleft square root\textquotedblright\ of a translation,
$\{Q,Q\}\sim P$. In a conformal field theory, we introduce additional
generators, the special conformal transformations $K$. The \textquotedblleft
square root\textquotedblright\ of these generators are also fermionic
generators $S$, and satisfy the schematic relation $\{S,S\}\sim K$. The
combined conditions of conformal symmetry and supersymmetry can all be
packaged in terms of a corresponding superconformal algebra. The
classification of superconformal algebras was carried out in 1977 by Nahm
\cite{Nahm:1977tg}, who found that such superalgebras only exist for $D\leq6$.

As reviewed for example in \cite{Minwalla:1997ka}, the main issue is that there is a general
classification of Lie superalgebras, and the additional condition of a
superconformal theory is that we need $G_0$, the bosonic (even) part of the superalgebra
to have a spinorial representation on $G_1$, the fermionic (odd) part of the superalgebra.\footnote{Recall that in a
superalgebra, we have a $\mathbb{Z}_2$ grading
of all elements. The $\mathbb{Z}_2$ even parts are the bosonic ``diagonal blocks'' and the $\mathbb{Z}_2$
odd parts are the fermionic ``off-diagonal blocks.''} This can only be accomplished when $D\leq6$, and in the special case
$D=6$ it \textquotedblleft just barely\textquotedblright\ happens because of
the triality automorphism of $\mf{so}(8,\mathbb{C})$, which allows us to exchange a vectorial
representation of $G_{0}$ on $G_{1}$ (the supercharges) with a spinorial representation.
Starting at $D>6$, no miracles occur and we cannot realize a superconformal algebra. There is no such constraint on
interacting CFTs without supersymmetry, but on the other hand, there are no
known examples either. This already indicates the privileged role of
$D=6$:\ it is the highest spacetime dimension in which we can combine
supersymmetry and conformal symmetry. We now specialize further to this case.

Appendix \ref{app:SCA} provides a brief review of the 6D superconformal algebra.
Here, we summarize the more general properties which will feature in later discussions of this review.
In six dimensions, the available options for superconformal algebras are
$\mf{osp}(6,2|\mathcal{N})$, which has a bosonic subalgebra:%
\begin{equation}
\mf{osp}(6,2|\mathcal{N})\supset \mf{so}(6,2)\times \mf{sp}(\mathcal{N}).
\end{equation}
The second factor denotes the R-symmetry of the SCFT. In our conventions,
$\mf{sp}(1)\simeq \mf{su}(2)$ and $\mf{sp}(2)\simeq \mf{so}(5)$. The $Q$'s and $S$'s transform in
the \textquotedblleft off-diagonal\textquotedblright\ fermionic (i.e., odd) blocks of the
superalgebra. Under the Lorentz subalgebra $\mathfrak{so}(5,1)$, $Q$ transforms with
positive chirality while $S$ transforms with negative chirality. These combine
to form a single spinor of $\mf{so}(6,2)$ which transforms with positive chirality.
Additionally, these spinors rotate in the fundamental representation of
$\mf{sp}(\mathcal{N})$.

In six dimensions, the existence of a supermultiplet containing the stress
tensor imposes the important restriction that $\mathcal{N}\leq2$, so the theory has at most sixteen real supercharges \cite{Cordova:2016emh}.\footnote{This is a
somewhat different argument from the \textquotedblleft standard
lore\textquotedblright\ that a theory with more than sixteen real supercharges
necessarily must contain a graviton multiplet. The loophole in such an
argument is that it presumes the existence of weakly coupled particle-like
states, and this condition is definitely \textit{not} satisfied in any known
6D\ SCFT!} The SCFTs in six dimensions thus come in two types:
$\mathcal{N}=2$ or $\mathcal{N}=1$, which respectively denote the $(2,0)$
and $(1,0)$ theories. The reason for the notation is that in six spacetime
dimensions, we can simultaneously impose a pseudo-Majorana condition and a
chirality condition, giving a pseudo-Majorana-Weyl spinor:%
\begin{equation}
Q_{\alpha,i}=\Omega_{ij}C_{\alpha\beta}Q^{\beta,j\dag}.
\end{equation}
Here, $C_{\alpha\beta}$ is a suitable charge conjugation matrix generated from
the Clifford algebra for $\mf{so}(5,1)$, and $\Omega_{ij}$ is the analogous
quantity for the R-symmetry algebra factor. So, we can either have one chiral
$Q$ or two.

It is natural to ask whether we can actually realize concrete examples of 6D CFTs. A trivial
example of a 6D\ CFT is the free field theory of a real scalar with Lagrangian
density:\footnote{We use metric conventions with mostly $+$'s.}%
\begin{equation}
L=-\frac{1}{2}\left(  \partial\phi\right)  ^{2}.
\end{equation}
The scaling dimension for the scalar is precisely two; it is a free
field and saturates the unitarity bound for a scalar operator of a 6D\ CFT.

To realize an interacting CFT, we might attempt to perturb this system by a
real scalar potential. Since we want interactions in the infrared, we cannot
perturb by a quadratic term as this would give a mass to the field $\phi$. So,
we can start with cubic or higher order terms.
For quartic and higher terms, however, we see that such perturbations are
irrelevant as they have scaling dimension at least eight. Thus, the best we can
hope for is a cubic potential energy density $V(\phi)\sim\phi^{3}$. This is
perfectly suitable at the level of perturbation theory, but there is clearly a big problem
with the theory non-perturbatively: taking $\phi<0$ we
can make the potential energy density arbitrarily negative. This is
problematic since a CFT certainly requires the existence of a stable ground
state. A possible \textquotedblleft way out\textquotedblright\ is to then
reintroduce the higher order interaction terms. For these to play a role in
the analysis, however, we need to pass beyond perturbations of the free field
fixed point. This is again problematic since we are dealing with strong
coupling effects over which we have little control. Effectively, we face the
quandary that a Lagrangian description (if it even exists) needs to include
irrelevant operators with very large coefficients \cite{Seiberg:1996qx}.

The situation changed dramatically with the second superstring revolution and
the appearance of BPS\ solitons which are exactly stable. Using these new
ingredients, it became possible for the first time to argue for the existence
of new strongly coupled six-dimensional theories decoupled from gravity. To
see how this comes about, it is helpful to consider a few explicit examples.
The first constructions appeared in references \cite{Witten:1995zh, Strominger:1995ac}
and their interpretation as a \textquotedblleft conventional\textquotedblright%
\ quantum field theories appeared in reference \cite{Seiberg:1996qx}.

The main idea in all of these constructions is to isolate the quantum field
theory sector of a string construction. Since we are dealing with a
six-dimensional quantum field theory, this means there are four extra
dimensions in the context of type IIB\ string theory as well as F-theory, and
five extra dimensions in the context of M-theory. For example, when there are
four extra dimensions described by a manifold $M_{\text{extra}}$, the value of the
six-dimensional Newton's constant is set by the volume of these extra
dimensions according to the scaling relation:%
\begin{equation}
\frac{\text{Vol}(M_{\text{extra}})}{\ell_{\ast}^{8}}\sim\frac{1}{G_{6D}},
\end{equation}
with $\ell_{\ast}$ a short distance scale. To decouple gravity we need
to take the limit where the volume of the extra
dimensions is extremely large compared to $\ell_{\ast}$. In such a limit,
there is no coupling between gravity and the field theory degrees of freedom
localized on various compact subspaces.

With this limit in mind, we now turn to some examples. We first discuss
the realization of all known $\mathcal{N} = (2,0)$ 6D\ SCFTs, and then turn to examples of
$\mathcal{N} = (1,0)$ 6D\ SCFTs. The main hallmark of all these constructions is that by
appropriate tuning in the moduli space of vacua, it is possible to reach a
limit where effective strings become tensionless. The absence of a mass scale
provides strong evidence that we have a theory without distance scales, and thus a conformal fixed point. For such an interpretation to be compatible with the
existence of a local quantum field theory, this also suggests that the
effective strings are simply emergent objects which only appear at
long distances. Indeed, one of the hallmarks of a CFT is
that it is a \textit{local} theory. We will return to this issue in subsequent sections.

\subsection{The $\mathcal{N}=(2,0)$ Theories \label{ssec:OLDIES}}

One of the first examples from reference \cite{Witten:1995zh} involves type IIB\ string
theory on a non-compact Calabi-Yau twofold. In some cases, this can be viewed
as a local patch in a compact K3 surface, though there are other examples which do
not necessarily embed in a compact geometry. The Calabi-Yau condition ensures
that we retain half of the supersymmetry of IIB\ in flat space so since we
started with the IIB string theory with $32$ real supercharges preserved in
flat space, we have a six-dimensional effective theory with $16$ real
supercharges. These supercharges assemble into two spinors of the same
chirality, so we have an $\mathcal{N}=(2,0)$ supersymmetric theory.

Now, inside this K\"{a}hler surface, we suppose that there is some collection
of $\mathbb{CP}^{1}$'s, which are sufficiently small that all details of the
geometry far from these curves can be neglected. In such a limit, we can
parameterize the geometry in terms of the curves and the line bundles over each of them.
For the case of a single curve $\Sigma$, we express the line bundle
as $\mathcal{O}(-n)\rightarrow\mathbb{CP}^{1}$, and we refer to $\Sigma$ as a
\textquotedblleft$-n$ curve\textquotedblright\ since it has self-intersection number
$\Sigma \cap \Sigma = -n$. The Calabi-Yau condition locally requires us to cancel the curvature of
the line bundle against that of the base curve so that $n=2$. For additional
details on the local geometry of $-n$ curves, see Appendix \ref{app:LOCO}.

Starting with the local geometry $\mathcal{O}(-2)\rightarrow
\mathbb{CP}^{1}$, we can ask what sort of quantum field theory we expect
to generate. Kaluza-Klein reduction of the type IIB\ chiral four-form yields
(upon integrating over the curve) an anti-chiral two-form in the
six-dimensional theory. That is to say, the three-form field strength $H$ of
this two-form $B$ is anti-self-dual:%
\begin{equation}
H=-\ast H.
\end{equation}
Additionally, there is the ``breathing mode" which parameterizes the volume of the curve. Though we are
decoupling gravity, this mode survives and is a dynamical field.\footnote{The rule of thumb
is that the volume of cycles with dimension equal to half or more than the dimension of the
internal space remain dynamical even in a decoupling limit. This is due to the fact that they have
normalizable wave functions.}
Together, the two-form and the breathing mode comprise the bosonic components of an $\mathcal{N} = (1,0)$ supermultiplet known as the
\textquotedblleft tensor multiplet,\textquotedblright\ and we shall present a
more systematic discussion of these considerations in Section \ref{sec:BOTTOM}.

Now, as is well-known from string theory, the chiral four-form couples to a
dynamical BPS\ soliton, the D3-brane. The reduction of this object on our
compact curve leaves us with an effective string in the six-dimensional
effective theory, as is clear from the following picture of the dimensions:%
\begin{equation}%
\begin{tabular}
[c]{|c|c|c|c|c|c|c|c|c|c|c|}\hline
&\multicolumn{6}{|c|}{$\mathbb{R}^{5,1}$}&\multicolumn{4}{|c|}{$\mathbb{CP}^1\leftarrow \mathcal{O}(-n)$}\\\hline
& $0$ & $1$ & $2$ & $3$ & $4$ & $5$ & $6$ & $7$ & $8$ & $9$\\\hline
D3 & $\times$ & $\times$ & $\cdot$ & $\cdot$ & $\cdot$ & $\cdot$ & $\times$ &
$\times$ & $\cdot$ & $\cdot$\\\hline
\end{tabular}
,
\end{equation}
where here, a $\times$ indicates a direction filled by the D3-brane, and a
$\cdot$ denotes a direction where it is localized at a particular value of the
coordinate. The $\mathbb{R}^{5,1}$ coordinates are $0,...,5$, the local
coordinates of the $\mathbb{CP}^{1}$ are $6,7$ and the (real) coordinates
of the fibers of the line bundle are $8,9$. So in other words, the reduction of the chiral
four-form to an anti-chiral two-form directly couples to
effective strings in the six-dimensional theory.

Such strings have tension set by the volume of the $\mathbb{CP}^{1}$ they wrap:%
\begin{equation}
T_{\text{eff}}=\frac{\text{Vol}(\mathbb{CP}^{1})}{\ell_{\ast}^{4}}.
\label{eq:Tension}
\end{equation}
So, for a large volume $\mathbb{CP}^{1}$, these heavy objects are non-dynamical. Let us note that the
tension formula used here is exact due to the large amount of supersymmetry
preserved by this background.

Now, in the limit where the volume of the $\mathbb{CP}^{1}$ collapses, we see
that the tension of this effective string vanishes. This suggests the presence
of a \textquotedblleft tensionless string,\textquotedblright\ indicating a new
set of light degrees of freedom have entered the low energy effective field
theory. In this limit, the geometry of the four extra dimensions also becomes
singular, being replaced by $\mathbb{C}^{2}/\mathbb{Z}_{2}$. It is well-known
that string theory on this sort of orbifold singularity is still sensible
\cite{Dixon:1985jw, Dixon:1986jc}, so we can still trust our physical picture.

Continuing in this vein, we can ask what happens if we have more than one
collapsing $\mathbb{CP}^{1}$, each with self-intersection $-2$. The
intersection pairing between a basis of such cycles is captured by a symmetric
\textquotedblleft adjacency matrix\textquotedblright\ with $-2$'s along the
diagonal, and positive integers on the off-diagonal entries:%
\begin{equation}
A_{ji}=A_{ij}=\Sigma_{i}\cap\Sigma_{j}.
\end{equation}
In this case, the condition that we can simultaneously contract all curves is
that the normal bundle to the full configuration of curves is negative
definite:%
\begin{equation}
A<0\text{,}%
\label{eq:Anegdef}
\end{equation}
and that all off-diagonal entries are zero or one.

\begin{figure}[t!]
\centering
\includegraphics[trim = 0cm 2.5cm 0cm 2.5cm, clip=true, scale=0.5]{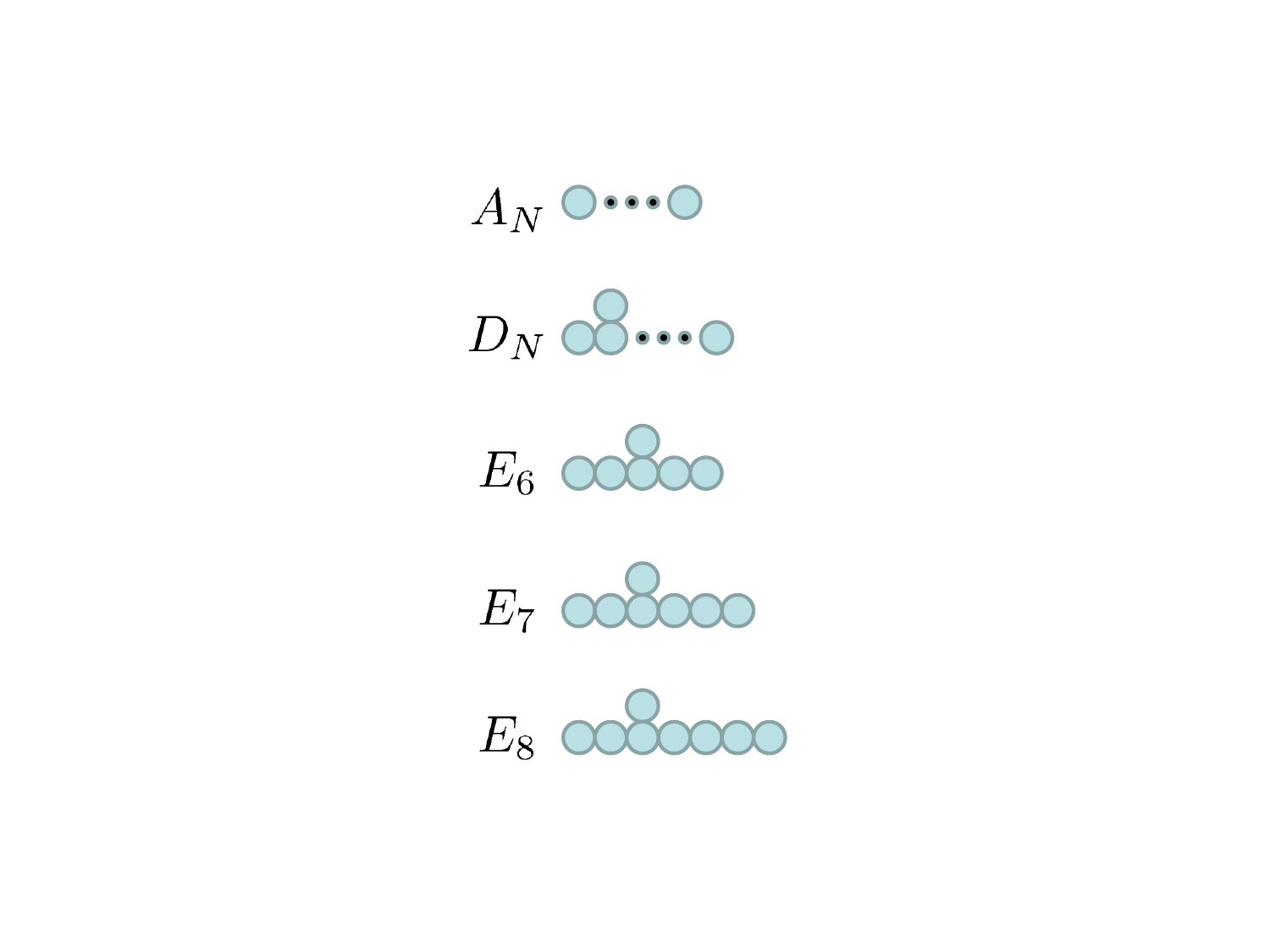}
\caption{Depiction of the ADE Dynkin diagrams. Geometrically, each circle denotes a curve of
self-intersection $-2$, and neighboring curves intersect once. For each graph, the subscript indicates the total number
of independent curves appearing in the resolution of the corresponding ADE singularity.}%
\label{fig:ADE}%
\end{figure}

Such matrices have all been classified and are associated
with the Dynkin diagrams of the simply laced algebras, namely $A_{n}$, $D_{n}$
and $E_{n}$. See figure \ref{fig:ADE} for a depiction of these Dynkin diagrams.

In this description, we visualize each node of the Dynkin diagram as
one of our $\mathbb{CP}^{1}$'s, and the links between nodes correspond to
curves which have intersection number one. In other words, the off-diagonal
entries of $A_{ij}$ are $+1$ or $0$. In the limit where the $\mathbb{CP}^{1}%
$'s collapse to zero size, we again reach a six-dimensional theory with
tensionless strings, and the IIB\ background is best described as an orbifold
singularity $\mathbb{C}^{2}/\Gamma$, where $\Gamma$ is a discrete subgroup of
$SU(2)$. Such discrete subgroups are in one to one correspondence with the
ADE\ classification of simply laced algebras, a fact which is part of the
McKay correspondence \cite{MR604577}.

What sort of theory have we produced in taking such a limit? At first glance
it appears problematic to have generated a theory with tensionless
strings, since one might worry that the resulting theory is then non-local. On
the other hand, the absence of any distance scales in this decoupling limit
suggests instead that we may have simply realized a quantum field theory which
is conformally invariant. We will see additional evidence for the latter point
of view shortly. From this perspective, we see that the classification of
IIB\ backgrounds which can produce an SCFT with $\mathcal{N}=(2,0)$
supersymmetry is neatly summarized by ADE\ Dynkin diagrams.

At this point one might naturally ask whether there might be other ways to
generate a 6D SCFT with $\mathcal{N}=(2,0)$ supersymmetry. First of all, we
see that type IIA\ on the same background geometries will fail to produce any
interesting examples. Indeed, although this background preserves sixteen real
supercharges, they do not assemble into spinors of the same chirality. Rather,
we obtain a theory with $\mathcal{N}=(1,1)$ supersymmetry. Additionally, note
that in type IIA\ string theory, there is no D3-brane wrapping the
two-cycles. Instead, we have D2-branes wrapping the curves, which
result in point particles in the six-dimensional effective theory. This
provides a stringy way to engineer a 6D gauge theory with gauge group of ADE\ type,
but not a 6D SCFT.

We can, however, engineer examples by either using NS5-branes (the T-dual of a
$\mathbb{C}^{2}/\mathbb{Z}_{n}$ singularity \cite{Ooguri:1995wj}) or by working in terms of
coincident M5-branes in M-theory \cite{Strominger:1995ac}, the strong coupling lift of type
IIA\ strings. Along these lines, consider a collection of $n$
M5-branes filling the first factor of the background $\mathbb{R}^{5,1}%
\times\mathbb{R}^{5}$. Separating each M5-brane from one another, we see that
there can be M2-branes suspended between these M5-branes, as in the following
picture:%
\begin{equation}%
\begin{tabular}
[c]{|c|c|c|c|c|c|c|c|c|c|c|c|}\hline
& $0$ & $1$ & $2$ & $3$ & $4$ & $5$ & $6$ & $7$ & $8$ & $9$ & $10$\\\hline
M5 & $\times$ & $\times$ & $\times$ & $\times$ & $\times$ & $\times$ & $\cdot$
& $\cdot$ & $\cdot$ & $\cdot$ & $\cdot$\\\hline
M5$^{\prime}$ & $\times$ & $\times$ & $\times$ & $\times$ & $\times$ &
$\times$ & $\cdot$ & $\cdot$ & $\cdot$ & $\cdot$ & $\cdot$\\\hline
M2 & $\times$ & $\times$ & $\cdot$ & $\cdot$ & $\cdot$ & $\cdot$ & $\times$ &
$\cdot$ & $\cdot$ & $\cdot$ & $\cdot$\\\hline
\end{tabular}
,
\end{equation}
where in the above, we have separated the M5-branes in the direction with local
coordinate labelled by the number $6$. This again looks like an effective string in the six
directions spanned by $0,...,5$, with tension:%
\begin{equation}
T_{\text{eff}}=\frac{\text{dist}(M5,M5^{\prime})}{\ell_{\ast}^{3}},
\end{equation}
where dist$(M5,M5^{\prime})$ refers to the distance between the M5-branes. In
the limit where the M5 branes become coincident, this string becomes
tensionless. This provides another way to realize the $A_{1}$ theory, as can
be seen, for example, by compactifying on a transverse circle and dualizing to
the geometry $\mathbb{C}^{2}/\mathbb{Z}_{2}$.\footnote{There is a subtlety
here which we are glossing over:\ what becomes of the center of mass degree of
freedom on the M5-branes in the IIB\ picture? This has to do with the global
structure of the resulting 6D theory and the existence / absence of a
partition function for the theory when placed on various background
geometries. We will return to this point later in Section \ref{sec:INDICES}.} Continuing in
this way, one can realize the $A_{n}$ series of $\mathcal{N}=(2,0)$ 6D SCFTs.
If one entertains the possibility of orientifold M5-branes (see e.g. \cite{Hanany:1999jy, Hanany:2000fq, Bergman:2001rp})
one can also engineer the D-type series. The E-type series cannot be engineered using this approach, however.

Even so, a particularly elegant feature of the M5-brane construction is the
geometric realization of the $\mf{sp}(2)$ R-symmetry algebra of the superconformal
field theory. Observe that $\mf{sp}(2)\simeq \mf{so}(5)$ is nothing but the isometry
algebra for the $\mathbb{R}^{5}$ factor transverse to the M5-branes. Note also
that it is a property which only emerges when we tune to the putative SCFT
point: we need to bring all M5-branes to the same point in the $\mathbb{R}%
^{5}$ factor to preserve this symmetry. This is harder to
see in the IIB\ picture, which speaks to the relative merits of the two descriptions.

Now, as we have already mentioned, the constructions of references \cite{Witten:1995zh, Strominger:1995ac}
present the intriguing possibility of realizing a superconformal
field theory in six dimensions. Indeed, in the scaling limit just discussed,
the absence of any mass scales provides quite suggestive evidence in favor of
this proposal, as noted in \cite{Seiberg:1996qx}. In the theories of reference \cite{Seiberg:1996qx},
a rather conventional gauge theory description emerges away from the fixed point (by passing to the tensor branch, a point
we return to later). Passing to the point
of strong coupling in the moduli space then takes us back to the UV fixed point, providing compelling evidence that we are
dealing with a conventional local field theory.

Additional evidence in support of this proposal comes from the AdS/CFT correspondence \cite{Maldacena:1997re, Gubser:1998bc, Witten:1998qj},
at least for the A-type theories. The reason is that the near horizon limit of $N$ M5-branes leaves us
with a supergravity background in M-theory $AdS_{7}\times S^{4}$ with $N$
units of four-form flux threading the $S^{4}$. The dual is nothing but our
theory of M5-branes in a suitable decoupling limit, so it again strongly
suggests that we have engineered a 6D\ SCFT. Additional evidence for this
interpretation has also been found using methods from the conformal bootstrap
\cite{Beem:2015aoa}. This provides additional supporting evidence that such conformal
fixed points truly do exist, and are correctly engineered by the string constructions.

\subsection{Examples of $\mathcal{N}=(1,0)$ Theories \label{ssec:EXAMPLES}}

So far, we have focused on 6D\ SCFTs with maximal supersymmetry. From the
classification of superconformal algebras we should also expect to generate
examples with reduced supersymmetry. A conceptual way to produce examples with
reduced supersymmetry is to take our M5-brane examples and place them on
singular backgrounds which already break half of the supersymmetry. The non-trivial background geometry (such as a non-compact
Calabi-Yau twofold) breaks half the supersymmetry of M-theory, and the M5-branes break an additional half, leaving us with eight real
supercharges, as required for $\mathcal{N} = (1,0)$ supersymmetry. Here, we discuss
some illustrative examples which will show up repeatedly in later discussions.

Perhaps the best studied example of such a theory is given by $N$ M5-branes
probing an $E_{8}$ nine-brane in M-theory. Recall that in M-theory, the
$E_{8}$ wall arises as a localized defect in the 11D spacetime at the boundary
of $\mathbb{R}^{9,1}\times S^{1}/\mathbb{Z}_{2}$, as required by anomaly
cancellation considerations \cite{Horava:1995qa, Horava:1996ma}. Since we are interested in
decoupling gravity, we shall primarily focus on the local geometry as
described by $\mathbb{R}/\mathbb{Z}_{2}$, where the $\mathbb{Z}_{2}$ acts by
reflection about the origin. On this nine-brane we have at low energies a
ten-dimensional super Yang-Mills theory with gauge group $E_{8}$. This appears as
a flavor symmetry in six dimensions because it wraps a non-compact four-manifold.

To get a 6D\ SCFT, we now introduce $N$\ M5-branes in the vicinity of the
wall (see e.g. \cite{Ganor:1996mu, Seiberg:1996vs,
Witten:1996qb, Morrison:1996pp, Witten:1996qz}).
Far from the wall, we see the same spectrum of M2-branes stretched
between the M5-branes. However, there are additional states that arise as we
bring the M5-branes close to the wall, as can be seen from the following
picture:%
\begin{equation}%
\begin{tabular}
[c]{|c|c|c|c|c|c|c|c|c|c|c|c|}\hline
& $0$ & $1$ & $2$ & $3$ & $4$ & $5$ & $6$ & $7$ & $8$ & $9$ & $10$\\\hline
M5 & $\times$ & $\times$ & $\times$ & $\times$ & $\times$ & $\times$ & $\cdot$
& $\cdot$ & $\cdot$ & $\cdot$ & $\cdot$\\\hline
M9 & $\times$ & $\times$ & $\times$ & $\times$ & $\times$ & $\times$ &
$\times$ & $\times$ & $\times$ & $\times$ & $\cdot$\\\hline
M2 & $\times$ & $\times$ & $\cdot$ & $\cdot$ & $\cdot$ & $\cdot$ & $\cdot$ &
$\cdot$ & $\cdot$ & $\cdot$ & $\times$\\\hline
\end{tabular}
\ .
\end{equation}
Namely, we have an M2-brane which can stretch from the M5-branes to the
nine-brane (denoted as M9 above). In the limit where the M5-branes sit on top
of the nine-brane, the tension of these effective strings again vanishes, so
by the same sort of scaling arguments presented earlier, we conclude
that we have a 6D\ SCFT with $\mathcal{N}=(1,0)$ supersymmetry. The
$\mf{sp}(1)\simeq \mf{su}(2)$ R-symmetry of the SCFT is again cleanly realized as the
$\mf{su}(2)_{R}$ factor in the $\mf{so}(4)\simeq \mf{su}(2)_{L}\times \mf{su}(2)_{R}$ isometries
of the $6789$ geometry. In addition to the appearance of the global symmetry
$\mf{su}(2)_{L}$, we also see that the theory enjoys an $E_{8}$ global symmetry
realized on the nine-brane. This theory is often referred to as the
\textquotedblleft E-string theory\textquotedblright\ because, away from the
fixed point, it involves strings which enjoy a global $E_{8}$ symmetry.

As another example (see e.g. \cite{Brunner:1997gf, Blum:1997mm, Blum:1997fw, Intriligator:1997dh}
and more recently \cite{Gaiotto:2014lca, DelZotto:2014hpa}), we can consider M5-branes probing the orbifold singularity
$\mathbb{C}^{2}/\Gamma$ for $\Gamma$ a discrete subgroup of $SU(2)$. Whereas IIB\
string theory on this orbifold yields a 6D\ SCFT, IIA\ string theory instead realizes a 6D gauge
theory with $\mathcal{N}=(1,1)$ supersymmetry. The lift to M-theory realizes a 7D super Yang-Mills theory with
$\mathcal{N}=1$ supersymmetry and an ADE\ gauge algebra, with W-bosons of the
gauge algebra arising from M2-branes wrapped on the collapsing $\mathbb{CP}^{1}$'s
of the geometry.

\begin{figure}[t!]
\centering
\includegraphics[trim = 0cm 2.5cm 0cm 2.5cm, clip=true, scale=0.5]{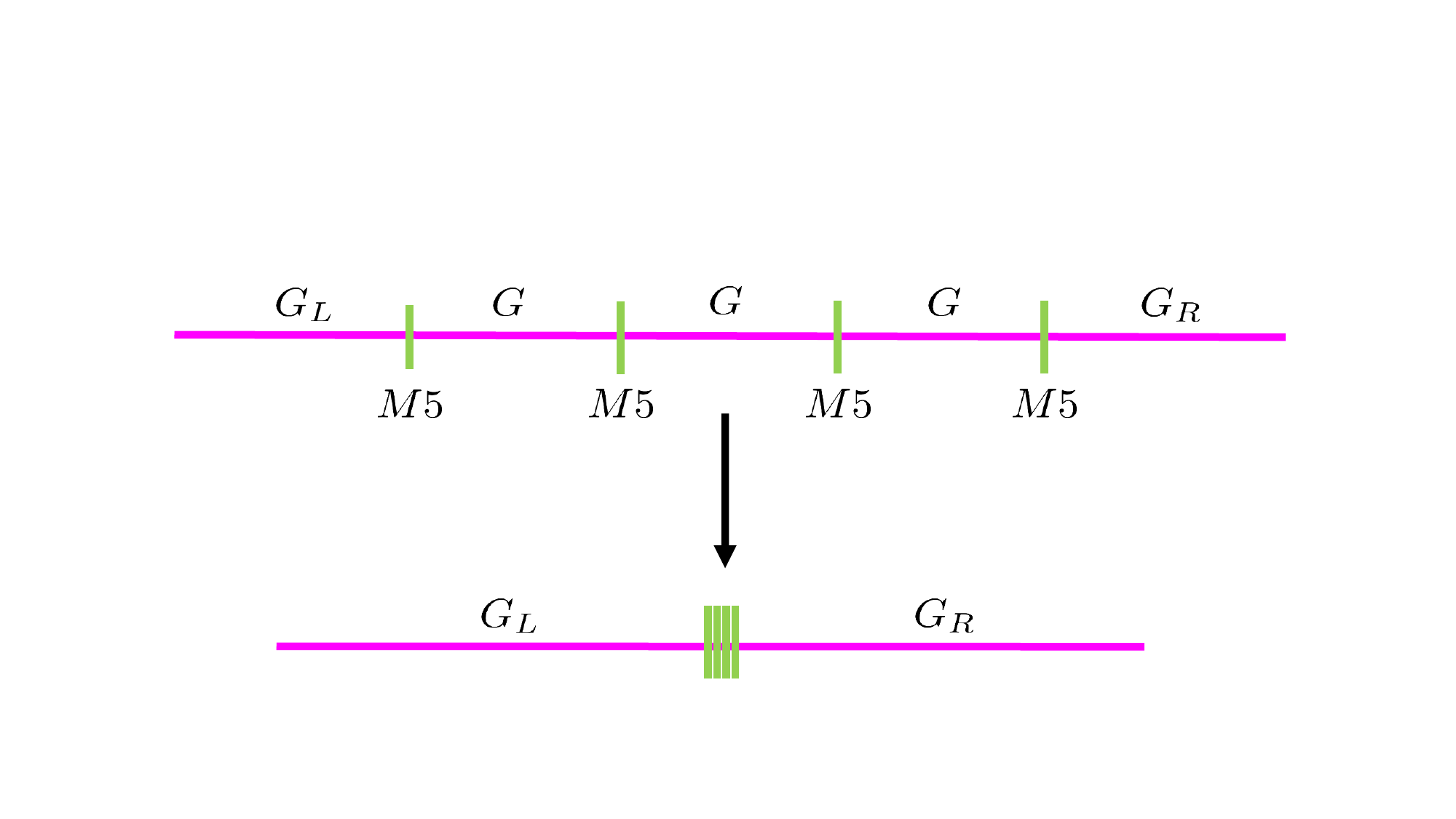}
\caption{Depiction of the 6D SCFTs obtained from M5-branes probing an ADE singularity. On the partial
tensor branch we keep the M5-branes at the orbifold singularity but separate them along a common
real line. The conformal fixed point is reached by sending all the M5-branes to the same point on this
interval. On the partial tensor branch, we have $\mathcal{N} = 1$ 7D super Yang-Mills theory with gauge group $G$
coupled to a collection of domain walls as given by M5-branes. The M5-branes chop up the transverse line into finite intervals,
as well as two semi-infinite intervals. The left and right therefore produce flavor symmetries $G_L \simeq G$ and $G_R \simeq G$ on the left
and right of the brane configuration.}%
\label{fig:6Dedge}%
\end{figure}

To realize a 6D\ SCFT, we now introduce M5-branes into this geometry. These
objects appear as domain walls in the 7D effective theory, cutting the transverse space
\textquotedblleft in half\textquotedblright\ along the $\mathbb{R}_{\bot}$
factor of the geometry $\mathbb{R}^{5,1}\times\mathbb{R}_{\bot}\times
\mathbb{C}^{2}/\Gamma$. The description is conveniently summarized by the
following picture:%
\begin{equation}%
\begin{tabular}
[c]{|c|c|c|c|c|c|c|c|c|c|c|c|}\hline
& $0$ & $1$ & $2$ & $3$ & $4$ & $5$ & $6$ & $7$ & $8$ & $9$ & $10$\\\hline
M5 & $\times$ & $\times$ & $\times$ & $\times$ & $\times$ & $\times$ & $\cdot$
& $\cdot$ & $\cdot$ & $\cdot$ & $\cdot$\\\hline
7D SYM & $\times$ & $\times$ & $\times$ & $\times$ & $\times$ & $\times$ &
$\times$ & $\cdot$ & $\cdot$ & $\cdot$ & $\cdot$\\\hline
\end{tabular}
.
\end{equation}
We expect to realize a 6D\ SCFT for the same reasons previously
outlined:\ There are strings coming from M2-branes stretched between one M5-brane and another, and in the limit where they are all coincident, these become
tensionless. Note that this argument works for $N>1$ M5-branes. In fact, it is
natural to expect that at least for the D- and E-type orbifold singularities,
there is always an \textquotedblleft image M5-brane\textquotedblright\ so in
this case an M2-brane stretched from an M5-brane to its \textquotedblleft
image\textquotedblright\ ought to also produce a 6D\ SCFT. See figure \ref{fig:6Dedge}
for a depiction of this construction. We shall expand on
this heuristic analysis when we turn to the F-theory description of such theories.

In this geometry, we observe that the isometries of $\mathbb{R}^{4}%
\simeq\mathbb{C}^{2}$ are $\mf{so}(4)\simeq \mf{su}(2)_{L}\times \mf{su}(2)_{R}$, where we
have embedded the discrete subgroup $\Gamma$ in the $\mf{su}(2)_{L}$ factor. So, we
can identify the $\mf{su}(2)_{R}$ factor with the R-symmetry of the theory.
Additionally, we see that in the special case where $\Gamma$ is abelian, the
commutant subgroup inside $\mf{su}(2)_{L}$ includes an additional $\mathfrak{u}(1)$ factor. We
also see that although the gauge theory on the 7D super Yang-Mills sector decouples,
it nevertheless contributes a flavor symmetry to the low energy
effective theory, namely a $G_{L}\times G_{R}$ global symmetry.

Another interesting feature is that we can keep the M5-branes on top of the
orbifold singularity in the $\mathbb{C}^{2}/\Gamma$ direction, but separate
them along the $\mathbb{R}_{\bot}$ direction. Doing so, we see that the
interval will be chopped up into two semi-infinite intervals (one on the
left and one on the right)\ as well as $N-1$ finite size intervals. Each one
of these finite size intervals yields a 6D gauge theory with gauge coupling%
\begin{equation}
\frac{1}{g_{\text{6D}}^{2}}=\frac{\text{dist}}{g_{\text{7D}}^{2}},
\label{sixdcoup}%
\end{equation}
which is the dimensional reduction of the 7D super Yang-Mills theory. Note that
the presence of the domain walls breaks half of the supersymmetry in this configuration, leaving us
with just eight real supercharges.

Separating all the M5-branes from one another, we wind up with a
\textquotedblleft generalized quiver,\textquotedblright\ which we can
schematically depict as:%
\begin{equation}
\lbrack G_{L}]-\underset{N-1}{\underbrace{G-...-G}}-[G_{R}],
\end{equation}
where each horizontal line denotes the presence of an M5-brane. Here, following
\cite{Gaiotto:2009we} we use the notation that a symmetry in square brackets indicates a flavor symmetry
of the system. In reference
\cite{DelZotto:2014hpa, Heckman:2014qba} these links were interpreted as \textquotedblleft6D
conformal matter,\textquotedblright\ a generalization of ordinary
matter fields.

Observe that bringing all the M5-branes on top of one another amounts to going
to infinite coupling in line (\ref{sixdcoup}). From an effective field theory
point of view, it is as if we are attempting to tune an infinite number of
irrelevant operators to reach a conformal fixed point. This illustrates again
the reason why the use of string theory methods is so crucial for establishing
existence of such fixed points.

Even so, bottom up considerations impose strong constraints on the structure of these theories.
Let us now turn to some of these consistency conditions.

\section{Bottom Up Approach to 6D SCFTs \label{sec:BOTTOM}}

In this section, we attempt a purely bottom up approach to 6D SCFTs, though as we have already mentioned,
we will need to supplement these considerations by stringy considerations to ensure that we reach a genuine
conformal fixed point. We begin with a discussion of 6D supersymmetry, followed by a discussion of
moduli spaces and anomalies in 6D supersymmetric theories. Finally, we
study the complications involved at strong coupling, which will lead us to our
top down approach to the subject.

\subsection{Supersymmetry in Six Dimensions}

Even though we are interested in SCFTs, it is helpful to first list some general properties of
6D theories with $\mathcal{N} = (1,0)$ supersymmetry. We can always use this symmetry as an organizational principle,
whether or not we remain at the fixed point. Indeed, away from a conformal fixed point, we expect in many cases
to realize a more conventional field theory, which may flow to a trivial fixed point in the infrared. Additionally,
we note that since a $\mathcal{N} = (2,0)$ supermultiplet can always be decomposed into $\mathcal{N} = (1,0)$ supermultiplets,
it suffices to study the case of minimal supersymmetry.

Massless states are labeled by representations of their \emph{Spin}$(4)\sim SU(2) \times SU(2)$ little group,
and there are four massless $(1,0)$ supermultiplets with low spin which generically appear in
string compactification:\footnote{Here we neglect some possibilities which are less common in
the study of 6D SCFTs. This includes the gravitino multiplet (which in spite of its name contains no graviton) and
the linear multiplet (common in the study of spontaneously broken symmetries).}

\begin{itemize}
\item \textbf{The gravity multiplet}, which has one graviton of spin $(1,1)$,
two gravitinos of spin $(\frac{1}{2},1)$, and one additional field of spin
$(0,1)$, a chiral two-form with self-dual field strength.

\item \textbf{The tensor multiplet}, which has one tensor field $B$ (with
anti-self-dual field strength $H$) of spin $(1,0)$, two fermions of spin $(\frac
{1}{2},0)$, and one scalar of spin $(0,0)$.

\item \textbf{The vector multiplet}, which has one vector field of spin
$(\frac{1}{2},\frac{1}{2})$ and two fermions of spin $(0,\frac{1}{2})$.

\item \textbf{The hypermultiplet}, which has two fermions of spin $(\frac
{1}{2},0)$ and four scalars of spin $(0,0)$.
\end{itemize}

6D SCFTs are non-gravitational theories, so they do not contain a gravity
multiplet. This leaves us with tensor multiplets, vector multiplets, and
scalar multiplets as the low-spin massless degrees of freedom.

$\mathcal{N} = (2,0)$ SCFTs, on the other hand, have superconformal algebra $\mf{osp}%
(6,2|2)$, with R-symmetry $ \mf{sp}(2)_{R} \simeq \mf{so}(5)_{R}$.
The only massless multiplet of low-spin is the $(2,0)$ tensor multiplet, which
is a combination of a $(1,0)$ tensor multiplet and a $(1,0)$ hypermultiplet.
This multiplet now has five scalars, which transform in the $\mathbf{5}$
of the $\mathfrak{sp}(2) \simeq \mathfrak{so}(5)$ R-symmetry.

Interacting field theories with tensor multiplets do not have a known Lorentz covariant formulation,
and even for free classical fields it is rather unwieldy (see e.g. \cite{Pasti:1996vs, Pasti:1997gx, Aganagic:1997zq}).
To see some of the issues involve, consider what happens
when we try to write down a kinetic term with 3-form field
strength $H$ of the form
\begin{equation}
\mathcal{L} \supset- \frac{1}{2} H \wedge* H.
\end{equation}
However, the anti-self-duality condition $H = - *H$ implies this term vanishes.
This makes these theories quite difficult to study from a field theory
perspective. Nevertheless, some aspects of these theories can still be
understood and have direct analogs in ordinary gauge theories in four
dimensions. Our discussion follows that presented in
reference \cite{Seiberg:1996vs}.

For instance, in four-dimensional electromagnetism, electric and
magnetic currents are one-forms, and the associated charged objects are
conventional point particles. The charge of such a particle is expressed in terms
of an integral over all of space (i.e. a three-dimensional hypersurface
$M_{3}^{\perp}$ transverse to the particle's worldline):
\begin{align}
n_{e}  &  = \int_{M_{3}^{\perp}} J_{e}\\
n_{m}  &  = \int_{M_{3}^{\perp}} J_{m}.
\end{align}
In the present situation, the electric current is a 2-form, so the associated
charged objects must be one-dimensional strings. Furthermore, the anti-self-duality
condition on the field strength $H$ implies that the electric currents and
magnetic currents are the same. Thus, all charged strings in six-dimensional
theories are dyonic, with electric charge equal to their magnetic charge. The
charge $n=n_{e}=n_{m}$ of such a string is computed by integrating the current
$J:= dH$ over a four-dimensional hypersurface $M_{4}^{\perp}$ transverse to
the worldvolume of the string:
\begin{align}
n=\int_{M_{4}^{\perp}} J.
\end{align}

In 4D theories with $\mathcal{N}=2$ supersymmetry, the supersymmetry algebra
can be extended by a central charge $Z$:
\begin{equation}
\{Q_{\alpha}^{A}, Q_{\beta}^{B} \} = 2 \epsilon_{\alpha\beta}\epsilon^{AB}%
\bar{Z}.
\end{equation}
This central charge commutes with the supercharges, and the objects charged
under it are simply charged particles. Furthermore, the
Bogomol'nyi-Prasad-Sommerfield (BPS) bound requires that the mass $M$ of a
particle of central charge $Z$ must satisfy $M \geq|Z|$, and states saturating
the bound are called BPS states. For an electrically charged particle, this central charge is linear in the vev $\langle \phi \rangle$of the scalar in the vector multiplet,
\begin{equation}
M = |Z| \propto  n_{e} \langle \phi \rangle  ,
\end{equation}
Coulomb branch singularities arise when a central charge vanishes, and the corresponding BPS states become massless.

The 6D $(1,0)$ supersymmetry algebra can also be extended by a central charge.
The 6D central charge, however, is not a scalar, but rather a vector. The
objects charged under this vector central charge are the aforementioned
charged strings. The string tension obeys a similar BPS bound, growing linearly with the vev of the scalar $t$
in the tensor multiplet,
\begin{equation}
T \propto n \langle t \rangle ,
\end{equation}
in accordance with (\ref{eq:Tension}).
Singularities arise when the central
charge of some string tends to zero and the string becomes tensionless,
resulting in a 6D SCFT.

In 4D electromagnetism, there is an antisymmetric Dirac pairing on the charge
lattice. Given two particles with dyonic charges $q=(e,m)$ and $q^{\prime
}=(e^{\prime},m^{\prime})$, one defines
\begin{equation}
\langle q, q^{\prime}\rangle= e m^{\prime}- e^{\prime}m \in\mathbb{Z}.
\end{equation}
The requirement that this must be an integer is the statement of Dirac
quantization. In 6D, there is similarly a Dirac pairing on the string charge
lattice. However, since the electric and magnetic charges are identified in
this case, the lattice for a theory with $n$ tensor multiplets is
$n$-dimensional rather than $2n$-dimensional. Further, the Dirac pairing in 6D
is symmetric rather than antisymmetric. We may thus express it in terms of a
symmetric $n \times n$ matrix $A_{ij}$,
\begin{equation}
\langle q, q^{\prime}\rangle_{6D} = A_{ij} q^{i} q^{\prime j}.
\label{eq:Diracpairing}%
\end{equation}
Dirac quantization amounts to the statement that $A_{ij}$ must be
integral.

As we shortly explain, in a 6D SCFT, $A_{ij}$ must be negative-definite.
Note also the similarity with line (\ref{eq:Anegdef}): indeed, in a 6D SCFT coming from a
type IIB compactification, the Dirac pairing $A_{ij}$ is precisely the
intersection pairing of the compactification geometry.

\subsection{Moduli Spaces and Anomalies}

\label{ssec:anomalies}

Among the three types of multiplets (tensor, vectors, and hypers) that can
arise in a $(1,0)$ 6D SCFT, only the tensor multiplet and hypermultiplet contain
scalar fields. The moduli space of the theory then splits into two branches: the
``tensor branch," in which scalars in the tensor multiplets acquire vevs, and
the ``Higgs branch," in which scalars in the hypermultiplets acquire vevs.
Note that the ``tensor branch" is sometimes referred to as the ``Coulomb
branch" in the literature, since under reduction to four or five dimensions,
these tensor multiplets become vector multiplets, and the tensor branch
descends to part of the Coulomb branch of the lower-dimensional theory. Moving onto
the tensor branch preserves the $\mathfrak{su}(2)$ R-symmetry of the theory, but moving onto the
Higgs branch breaks the R-symmetry. Viewed as a $(1,0)$ theory, all $(2,0)$
theories have a tensor branch. All known \textit{interacting}\footnote{A free hypermultiplet is an example
of a trivial CFT with no tensor branch.} (1,0) theories have a tensor
branch, and many have a Higgs branch as well.\footnote{Assuming we have a
gauge theory description away from the conformal fixed point,
we can understand the structure of the Higgs branch of moduli space in
terms of an $\mathfrak{su}(2)_{\mathcal{R}}$ triplet of
D-flatness conditions for hypermultiplets coupled to the vector multiplets (as may be
familiar to the reader from 4D $\mathcal{N} = 2$ supersymmetry).
This triplet of conditions is constructed from sums of bilinears in the hypermultiplet scalars.
Note that in some cases, having matter charged under a vector multiplet is
not enough to move onto a Higgs branch. For example, when we have a half-hypermultiplet in the
fundamental representation, there is a reality condition, so one needs at least two half-hypermultiplets to
have a genuine Higgs branch. Additional constraints are possible
depending on the details of the gauge group and matter content. We omit
the precise formulae for the general case since we will not make use of it in
what follows anyway.}

For a 6D SCFT, the metric on the tensor branch moduli space is controlled by the
same pairing $A_{ij}$ which appears in the Dirac pairing. The reason this
must be so is that by supersymmetry, the tension of the BPS strings
on the tensor branch is related to the charge of these strings. In particular,
with respect to a suitable raising / lowering convention for tensor branch coordinates, the
metric on tensor branch moduli space is:
\begin{equation}\label{tensormetric}
ds^2 = - A_{ij} dt^i dt^j.
\end{equation}
In this coordinate system, the CFT point corresponds to $t^i = 0$ for all $i$. Note that
the main physical condition we need to impose is that if we are at a generic point of moduli space away
from the SCFT, it must be at finite distance, which in turn requires $A$ to
be negative definite. One can also have a Dirac pairing $A$ which is not negative definite,
as happens in supergravity theories (as well as little string theories \cite{Bhardwaj:2015oru}).
In this case, the correspondence between the Dirac pairing and
metric is different.\footnote{In the non-conformal case such as a supergravity theory
we treat the $t^i$ as local coordinates in the coset $\mathfrak{so}(N_T , 1) / \mathfrak{so}(N_T)$, with
$N_T$ the number of tensor multiplets. For additional details on the metric on moduli space for
6D supergravity theories, see e.g. reference \cite{Sagnotti:1992qw}. An analogous issue
appears in the moduli space of metrics for Calabi-Yau threefolds,
a point we return to in section \ref{sec:FAGAIN}.}

Assuming we have given a vev to some operators, we can expect to move away from the conformal
fixed point, resulting in a theory which makes explicit reference to some mass scales. Even so,
the general principles of 't Hooft anomaly matching provide a way to match the anomalies of this theory
to that of the conformal fixed point \cite{tHooft:1979rat}. With this in mind, let us now turn to the constraints imposed by
anomalies.

Anomalies play a crucial role in our understanding of 6D SCFTs. Chiral
anomalies exist in any even number of dimensions. In 6D, these anomalies are
encoded in a four-point function $\langle j^{\mu_{1}}_{\alpha_{1}}
j^{\mu_{2}}_{\alpha_{2}} j^{\mu_{3}}_{\alpha_{3}} j^{\mu_{4}}_{\alpha_{4}}
\rangle$, where $j^{\mu_{i}}_{\alpha_{i}}$ is the current of some symmetry
$\alpha_{i}$ of the theory. There are several ways to produce a non-vanishing
anomaly. The first is to take all of the currents identical ($\alpha
_{1}=\alpha_{2}=\alpha_{3}=\alpha_{4}$). There are three types of anomalies of
this form:

\begin{itemize}
\item \textbf{Flavor anomalies.} These anomalies involve continuous global symmetries of
the theory and are thus benign. Here we shall, by abuse of notation refer to an R-symmetry
as a global symmetry of this type, though in some cases it is helpful to distinguish the R-symmetry from
all continuous global symmetries which commute with it.

\item \textbf{Gauge anomalies.} These anomalies involve gauge symmetries of
the theory and are therefore dangerous. They must be cancelled in any
consistent six-dimensional theory.

\item \textbf{Gravitational anomalies.} These anomalies involve the $\mathfrak{so}(5,1)$
Lorentz symmetry. In the case of a 6D supergravity theory, these anomalies are
dangerous and must be cancelled, leading to strong constraints on the massless
spectrum of a supergravity. In the case of a non-gravitational theory such as
a 6D SCFT, however, they are benign.
\end{itemize}

The second possibility is to take distinct choices for the external currents.
Restricting to the case of non-abelian (and traceless)
symmetry generators, we need two pairs of distinct currents ($\alpha
_{1}=\alpha_{2} \neq\alpha_{3}=\alpha_{4}$). These are called ``mixed
anomalies," and there are five types of mixed anomalies, corresponding to a
choice of any two of the above symmetry currents:

\begin{itemize}
\item \textbf{Mixed flavor-flavor anomalies.} These anomalies involve two
insertions of one global continuous symmetry current and two insertions of another. They are
benign in 6D SCFTs.

\item \textbf{Mixed flavor-gravitational anomalies.} These anomalies involve
two insertions of some global continuous symmetry current and two insertions of the
Lorentz symmetry current. They are benign in 6D SCFTs.

\item \textbf{Mixed gauge-gauge anomalies.} These anomalies involve two
insertions of one gauge symmetry current and two insertions of another.
They are dangerous in 6D SCFTs and must be cancelled.

\item \textbf{Mixed gauge-global anomalies.} These anomalies involve two
insertions of some global continuous symmetry current and two insertions of a gauge
symmetry current. They are allowed in a general 6D theory, but they cannot
arise in an SCFT due to constraints imposed by superconformal invariance \cite{CordovaMixer},
as referred to in \cite{Cordova:2018cvg}.

\item \textbf{Mixed gauge-gravitational anomalies.} These anomalies involve
two insertions of some gauge symmetry current and two insertions of the
Lorentz symmetry current. They are allowed in a general 6D theory, but they cannot
arise in an SCFT due to constraints imposed by superconformal invariance \cite{CordovaMixer},
as referred to in \cite{Cordova:2018cvg}.
\end{itemize}

For non-Abelian symmetries, this is the full list of possibilities.
For Abelian symmetries, however, the possibilities are much richer:
we can take any number of insertions of abelian currents--gauge or global--and either
zero, two, or three insertions of one of the non-abelian symmetry currents listed above
(insertions of a single non-Abelian current always vanish). As in the non-Abelian case,
anomalies for a 6D theory involving any insertion of a \emph{gauge} current must vanish.
Superconformal invariance actually forbids the appearance
of abelian gauge symmetries in 6D SCFTs \cite{Cordova:2015fha}, and this is
also apparent in F-theory because such phenomena only arise in models coupled to gravity.
However, one can have abelian flavor symmetries,
and then each such symmetry generator can appear in various mixtures with other abelian and non-abelian
symmetry generators. In most of the
6D SCFT literature, anomalies for Abelian symmetries have been ignored, as they are more difficult
to analyze from the F-theory perspective, and we will largely ignore them as well.
See however the recent work \cite{Lee:2018ihr} for a discussion of Abelian anomalies
in 6D SCFTs.

All of these anomalies for continuous symmetries are
encoded in a formal 8-form anomaly polynomial $I_{8}$,
which is built out of the curvatures of the flavor, gauge, and gravitational
symmetries. $I_{8}$ is related to the anomalous variation
of the action $I_{6}$ via the ``descent equations,"
\begin{align}
I_{8}  &  = d I_{7}\\
\delta I_{7}  &  = d I_{6}.
\end{align}

Much as in the case of fundamental strings and the celebrated Green-Schwarz mechanism \cite{Green:1984sg},
anomalies receive both one loop contributions
as well as contributions from the variation of the two-form potentials
associated with the tensor multiplets. In six dimensions
this is often referred to as the Green-Schwarz-Sagnotti-West mechanism (see references \cite{Green:1984bx, Sagnotti:1992qw}). Thus we may write:
\begin{equation}
I_{\text{tot}} = I_{\text{1-loop}}+ I_{\text{GS}} .
\end{equation}
The Green-Schwarz term comes from a coupling of the form,
\begin{equation}
\mathcal{L}_{\text{GS}} \propto\sum_{i=1}^{N_{T}} B^{(i)} \wedge I_{i},
\end{equation}
where the sum runs over the number of anti-self-dual 2-forms $B^{(i)}$ (which is
also the number of tensor multiplets $N_{T}$), and $I_{i}$ is some 4-form
constructed from various characteristic classes of the gauge and gravitational field.

The associated contribution to the anomaly polynomial is then
\begin{equation}
I_{\text{GS}} = \frac{1}{2} A^{ij} I_{i} I_{j},
\end{equation}
where $A^{ij}$ is the inverse of the Dirac pairing on the string charge
lattice from (\ref{eq:Diracpairing}). $I_i$ can be written as
\begin{equation}
I_{i} = a_{i} c_{2}(R) + b_{i} p_{1}(T) + \sum_{j} c_{ij} \, \mathrm{Tr}
F_{j}^{2},
\end{equation}
where $c_{2}(R)$ is the second Chern class of the $\mf{su}(2)$ R-symmetry, $p_1(T)$ is
the first Pontryagin class of the tangent bundle, and $F_{i}$ is the field
strength of the $i$th symmetry, where $i$ and $j$ run over both the gauge and
global symmetries of the theory.

In the conventions of this review (obtained from \cite{Ohmori:2014pca, Ohmori:2014kda}), we write:
\begin{equation}
\mathrm{Tr} F^2 = \frac{1}{h_G^{\vee}}\mathrm{Tr}_{\text{adj}} F^2,
\end{equation}
for a normalized trace in the adjoint representation. Here, $h_{G}^{\vee}$ is the dual Coxeter number of the group. A convenient feature of this
normalization convention is that for a field configuration describing a single instanton, $\frac{1}{4} \mathrm{Tr} F^2$ integrates to one.
This is especially helpful in studies of compactifications of 6D SCFTs on manifolds. Other conventions which do not obscure integrality
conditions of the anomalies are also common, see reference \cite{Intriligator:2014eaa} for an example along these lines.

We will describe the computation of the anomaly polynomial and the Green-Schwarz 4-form $I_i$ in more detail in Section \ref{sec:ANOMPOLY}. For now, we simply present the constraints from anomaly cancellation on charged matter.
First, for a given representation $\rho$ of a simple gauge algebra, we define group theory constants $x_\rho$, $y_\rho$, and Ind$_\rho$ by
\begin{align}
\tr_{\rho} F^2 = \text{Ind}_\rho \Tr F^2, ~~\tr_{ \rho }F^4 = x_\rho \Tr F^4 + y_\rho (\Tr F^2)^2.
\end{align}
The values of these constants can be found in Appendix \ref{sec:GROUPTHEORY}. For a given gauge algebra $\mf{g}_k$, the requirement that gauge anomalies cancel implies
\begin{align}
x_{\text{adj}}-\sum_{\rho}n_{\rho}x_{\rho}&=0, \label{eq:gaugeanom1a} \\
y_{\text{adj}}-\sum_{\rho}n_{\rho}y_{\rho}  &  = -  12 A^{ij} c_{ik} c_{jk}. \label{eq:gaugeanom2a}
\end{align}
with $n_{\rho}$ the number of
hypermultiplets charged under $\mathfrak{g}_k$ in the representation $\rho$. In addition, cancellation of
gauge-gravitational anomalies requires
\begin{align}
\text{Ind}_{\text{adj}}-\sum_{\rho}n_{\rho}\text{Ind}_{\rho}  &  = - 48 A^{ij} b_{i} c_{jk}.
\label{eq:gaugeanom3a}%
\end{align}
For theories with multiple gauge algebras $\mf{g}_k$, $\mf{g}_{k^\prime}$, mixed gauge anomaly cancellation implies
\begin{equation}
\sum_{\rho}n_{\rho, \rho^{\prime}} \text{Ind}_{\rho} \text{Ind}_{\rho^{\prime
}} = 16 A^{ij} c_{ik} c_{jk^\prime} . \label{eq:mixeda}%
\end{equation}
Note that the right-hand side of (\ref{eq:gaugeanom1a})-(\ref{eq:mixeda}) encodes the contribution of the Green-Schwarz term. We will see that these terms are fixed geometrically in the F-theory construction of 6D SCFTs.

\subsubsection{Global Discrete Anomalies}

In addition to the anomalies associated with continuous symmetries, there can also be important restrictions imposed by the global
structure of the gauge group, as specified by the homotopy group $\pi_{D}(G)$ in a $D$-dimensional field theory. For example,
in four dimensions this is the statement that $SU(2)$ gauge theory with $N_f$ doublets must have $N_f$ even \cite{Witten:1982fp}.
In six dimensions, the relevant question has to do with which gauge groups have non-trivial $\pi_6(G)$. Following reference \cite{Bershadsky:1997sb},
for $SU(2)$ gauge theory with $n_2$ doublets, $SU(3)$ gauge theory with $n_3$ fundamentals and $n_6$ sextics, and $G_2$ gauge
theory with $n_7$ fundamentals, the relevant constraint is:
\begin{align}
SU(2) & :  4 - n_2 \equiv 0 \, \text{mod} \, (6) \\
SU(3) & :  n_3 - n_6 \equiv 0 \, \text{mod} \, (6) \\
G_2   & :  1 - n_7 \equiv 0 \, \text{mod} \, (3) \\
\end{align}

It is natural to also ask about anomalies for global discrete symmetries
for 6D SCFTs. This question is to a large extent still unexplored,
but it would be interesting to pursue in future work.

\subsection{Top Down versus Bottom Up}

Given the rather stringent nature of the constraints imposed by bottom up considerations,
it is natural to ask whether this sort of approach actually ``misses'' any additional structure
imposed by the stringy realization of 6D SCFTs.

The first point is a philosophical one: To actually reach the conformal fixed point, the effective field theory
on the tensor branch will necessarily break down. Another way of saying this is it would require one to catalog the
effects of an infinite collection of irrelevant operators. From this perspective, the string
construction comes to the rescue at this strong coupling point, since there are well established methods for consistently constructing such vacua.

From this perspective, the more appropriate question is whether consistency conditions imposed on the tensor branch are sufficient
to realize a consistent interacting fixed point, or whether the string construction imposes some additional still poorly understood conditions.

As we will see in subsequent sections, it is possible to write down models with a negative-definite Dirac
pairing $A$ that satisfy anomaly cancellation but do not produce a
consistent 6D SCFT. For instance, one well-studied example is the theory with
Dirac pairing
\begin{equation}
A = \left[
\begin{array}
[c]{cc}%
-3 & 1\\
1 & -2
\end{array}
\right]  ,
\end{equation}
gauge algebra $\mf{so}(8) \oplus \mf{su}(2)$, and charged
hypermultiplets
\begin{equation}
\frac{1}{2}(\mathbf{{8}_{v}, {2}) \oplus({8}_{s}, {1})\oplus({8}_{c}, {1}).}%
\end{equation}
One can check that $A < 0$, and anomalies cancel
for this theory. Nonetheless, it is not a consistent 6D SCFT: at the
superconformal fixed point, the symmetry under which the $8$ $\mf{su}%
(2)$ half-hypermultiplets transform is reduced from $\mf{so}(8)$ to
$\mf{so}(7)$ \cite{Ohmori:2015pia}, so it is not possible to couple an
$\mf{so}(8)$ gauge algebra to this theory.

Similarly, the theory with
Dirac pairing
\begin{equation}
A_{ij} = \left[
\begin{array}
[c]{cc}%
-2 & 1\\
1 & -1
\end{array}
\right]  ,
\end{equation}
gauge algebra $\mf{g}_2 \oplus \mf{sp}(4)$, and charged
hypermultiplets
\begin{equation}
\frac{1}{2}(\mathbf{{7}, {8}}) \oplus 12 \frac{1}{2} \mathbf{(1, 8),}%
\end{equation}
appears to be consistent from a bottom up perspective, but so far an explicit F-theory
construction has not been found \cite{Johnson:2016qar}. The same is true for a theory
with this Dirac pairing but gauge algebra $\mf{su}(10) \oplus \mf{su}(3)$ and charged
hypermultiplets
\begin{equation}
(\mathbf{{10}, {3}}) \oplus 17 \mathbf{(10, 1)} \oplus 2 \mathbf{(1, 3),}%
\end{equation}

In these cases, it is not yet clear whether this discrepancy represents a failure of
the top down perspective or the bottom up perspective. A catalog of examples of this
type can be found in Section 6.2 of \cite{Johnson:2016qar}.

As an even simpler example, one could consider a simple theory with a single
tensor multiplet, namely the matrix $A$ is just the $1 \times 1$ matrix $-3$.
This theory clearly has negative-definite Dirac pairing, and there are no gauge
anomalies to cancel because we have purposefully omitted the appearance of any vector multiplets.
Nevertheless, for reasons that are not well understood at present, this theory does not seem to give
rise to a consistent 6D SCFT.

In light of these issues, we shall instead turn to a top down construction of 6D SCFTS.
In the following section, we will give examples of 6D SCFTs constructed using string/M-theory.
Then, we will see that F-theory provides a powerful means for constructing 6D
SCFTs, encompassing the known string/M-theory constructions and supplementing
the aforementioned bottom up constraints with a precise set of top down
conditions, which exclude each of the possibilities discussed in this subsection.


\section{F-theory Preliminaries \label{sec:FTHEORY}}

In the previous sections we illustrated the highly non-trivial fact that
string constructions provide substantial evidence for the existence of
conformal fixed points in six dimensions. Additionally, we have seen that
bottom up considerations impose remarkably tight constraints on candidate
SCFTs with a tensor branch. In this section we take our first steps toward a top down approach by giving a brief introduction to the
relevant parts of F-theory necessary to construct 6D\ SCFTs. Some resources
for additional background can be found for example in the book \cite{Johnson:2005mqa},
and in the context of particle physics model building applications in references
\cite{Heckman:2010bq, Weigand:2010wm}.

This section is organized as follows. First, we present the general ideas of
F-theory. After this, we turn to some
additional details on the construction of six-dimensional vacua.

\subsection{F-theory Preliminaries}

Let us begin with the general ideas of F-theory. From a practical standpoint,
the main idea in this approach is to systematically construct consistent
background solutions for type IIB\ string theory in cases where the
axio-dilaton,%
\begin{equation}
\tau=C_{0}+ie^{-\phi},
\end{equation}
has non-trivial position dependence on the spacetime coordinates and
potentially order-one values for the couplings.

To access these non-trivial solutions, the main observation made in
\cite{Vafa:1996xn} is that in the type IIB\ string theory, there is a well-known duality
invariance of the 10D\ type IIB\ supergravity action, with the axio-dilaton
transforming as:%
\begin{equation}
\tau\mapsto\frac{a\tau+b}{c\tau+d}\text{ \ \ with \ \ }\left[
\begin{array}
[c]{cc}%
a & b\\
c & d
\end{array}
\right]  \in SL(2,\mathbb{Z}).
\end{equation}
This is the same redundancy present in specifying the complex structure
modulus of a complex $T^{2}$, namely an \textquotedblleft elliptic
curve.\textquotedblright\ The main idea in F-theory is to visualize the
position dependence of the axio-dilaton directly in terms of a family of
elliptic curves. So, rather than dealing directly with a ten-dimensional
spacetime, we can geometrize this profile in terms of a twelve-dimensional
geometry. Let us stress that physical degrees of freedom still propagate on ten spacetime dimensions.
For example, the volume of this elliptic curve has no
physical meaning, so it is customary to work in a limit where it has collapsed
to zero size. There have been various attempts to interpret these two
additional directions in physical terms, (see e.g. \cite{Vafa:1996xn, Bars:1996ab,
Hewson:1997wv, Choi:2015gia, Linch:2016ipx, Heckman:2017uxe}),
but this will not be necessary in what follows.

\begin{figure}[t!]
\centering
\includegraphics[trim = 0cm 6cm 0cm 6cm, clip=true, scale=0.5]{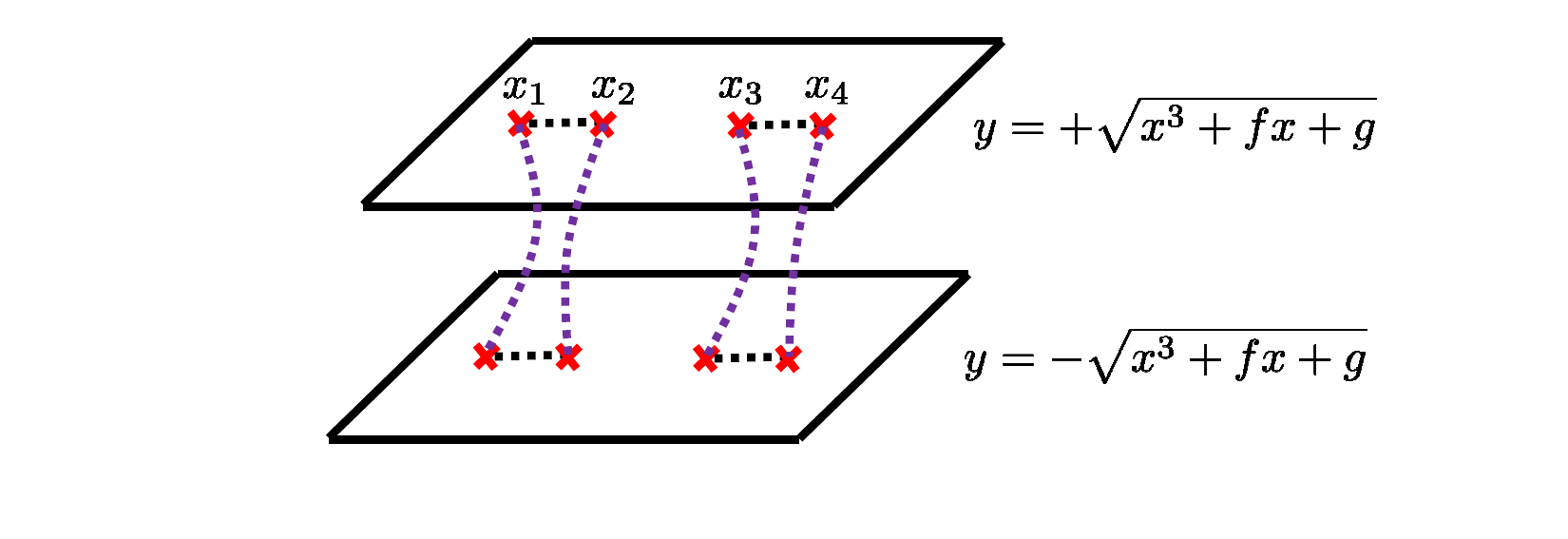}
\caption{Depiction of the elliptic curve as a two-sheeted cover, as specified by the
Weierstrass equation $y^2 = x^3 + fx + g$. Each sheet is spanned by the $x$-coordinate, which has four roots,
three of which are ``visible'' in the cubic, with the fourth a point at infinity. Pairing these roots and joining the corresponding
branch cuts, we realize a genus one curve, or $T^2$. The tube of this $T^2$ can be seen as the purple segment joining the two sheets.}%
\label{fig:doughnut}%
\end{figure}

It is a beautiful fact from the theory of elliptic curves that any such
curve can be placed in the form:%
\begin{equation}
y^{2}=x^{3}+fx+g,
\end{equation}
where we interpret the curve as a hypersurface in the complex space
spanned by the coordinates $x$ and $y$, with $f$ and $g$ as fixed
coefficients. There are various ways to see that this does indeed describe a
$T^{2}$. Pictorially, we can use the approach of Riemann to visualize Riemann
surfaces as branched covers of the complex plane. Factorizing the cubic in
$x$, we see that there are three roots, with an additional root
\textquotedblleft at infinity.\textquotedblright\footnote{This latter root is
most clearly seen upon projectivizing the coordinates $x$ and $y$.} There are
thus a pair of branch cuts (grouping the four roots into two cuts), and two
sheets, since we have two solutions to the equation:%
\begin{equation}
y=\pm\sqrt{x^{3}+fx+g}.
\end{equation}
As can be seen from figure \ref{fig:doughnut}, joining the two sheets produces a
\textquotedblleft doughnut,\textquotedblright\ namely a $T^{2}$.

From a physical point of view, our primary interest is not in the
case where the Weierstrass model coefficients $f$ and $g$ are constant, but
instead in situations where there is non-trivial position dependence over the ten-dimensional spacetime. At this point, the
question is clearly more subtle, since we now need to solve the supergravity
equations for the 10D\ metric, as well as a position dependent profile for the
axio-dilaton system.

Thankfully, this is precisely where F-theory starts to exhibit its full
strength. Suppose we are interested in constructing a consistent solution to
the IIB\ supergravity equations of motion for a 10D spacetime: $\mathbb{R}%
^{D-1,1}\times\mathcal{B}$. Here, $\mathcal{B}$ has real dimension $10-D$. In
this case we have a $D$-dimensional Minkowski spacetime, and $10-D$ \textquotedblleft
internal directions\textquotedblright\ for our compactification. We refer to
this $\mathcal{B}$ as the \textquotedblleft base\textquotedblright\ of an
F-theory model. Each point of the base is decorated by an elliptic curve, so
we shall be interested in F-theory background geometries $X$ of real dimension
$12-D$.

The most well-studied case corresponds to the situation where we can take full
advantage of methods from algebraic geometry. We therefore restrict to
the case where $D$ is even. Specializing further to situations where we retain
minimal supersymmetry in the uncompactified directions, the total space $X$
must be a Calabi-Yau space, and moreover, the base $\mathcal{B}$ must be a
K\"{a}hler surface. To see why this is the correct condition to impose on $X$,
we observe that by packaging the axio-dilaton in terms of an elliptic fiber,
we can rephrase the supergravity equations of motion on $\mathcal{B}$ with
position dependent axio-dilaton in terms of conditions on $X$. In
that context, it is well known that type II\ strings compactified on $X$
preserve supersymmetry provided $X$ is Calabi-Yau. The same condition thus
follows for F-theory backgrounds as well.

Indeed, the standard T-duality between circle compactifications of IIA\ and
IIB\ extends to M-theory and F-theory \cite{Vafa:1996xn, Morrison:1996na, Morrison:1996pp}.
In this vein, F-theory on the background $S^{1}\times X$ is, at low energies, nothing but M-theory on
$X$. In passing from the M-theory description to the F-theory description, we
also must collapse the volume of the elliptic curve on the M-theory side to
zero size \cite{Vafa:1996xn}. Observe that M-theory on $X$ yields a minimally
supersymmetric theory on $\mathbb{R}^{D-2,1}$. In this correspondence, the
radius of the circle on the F-theory side is related to the volume of the
elliptic fiber on the M-theory side as:%
\begin{equation}
\text{Radius}\sim\frac{1}{\text{Vol}\left(  \text{Elliptic Fiber}\right)  ^{\nu}},
\end{equation}
where the specific power of $\nu>0$ depends on the dimension of the
uncompactified directions. Note that this is consistent with the fact that the
volume of the elliptic fiber in the F-theory picture has no physical meaning.
Indeed, one way to define F-theory vacua is to first start with M-theory on
$X$ and then use this to construct F-theory in the adiabatic limit where the
$S^{1}$ expands to infinite size. This is the same limit previously
mentioned where the elliptic curve has collapsed to zero size.

Thus, for supersymmetric backgrounds of even-dimensional Minkowski spacetimes, we see that
F-theory necessarily involves the study of elliptically fibered Calabi-Yau
manifolds. With this in mind, let us turn to the conditions which must be
imposed on $f$ and $g$ to satisfy these conditions.

Recall that a Calabi-Yau $n$-fold has a holomorphic $n$-form $\Omega_{(n,0)}$.
To construct this differential form, we return to the form of our Weierstrass
model:%
\begin{equation}
y^{2}=x^{3}+fx+g,
\end{equation}
where now, we allow non-trivial position dependence on the base $\mathcal{B}$
for the coefficients $f$ and $g$. Interpreting $f$ and $g$ as sections of a
line bundle, we can partially fix the bundle assignments using
homogeneity of the Weierstrass model:%
\begin{equation}
y\sim\mathcal{L}^{3}\text{, \ \ }x\sim\mathcal{L}^{2}\text{, \ \ }%
f\sim\mathcal{L}^{4}\text{, \ \ }g\sim\mathcal{L}^{6},
\end{equation}
for some choice of line bundle $\mathcal{L}$, where $\sim$ here means ``is a section of." Using the fact that the
holomorphic $n$-form has the local presentation:%
\begin{equation}
\Omega_{(n,0)}\sim\frac{dx}{y}\wedge\Omega_{\mathcal{B}},
\end{equation}
where $\Omega_{\mathcal{B}}$ is a section of the $(n-1)$-form on $\mathcal{B}$,
and the fact that the canonical class of a Calabi-Yau space is trivial, we
learn that $f$ and $g$ are sections of:%
\begin{equation}
f\sim K_{\mathcal{B}}^{-4}\text{, \ \ }g\sim K_{\mathcal{B}}^{-6}.
\label{fgsections}%
\end{equation}
This tells us the sort of polynomials
we need to write in terms of coordinates of the base in order to get an
elliptically fibered Calabi-Yau space.  Since the coefficients $f$ and~$g$ have
position dependence on the base coordinates, we see that the shape of our
elliptic fiber will vary from point to point of the base.

It can also happen that the elliptic curve becomes singular at
various locations of the base. This will occur whenever the roots $x_{i}$ of
our cubic in $x$ collide. From the theory of cubics, we know this happens
whenever the discriminant $\Delta$ of the polynomial vanishes, namely the
product over the differences of roots:%
\begin{equation}
\Delta=\underset{i\neq j}{%
{\displaystyle\prod}
}(x_{i}-x_{j})=4f^{3}+27g^{2}.
\end{equation}
The set of points on $\mathcal{B}$ where $(\Delta=0)$ is referred to as the
\textquotedblleft discriminant locus.\textquotedblright\ In general, $\Delta$
can factor into several irreducible components, so we can write:%
\begin{equation}
\Delta=\underset{a}{%
{\displaystyle\prod}
}\Delta_{a}.
\end{equation}
Each component of the discriminant locus $(\Delta_{a}=0)$ occurs along a
complex codimension one subspace of $\mathcal{B}$, and so we see that it fills
out one temporal direction and seven spatial directions of our
ten-dimensional spacetime. For this reason, it is common to refer to each such
component of the discriminant locus as a \textquotedblleft
seven-brane.\textquotedblright

There are clearly many different choices for how $f,g,\Delta$ can vanish
along a codimension one locus in the base
while still preserving the general conditions necessary to achieve an
elliptic fibration. Thankfully, these have already been classified by Kodaira
\cite{Kodaira} and are dictated by prescribed orders of vanishing along a
codimension one locus. We summarize these options in table \ref{tabooboo}.%
\begin{table}%
\centering
\begin{tabular}
[c]{|l|l|l|l|l|}\hline
Fiber & Ord$(f)$ & Ord$(g)$ & Ord$(\Delta)$ & Singularity Type\\\hline
$I_{N}$ & $0$ & $0$ & $N$ & $A_{N-1}$\\\hline
$II$ & $1$ & $1$ & $2$ & None\\\hline
$III$ & $1$ & $\geq2$ & $3$ & $A_{1}$\\\hline
$IV$ & $\geq2$ & $2$ & $4$ & $A_{2}$\\\hline
$I_{0}^{\ast}$ & $\geq 2$ & $\geq3$ & $6$ & $D_{4}$\\\hline
$I_{N}^{\ast}$ & $2$ & $3$ & $N+6$ & $D_{N+4}$\\\hline
$IV^{\ast}$ & $\geq3$ & $4$ & $8$ & $E_{6}$\\\hline
$III^{\ast}$ & $3$ & $\geq5$ & $9$ & $E_{7}$\\\hline
$II^{\ast}$ & $\geq4$ & $5$ & $10$ & $E_{8}$\\\hline
\end{tabular}
\caption{Kodaira's classification of fiber singularity types.}
\label{tabooboo}%
\end{table}
Here, \textquotedblleft singularity type\textquotedblright\ refers to the
fact that the local geometry will also be singular, with a classification akin
to what is found for the simply laced Lie algebras.\footnote{Recall that a
vanishing locus is said to have a singularity when both the polynomial and its
derivative vanish along the same point set since in this situation we cannot
set up a local spanning basis of tangent vectors.} To illustrate, the type
$IV^{\ast}$ singular fiber is locally presented as:%
\begin{equation}
y^{2}=x^{3}+z^{4}.
\end{equation}
This classification places some rather tight constraints on the local
structure of an elliptically fibered Calabi-Yau manifold. If the order
of vanishing for $f$,$g$,$\Delta$ is respectively $4,6,12$ or
worse, then the canonical bundle of the total space is no longer trivial, so
we cannot satisfy the supergravity equations of motion. When these order-of-vanishing constraints are satisfied, we say that the elliptic fiber is in
Kodaira-Tate form.

If the base $\mathcal{B}$ is one-dimensional, then the singularity type of
each elliptic fiber turns out to correspond to a choice of gauge group for the
corresponding seven-brane. However, if the base has dimension greater than
one, it is possible for the two-cycles of the resolved fiber to be permuted under a monodromy as we pass along a one-cycle in the discriminant locus. When this occurs, we refer to the fiber as
\textquotedblleft non-split\textquotedblright, and the actual gauge symmetry
realized in six dimensions is different from the singularity type of table
\ref{tabooboo}. In the physical theory, such non-split fibers amount to
quotienting by the outer automorphism of the algebra (more precisely reflection
symmetries of the affine twisted Dynkin diagram). The rules worked out in \cite{Bershadsky:1996nh, Katz:1996xe}
and revisited in \cite{Aspinwall:2000kf} are shown in table \ref{tab:taboo2}, where we have also indicated the presence of \textquotedblleft
matter.\textquotedblright
\begin{table}%
\centering
\begin{tabular}
[c]{|l|l|l|}\hline
Fiber & Split Algebra & Non-Split Algebra\\\hline
$I_{2N}$ & $\mf{su}(2N)$ & $\mf{sp}(N)$\\\hline
$I_{2N+1}$ & $\mf{su}(2N+1)$ & $\mf{sp}(N)+$matter\\\hline
$III$ & $\mf{su}(2)$ & $\mf{su}(2)+$matter\\\hline
$IV$ & $\mf{su}(3)$ & $\mf{sp}(1)$\\\hline
$I_{0}^{\ast}$ & $\mf{so}(8)$ & $\mf{so}(7)$ (semi-split), $\mathfrak{g}_{2}$ (fully non-split)\\\hline
$I_{n}^{\ast}$ & $\mf{so}(2n+8)$ & $\mf{so}(2n+7)$\\\hline
$IV^{\ast}$ & $\mathfrak{e}_{6}$ & $\mathfrak{f}_{4}$\\\hline
$III^{\ast}$ & $\mathfrak{e}_{7}$ & no automorphism\\\hline
$II^{\ast}$ & $\mathfrak{e}_{8}$ & no automorphism\\\hline
\end{tabular}
\caption{Monodromy of fiber singularity types.}
\label{tab:taboo2}
\end{table}

In addition to the locations where we have seven-branes, it can also
happen that seven-branes intersect each other. This occurs along a
codimension two subspace of the base, so it fills a six-dimensional
subspace of the ten-dimensional spacetime. At a general level, we refer to the matter at
such a collision as \textquotedblleft localized matter.\textquotedblright\

There is a rather intuitive way to understand the representation content of
localized matter which covers the vast majority of \textquotedblleft
non-exotic\textquotedblright\ situations: Starting from a seven-brane with
gauge algebra $\mathfrak{g}_{\text{parent}}$, we ask what happens when it is deformed to a
pair of intersecting seven-branes by activating an adjoint valued
scalar on the parent stack. This amounts to a Higgsing of the parent gauge
algebra to some subalgebra $\mathfrak{g}_{a}\times \mathfrak{g}_{b}\subset \mathfrak{g}_{\text{parent}}$ which
locally enhances back to $\mathfrak{g}_{\text{parent}}$ at the points of intersection
(see the review article \cite{Heckman:2010bq} for additional discussion). Decomposing the
adjoint representation of $\mathfrak{g}_{\text{parent}}$ into irreducible representations
of $\mathfrak{g}_{a}\times \mathfrak{g}_{b}$, the localized matter corresponds to those terms which
are in a non-trivial representation with respect to each gauge algebra factor.

In the F-theory literature, these are referred to as the \textquotedblleft
Katz-Vafa collision rules\textquotedblright\ \cite{Katz:1996xe}. Let us note that there are other
ways to analyze the resulting matter content which involve explicitly
analyzing the additional exceptional divisors for singular fibers of the
F-theory model \cite{Bershadsky:1996nh} (see also \cite{Grassi:2000wer, Hayashi:2014kca}).

As an illustrative example, we can understand matter in the $\mathbf{27}$ of $\mathfrak{e}_{6}$
(the fundamental representation) as descending from the decomposition
of the adjoint representation of $\mathfrak{e}_{7}$ to irreducible representations of
$\mathfrak{e}_{6}\times \mathfrak{u}(1)$:%
\begin{equation}
\mathbf{133}\rightarrow \mathbf{78}_{0}+\mathbf{1}_{0}+\mathbf{27}_{1}+\overline{\mathbf{27}}_{-1}\ .
\end{equation}
The Weierstrass model for $m$ localized fundamentals is:
\begin{equation}\label{e6matt}
y^2 = x^3 + q^{2}_{m}(v) u^4 + f_{m + 2}(v) u^3 x.
\end{equation}
with $q_m(v)$ and $f_{m+2}(v)$ polynomials of respective degrees $m$ and
$(m+2)$ in $v$, the local coordinate along the curve $(u = 0)$. The factorization
of the $u^4$ coefficient as a perfect square is necessary to have a split $IV^{\ast}$
fiber over $(u = 0)$.

It can also happen that ``matter'' is delocalized, namely the matter field wave function in the internal
directions is not concentrated at a single point. This occurs in many
situations with a non-split fiber. In these situations, we need to modify the adjoint decomposition
rules stated above by performing a projecting by the outer automorphism of the associated ``descendent algebra.''
Geometrically, it is most convenient to distinguish between the
discriminant locus and the branched cover with ramification at ``twist points.'' These twist points
account for the fact that the matter field wave function in the internal directions
is now shared across various points. For further discussion on how to analyze the resulting geometry
with branched covers, see e.g. \cite{Grassi:2000wer}.

An illustrative example along these lines is a model with $m$ matter fields in the
$\mathbf{26}$ of $\mathfrak{f}_4$:
\begin{equation}
y^2 = x^3 + q_{2m}(v) u^4 + f_{m + 2}(v) u^3 x,
\end{equation}
with notation similar to that in equation (\ref{e6matt}). Here we have a non-split $IV^{\ast}$
fiber over $(u = 0)$ because the coefficient of the $u^4$ term does not factorize as a perfect square.
It is described as a local collision of two components of the discriminant locus
with respective fiber types $I_1$ and type $IV^{\ast}$, though in this case, we cannot simply count each collision
point as contribution a hypermultiplet. Rather, these collisions collectively describe the internal profile of a matter
mode which is ``spread'' across these points. Though the geometric characterization of these cases is more subtle, it still describes a
weakly coupled hypermultiplet in the standard sense of 6D field theory.

But it can also happen that a collision of seven-branes cannot be interpreted in
terms of weakly coupled hypermultiplets. Indeed, this turns out to be the generic situation when analyzing
6D\ SCFTs. An example of this type is the collision of two $\mathfrak{e}_{8}$ (also known as type $II^*$)
singularities:%
\begin{center}
\includegraphics[width=50mm]{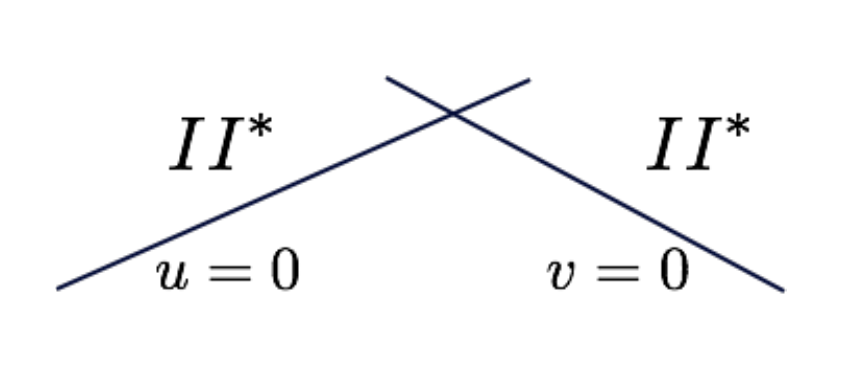}
\end{center}
The Weierstrass model that locally describes this collision is given by
\begin{equation}
y^{2}=x^{3}+u^{5}v^{5}\text{.}%
\end{equation}
If we restrict to the locus $u=v$, we see a Weierstrass model which is clearly
not in Kodaira-Tate form. This means that additional physical and mathematical
structure is localized along $u=v=0$. Indeed, this is an example of
\textquotedblleft6D\ conformal matter,\textquotedblright a phenomenon we will discuss at length in later sections.

For compactifications to four- or two-dimensional
theories, there can be additional triple and quartic intersections of
seven-branes. These will not play a role in the present review article, but
for completeness we note that such intersections correspond to interaction
terms between matter fields.

\subsection{10D and 8D\ Vacua}

To illustrate some of the general points, we proceed to the
explicit F-theory model associated with 10D, 8D and 6D\ vacua. As the
spacetime dimension decreases, the corresponding complexity of the internal
geometry increases, indicating a corresponding increase in the sorts of vacua
which can be realized.

To begin, we can consider the case of 10D\ vacua with F-theory on a constant
elliptic curve. In this case, $f$ and $g$ are simply constants. The particular
value of $f$ and $g$ dictates a fixed choice of axio-dilaton.\footnote{The specific value can be
read off from the $j$-invariant of the elliptic curve. It is given by the power series $j(q) = \frac{1}{q} + 744 + 196884 q + ...$ with $q = \exp(2 \pi i \tau)$, and with $f$ and $g$ related to $j(q)$ by: $j(q) = 1728 \times \frac{4 f^3}{4 f^3 + 27 g^2}$.}

Proceeding next to eight-dimensional vacua, we need to specify a complex
one-dimensional base $\mathcal{B}$ and an elliptic fibration over this base
so that the total space is a Calabi-Yau twofold. There is
precisely one compact K\"ahler surface available to us:\ an elliptically fibered K3 surface, in
which the base is a $\mathbb{CP}^{1}$. In this case, we have an
eight-dimensional spacetime, and the base of the F-theory model $\mathcal{B}$
is a $\mathbb{CP}^{1}$. The ten-dimensional
spacetime is $\mathbb{R}^{7,1}\times\mathbb{CP}^{1}$ which is clearly not
Ricci flat. Nevertheless, it is a consistent solution to the supergravity
equations of motion due to the backreaction of seven-branes placed at points
of the $\mathbb{CP}^{1}$. F-theory tells us precisely where these seven-branes
are located, as dictated by the Weierstrass model for the elliptically fibered
K3 surface. Returning to our general discussion around line (\ref{fgsections}%
), we note that since the canonical class of $\mathbb{CP}^{1}$ is the line
bundle $\mathcal{O}(-2)$, $f$ and $g$ are respectively sections of
$\mathcal{O}(8)$ and $\mathcal{O}(12)$, and the discriminant
polynomial $\Delta=4f^{3}+27g^{2}$ is a section of $\mathcal{O}(24)$.
Said differently, $f,g,\Delta$ are respectively homogeneous polynomials of degrees $8$,
$12$ and $24$ in the homogeneous coordinates $[u,v]$ of the $\mathbb{CP}^{1}$.
We can then present the K3 surface as:%
\begin{equation}
\text{Elliptic K3: }y^{2}=x^{3}+f_{8}x+g_{12}.
\end{equation}
To see the correspondence with the heterotic string, it is instructive to
consider the special case where we take:%
\begin{equation}
E_{8}\times E_{8}\text{ Limit:\ }y^{2}=x^{3}+(g_0 u^{5}v^{7}+ g_{\infty} u^{7}v^{5})
+ \left(\alpha u^{4}v^{4}x+ \beta u^{6} v^{6} \right).
\end{equation}
Returning to our list of singularities in table \ref{tabooboo}, we
see that in the patch where $v=1$, there is an $E_{8}$ singularity present at
$u=0$, and in the patch where $u=1$, there is an $E_{8}$ singularity where
$v= 0$. These are nothing but the two $E_{8}$ factors of the usual heterotic
string. More precisely, there is a duality between heterotic strings on
$T^{2}$ and F-theory on an elliptically fibered K3 surface \cite{Morrison:1996na,Morrison:1996pp}:%
\begin{equation}
\text{Heterotic / }T^{2}\leftrightarrow\text{F-theory / }K3, \label{HETF}%
\end{equation}
which is a lift of the well-established correspondence between heterotic
strings on $T^{4}$ and type II\ strings on a K3 surface \cite{Kachru:1995wm}. One can
perform a detailed match of the moduli on the two sides of this
correspondence. For example, deformations of the seven-branes to more generic
positions correspond to Wilson lines on the $T^{2}$ in the heterotic string description.

In the context of models decoupled from gravity, it is particularly helpful to
take a limit where we can \textquotedblleft zoom in\textquotedblright\ on just
one of these $E_{8}$ walls. This can be achieved in the so-called
\textquotedblleft stable degeneration limit.\textquotedblright\ It involves
taking a family of metrics for the K3 surface in which the base
degenerates to a long tubular cylinder, with the singularities of the elliptic
fiber localized at opposite ends of the cylinder. This can be viewed as
another elliptically fibered surface, namely a \textquotedblleft del Pezzo
nine surface\textquotedblright\ or $\frac{1}{2}K3$. In this model, the degrees
of $f$ and $g$ are half what they are for the K3 surface, and are given by:
\begin{equation}
dP_{9}\text{: }y^{2}=x^{3}+f_{4}x+g_{6}.
\end{equation}
Note that this is not a Calabi-Yau space, since the
putative homolorphic two-form of the surface has a pole along the
anti-canonical class of the $dP_{9}$, which is an elliptic curve. In the
$dP_{9}$, this amounts to picking a point of the base $\mathbb{CP}^{1}$ and
the corresponding elliptic fiber as well. One can produce a
\textit{non-compact} Calabi-Yau by deleting the offending region from the
space. From the perspective of the stable degeneration limit, we glue two such
$dP_{9}$'s along a common $T^{2}$. Observe that in this local geometry, we can
get just a single $E_{8}$ factor, since it is now possible to specialize to
the case $g_{6}\sim u^{5}v$, with a single $E_{8}$ singularity located at
$u=0$.

\subsection{6D\ Vacua \label{ssec:6Dvac}}

Let us now proceed to some general aspects of 6D vacua, as well as some examples
of F-theory models which realize this structure.

To begin, we can ask about the origin of the various $\mathcal{N} = (1,0)$ supermultiplets
already encountered in Section \ref{sec:BOTTOM}. The main idea is to proceed
by dimensional reduction of higher-dimensional forms.\footnote{Readers unfamiliar
with this procedure should consult \cite{Green:1987mn}.} For us, this involves
the decomposition of the higher-dimensional Laplacian into a 6D Laplacian and
an internal Laplacian:%
\begin{equation}
\nabla_{\text{high}}^{2}=\nabla_{\text{6D}}^{2}+\nabla_{\text{int}}^{2}.
\end{equation}
Massless states of the 6D theory therefore descend from harmonic forms on the
internal directions, i.e. differential forms which are annihilated by
$\nabla_{\text{int}}^{2}$. Using the relation:%
\begin{equation}
\nabla_{\text{int}}^{2}=\ast_{\text{int}}d_{\text{int}}\ast_{\text{int}}d_{\text{int}}+d_{\text{int}}\ast_{\text{int}}d_{\text{int}}\ast_{\text{int}},
\end{equation}
one can also show that such massless states are computed by a suitable
cohomology theory on the internal directions (closed forms modulo exact forms).

\begin{figure}[t!]
\centering
\includegraphics[trim=0mm 35mm 0mm 20mm, clip, width=140mm]{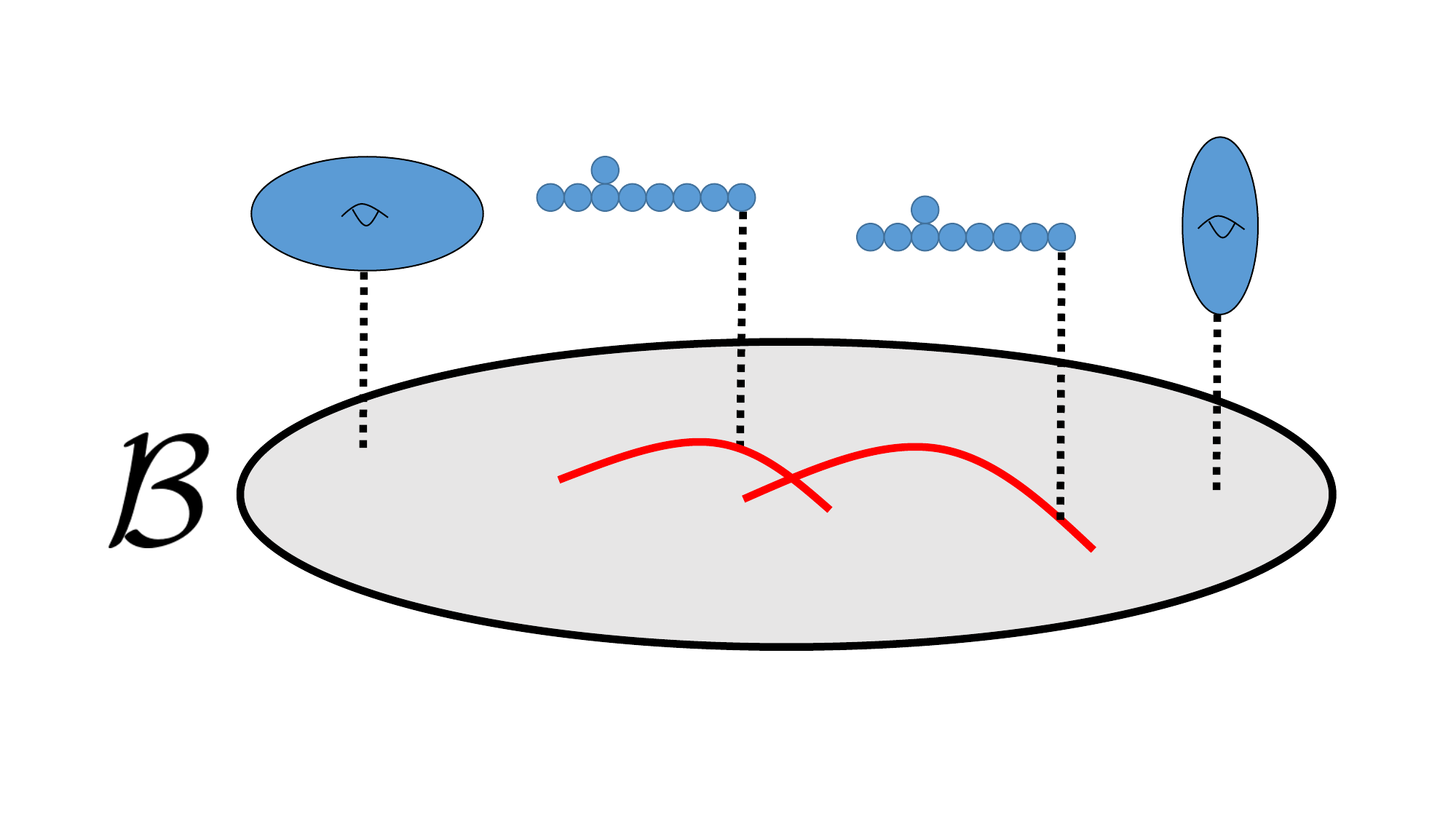}
\caption{Geometry for a 6D F-theory vacuum. The torus fiber degenerates over a codimension one curve (red),
producing vector multiplets in the 6D theory. Hypermultiplets are localized at the intersection point
of two such curves (codimension two). Tensor multiplets come from dimensionally reducing the fields of type IIB string
theory on these curves, and effective strings come from D3-branes wrapping the curves.}
\end{figure}

Let us apply this general prescription to now see how the various
supermultiplets of the 6D effective field theory are realized in F-theory.
First, we ask about the origin of the $\mathcal{N} = (1,0)$ vector multiplet. This comes
from seven-branes wrapping curves of the geometry. Indeed, as we have already
remarked, the singularity type of the elliptic fiber dictates the choice of
gauge group in six dimensions, so we recover a 6D\ vector multiplet with
corresponding gauge group $G$. In the special case where $G$ is simply laced,
one can directly see this by dimensional reduction of an 8D gauge field on the
curve. The 8D gauge field splits as:%
\begin{equation}
A_{8D}=A_{6D}+A_{\text{int}}, \label{A8D}%
\end{equation}
in the obvious notation. The $A_{6D}$ is a 6D gauge field that (with its
superpartners) fills out the 6D vector multiplet for gauge group
$G$. In the case where the elliptic fiber is not split so that we realize a
quotient by an outer automorphism, there is a marked \textquotedblleft twist
point\textquotedblright\ on the curve. This affects the dimensional reduction
and projects out some of the states of the simply laced \textquotedblleft
parent algebra.\textquotedblright

Consider next the F-theory origin of 6D hypermultiplets. In a general F-theory
vacuum, these can originate from three sources. First, there is the overall
volume modulus of the base $\mathcal{B}$ and its superpartners. This will play no role
in our discussion, as we shall always work on a non-compact base where gravity
is decoupled.

Another way to realize hypermultiplets is by compactifying a seven-brane on a
curve $\Sigma$ of genus $g>0$. This yields $h^{1,0}(\Sigma)$ complex scalars from the
dimensional reduction of $A_{\text{int}}$ of line (\ref{A8D}). These are
joined by the dimensional reduction of an adjoint valued $(1,0)$ form of the
seven-brane gauge theory, which is the 8D superpartner of the 8D gauge
field. This provides a way to get hypermultiplets in the adjoint
representation of a gauge group, and by construction, are completely non-localized (they are spread over the entire curve).
Moreover, much as in our discussion of delocalized matter in the presence of non-split fiber types, further non-localized
matter contributions in the adjoint representation arise when there is a difference in genus between the double cover of the cover ramified at the ``twist points'' and the genus of the component of the discriminant, the difference being half of ($b$ - 2), with $b$ the number
of branch points.\footnote{We thank D.R. Morrison for emphasizing this subtlety, and for patient explanations which we have summarized here.} See
reference \cite{Grassi:2000wer} for additional details.

In the specific context of 6D\ SCFTs where we always work with genus $g=0$ curves such subtleties will
play no role, and we therefore neglect them in what follows.

The last way to realize 6D hypermultiplets in more general representations
comes from the collision of distinct components of seven-branes. In the
specific context of 6D\ SCFTs where we always work with genus $g=0$ curves,
only these sort of multiplets will appear on the tensor branch.

Finally, there is the tensor multiplet. Let us consider the
dimensional reduction of the 10D metric on a compact curve. In this vein, it
is helpful to introduce the K\"ahler form $J$ of the base $\mathcal{B}$ and decompose
$J$ into a basis of harmonic two-forms on the internal geometry:%
\begin{equation}
J=t^{(i)}\wedge\omega_{(i)}.
\end{equation}
Here, $i=1,...,b_{2}^{\text{cpct}}$ runs over the basis of two-forms with
compact support with integral intersection pairing:%
\begin{equation}
A_{ij}=\underset{\mathcal{B}}{\int}\omega_{(i)}\wedge\omega_{(j)}.
\end{equation}
Dimensional reduction of the term $dJ\wedge\ast dJ$ yields a kinetic term for
these scalars:%
\begin{equation}
L_{kin}=-A_{ij}dt^{i}dt^{j}\text{,} \label{metric}%
\end{equation}
which is a natural metric on the K\"ahler moduli space.\footnote{Strictly speaking, we are considering
an expansion of the metric for the K\"ahler moduli near the conformal fixed point.
The form of the metric presented here follows the
discussion in reference \cite{Apruzzi:2017iqe}. For a more general
account of Weil-Petersson metrics in Calabi-Yau compactification, see e.g. reference
\cite{Candelas:1990pi}.} This is identical to the metric introduced in
line \ref{tensormetric}. Let us note that in this
presentation, the volume of the corresponding curve $\Sigma_{i}$ is:%
\begin{equation}
\text{Vol}(\Sigma_{i})=A_{ij}t^{j}. \label{VolVol}%
\end{equation}
The $t^i$ are the scalars of the 6D $\mathcal{N} = (1,0)$ tensor multiplets. They are
accompanied by anti-chiral two-forms coming from reduction of the chiral
four-form of type IIB\ string theory on these curves. The reduction of this
action is a bit subtle owing to the 10D\ self-duality condition, and the most
systematic way to read off various properties of this four-form and its
reduction to six dimensions involves extension to an 11D spacetime with
boundary \cite{Belov:2006jd} (see also \cite{Heckman:2017uxe}).

A byproduct of this discussion is that we can also readily identify the
effective strings of the 6D theory in terms of D3-branes wrapped over curves.
The tension of this effective string is simply the volume of the corresponding
two-cycle wrapped by the D3-brane. The tensionless string limit
corresponds to taking all volumes to zero, thus moving to the origin of the tensor branch. Note also that this
integral pairing $A_{ij}$ is nothing but the Dirac pairing matrix of (\ref{eq:Diracpairing}).

There are in general two sorts of deformations in the space of Calabi-Yau
metrics which both descend to physical operations in the 6D effective field
theory. First of all, given a Weierstrass model,%
\begin{equation}
y^{2}=x^{3}+fx+g,
\end{equation}
we consider complex structure deformations of the model. For elliptically fibered
Calabi-Yau spaces, this amounts to perturbations in the coefficients $f$ and
$g$. In particular, we can start from a very singular presentation of the
Calabi-Yau, and then switch on smoothing deformations to proceed to a less
singular model. An example of this sort is:
\begin{equation}
y^{2}=x^{3}+u^{5}\mapsto y^{2}=x^{3}+\varepsilon u^{3}x+u^{5},
\end{equation}
which describes the unfolding of an $E_{8}$ singularity to $E_{7}$. In the
6D\ effective field theory, the complex structure moduli partner with the
intermediate Jacobian to form 6D hypermultiplets
\cite{Aspinwall:1998he, Anderson:2013rka}.\footnote{For a review of the analogous statements obtained from
4D $\mathcal{N} = 2$ vacua obtained from compactifications of IIA string theory on
Calabi-Yau threefolds, see the review \cite{Alexandrov:2013yva} and references therein.}

We can also consider K\"ahler deformations of the Calabi-Yau threefold. In the
present context where we must always retain a singular elliptic fiber, this
amounts to changing the volumes of curves in the base $\mathcal{B}$. In
the 6D effective field theory, we associate this with adjusting
the background values of vevs for the tensor multiplet scalars $t^{i}$.

\subsection{Non-Higgsable Clusters \label{ssec:NHC}}

In this section, we introduce non-Higgsable clusters, which are the geometric building blocks of 6D F-theory geometries. The reader who is interested in more technical details should consult the discussion of Hirzebruch surfaces in Appendix \ref{sec:Hirzebruch}.

In general, it is clearly a daunting problem to provide an explicit list of
all possible bases, and all possible elliptic fibrations over such bases that are suitable for F-theory compactification. In \cite{Morrison:2012np}, a classification of bases was proposed. The essential idea in this
approach is to consider an elliptically fibered Calabi-Yau threefold with
generic complex structure moduli. In this situation, we have, in the 6D
effective field theory, performed all possible Higgsing operations of gauge
groups, so anything left over cannot be Higgsed further. The key observation
of reference \cite{Morrison:2012np} is that in this maximally Higgsed phase, we can
characterize the geometry of the base in terms of a configuration of minimal
building blocks. Moreover, the elliptic fiber can still be non-trivial, since
the non-trivial curvature of the base needs to be
compensated by wrapping seven-branes over such curves.

As shown in \cite{Morrison:2012np} via an intersection-theoretic argument
(and also in the earlier thesis \cite{grassi1990minimal}), curves of
self-intersection $-1$ and $-2$ may have a smooth fiber. Indeed, the former
produces the rank 1 E-string theory of a single M5-brane probing an $E_8$ wall,
discussed in Section \ref{ssec:EXAMPLES}, while the latter produces
the type $A_1$ (2,0) theory discussed in Section \ref{ssec:OLDIES}.
However, curves of self-intersection $-n$ for $n>2$ necessarily have
a nontrivial fiber, producing a non-Abelian gauge algebra in the 6D
effective theory.

These curves of self-intersection $-n$, $n > 2$ represent the simplest
examples of ``non-Higgsable clusters (NHCs)": configurations of curves in
Calabi-Yau geometries with no smoothing deformations, corresponding to a
field theory with trivial Higgs branch. As a result, the gauge algebras
associated with these curves cannot be further Higgsed.

These NHCs build up larger bases by joining them with a $-1$ curve.
The rough idea here is that the $-1$ curve carries an $E_{8}$ flavor
symmetry so we can weakly gauge parts of this flavor symmetry as:
\begin{equation}
1,[E_{8}]\rightarrow\overset{\mathfrak{g}_{L}}{a},1,\overset{\mathfrak{g}%
_{R}}{b}, \label{gluing}%
\end{equation}
where $\mf{g}_L$ and $\mf{g}_R$ must satisfy the ``$E_8$ gauging condition,"
\begin{align}
\mf{g}_L \times \mf{g}_R \subset \mf{e}_8.
\label{eq:E8condition}
\end{align}
here, we have adopted the following notation:%
\begin{equation}
m,n
\end{equation}
for curves of self-intersection $-m$ and $-n$ which intersect at a single
point. Additionally, the notation%
\begin{equation}
\overset{\mathfrak{g}}{a}
\end{equation}
refers to a curve of self-intersection $-a$ that has a seven-brane of gauge
algebra $\mathfrak{g}$ wrapped over it.

The full list of NHCs was determined in \cite{Morrison:2012np}.
Each such NHC consists of either one, two, or three curves. For the single curve
theories, we have:%
\begin{align}
&  \text{Single Curve NHCs}\\
&
\begin{tabular}
[c]{|c|c|c|c|c|c|c|c|}\hline
$n$ & $3$ & $4$ & $5$ & $6$ & $7$ & $8$ & $12$\\\hline
Theory & Pure $\mathfrak{su}(3)$ & Pure $\mathfrak{so}(8)$ & Pure
$\mathfrak{f}_{4}$ & Pure $\mathfrak{e}_{6}$ & $\mathfrak{e}_{7}+\frac{1}%
{2}\textbf{56}$ & Pure $\mathfrak{e}_{7}$ & Pure $\mathfrak{e}_{8}$\\\hline
\end{tabular}
,
\label{eq:NHCalgebras}
\end{align}
where in the above, the notation $\frac{1}{2}\textbf{56}$ for the $n=7$ theory refers
to having a half hypermultiplet in the fundamental representation. This is
possible because this representation is pseudoreal. In the $n=9,10,11$
theories, the elliptic fiber does not remain in Kodaira-Tate form over the
entire curve unless we perform further blowups in the base.
There are several ways to do this, depending on whether we have $n=9$, $10$, or $11$:%
\begin{align}
n  &  =9:1,\ov{1}{(12)},1 \mbox{ or }  1,(12),1,2 \mbox{ or } (12),1,2,2\\ \label{ninecurve}
n  &  =10:1,(12),1 \mbox{ or } (12), 1, 2\\
n  &  =11:(12),1. \label{elevencurve}
\end{align}
Where the relative positions in the above indicate how the various curves intersect pairwise.

For the two-curve theories, there is just one NHC:%
\begin{align}
&  \text{Two Curve NHCs}\\
&
\begin{tabular}
[c]{|c|c|}\hline
$n,m$ & $3,2$\\\hline
Theory & $\mathfrak{g}_{2}\times\mathfrak{su}(2)+$ $\frac{1}{2}(\mathbf{7},\mathbf{2})+\frac
{1}{2}(\mathbf{1},\mathbf{2})$\\\hline
\end{tabular}
.
\end{align}
Here, the $\mathfrak{g}%
_{2}$ comes from a non-split $I_{0}^{\ast}$ fiber localized on the $-3$ curve,
and the $\mathfrak{su}(2)$ factor comes from a type $III$ fiber localized on
the $-2$ curve. Additionally, the $(\mathbf{7},\mathbf{2})$ is a bifundamental representation
localized at the intersection of the $-3$ and $-2$ curve, and the $(1,2)$ is
matter localized at a distinct point of the $-2$ curve.

Finally, there are two possible three-curve configurations:%
\begin{align}
&  \text{Three Curve NHCs}\\
&
\begin{tabular}
[c]{|c|c|c|}\hline
$n,m$ & $3,2,2$ & $2,3,2$\\\hline
Theory & $\mathfrak{g}_{2}\times\mathfrak{sp}(1)\times\emptyset+$ $\frac{1}%
{2}(\mathbf{7},\mathbf{2},\mathbf{1})+\frac{1}{2}(\mathbf{1},\mathbf{2},\mathbf{1})$ & $\mathfrak{su}(2)\times\mathfrak{so}%
(7)\times\mathfrak{su}(2)+$ $\frac{1}{2}(\mathbf{2},\mathbf{8},\mathbf{1})+\frac{1}{2}(\mathbf{1},\mathbf{8},\mathbf{2})$\\\hline
\end{tabular}
\ .
\end{align}
In the $3,2,2~$configuration, the $\mathfrak{sp}(1)\simeq\mathfrak{su}(2)$ is
localized on the middle $-2$ curve, and the absence of a gauge algebra on the
rightmost $-2$ curve has been indicated by the symbol \textquotedblleft%
$\emptyset$\textquotedblright. The reason for the different notation for this
gauge algebra is that the $\mathfrak{sp}(1)$ factor arises from a non-split
type $IV$ fiber. In the $2,3,2$ configuration the $\mathfrak{su}(2)$ factors
come from type $III$ fibers localized on $-2$ curves and the $\mathfrak{so}%
(7)$ factor comes from a non-split $I_{0}^{\ast}$ fiber localized on the $-3$ curve.  Although it is important to keep this distinction in mind from a F-theory perspective, we will often use $\mf{su}(2)$ and $\mf{sp}(1)$ interchangeably.

A priori, one might have thought that a classification of bases comprised solely of NHCs joined by $-1$ curves would be insufficient, since a generic SCFT is, in fact, Higgsable. Na\"ively, one might have expected that once charged matter is added and fibers are enhanced, new base geometries would also arise. However, it turns out that this does not happen: all 6D F-theory SCFTs can be constructed by first building a base out of NHCs and then enhancing the fibers over this base. This is the magic of NHCs: the structure of a generic 6D SCFT is fixed almost entirely by the structure of a 6D SCFT built solely from NHCs glued with $-1$ curves.

Finally, we remark that in most applications to 6D SCFTs, it is customary to work with singular elliptic
fibrations. It is also helpful, especially in the context of compactification to lower-dimensional
theories to perform explicit resolutions of these geometries. This has been carried out for example
in references \cite{Hayashi:2017jze, DelZotto:2017pti, Esole:2017rgz, Esole:2018mqb}.

\subsection{Frozen Singularities \label{ssec:FROZEN}}

Our discussion in the previous subsections assumed the existence of a
geometric phase of F-theory. It is natural to ask whether there are other
non-geometric regimes which can produce additional examples of 6D\ SCFTs.

Some examples of this type have been referred to as \textquotedblleft frozen
singularities\textquotedblright\ in the context of M-theory and F-theory
compactification (see \cite{Witten:1997bs, deBoer:2001wca, Tachikawa:2015wka}) and amount to
singular geometries in the presence of a discrete (torsional) flux. This leads to a
configuration where the usual smoothing deformations of a geometry have been
projected out, so there is no modulus available in the effective field
theory to deform the geometry back to a smooth non-singular configuration.

The main idea can already be stated by considering M-theory on the background
$\mathbb{R}^{6,1}\times\mathbb{C}^{2}/\Gamma_{G}$ for $\Gamma_{G}$ a discrete
subgroup of $SU(2)$. In the absence of any additional fluxes, we know that
this realizes 7D\ super Yang-Mills theory with corresponding gauge group $G$.
Observe that the boundary of this geometry takes the form of a quotient
$S^{3}/\Gamma_{G}$, so one can ask what happens if there is
non-trivial torsion carried by a period of $C_{3}$, the three-form potential
of M-theory:%
\begin{equation}
r=\underset{S^{3}/\Gamma_{G}}{\int}C_{3},
\end{equation}
where $r$ is a rational number defined modulo $1$, namely we can view it as an
element $r\in\lbrack0,1)$. When this period is non-trivial, the actual gauge
algebra of the M-theory model changes, and one often encounters a non-simply
laced algebra instead. We refer the interested reader to \cite{Tachikawa:2015wka}
for further details.

Based on the close correspondence between M- and F-theory models, it is reasonable to
ask what becomes of this discrete torsional data in M-theory. For
the most part, this data is actually absorbed into geometric phases of
F-theory. For example, non-simply laced algebras abound in F-theory models,
and arise from having monodromy in the basis of two-cycles of an
elliptic fibration.

In reference \cite{Tachikawa:2015wka} it was pointed out that the only
Kodaira types of elliptic fibers which could remain frozen are those of type
$I_{n}^{\ast}$. Recall that in F-theory, we typically associate the
split case (no monodromy) with an $\mathfrak{so}(2n+8)$ gauge algebra. In
perturbative type IIB\ language, this comes from a stack of $n+4$
D7-branes near a single $O7^{-}$ orientifold plane. A frozen singularity
corresponds to the case where we change the form of the orientifold projection
so that instead of $O7^{-}$ planes we have $O7^{+}$ planes. An
important feature of changing the orientifold projection in this way is that
the effective monodromy of the type IIB\ axio-dilaton cannot distinguish
between the case of $k+8$ D7-branes and an $O7^{-}$ plane and $k$ D7-branes
and an $O7^{+}$ plane.

The question of how to consistently incorporate such frozen singularities in
F-theory is still an open problem, but for our present purposes, the primary
question is what impact this could have on the construction of 6D\ SCFTs.
Examples of 6D\ SCFTs using $O^{+}$-planes were presented in
\cite{Hanany:1997gh} and were also studied from the perspective of effective
field theory in \cite{Bhardwaj:2015xxa} so such objects can
appear in non-geometric phases of F-theory. Even so, the conditions imposed by
anomaly cancellation \cite{Bhardwaj:2015xxa} are sufficiently stringent that the
outlying examples of 6D\ SCFTs realized in F-theory are quite rare. Moreover,
in all known constructions involving such objects, the resulting model can
always be viewed as a \textquotedblleft quotient\textquotedblright\ of a
geometric phase of an F-theory background \cite{Bhardwaj:2015oru, Apruzzi:2017iqe}.
For the purposes of this review, then, we shall not discuss such non-geometric phases further.

\section{6D\ SCFTs via F-theory \label{sec:FAGAIN}}

Having now introduced the main elements of F-theory backgrounds, we now turn
to the explicit construction of 6D\ SCFTs in this approach. First, we provide
the general conditions for realizing 6D\ SCFTs in F-theory. After this, we present some
examples, illustrating how various stringy constructions appear in this description.

\subsection{SCFT Conditions \label{ssec:CONDITIONS}}

We begin with the F-theory construction of 6D\ SCFTs. Our starting point is an
elliptically fibered Calabi-Yau threefold $X$ with base $\mathcal{B}$, namely
$X\rightarrow\mathcal{B}$. Since we are dealing with a theory decoupled from
gravity, we always take $\mathcal{B}$ to be non-compact. Assuming that we have
successfully constructed such a Calabi-Yau threefold, we have a spectrum of
effective strings associated with D3-branes wrapped on compact two-cycles of
the base $\mathcal{B}$. To reach an SCFT, we need these cycles to collapse to
zero size, and we need this collapse to happen at finite distance in moduli space.

As one might suspect, this imposes severe limitations on the
configurations of curves which actually participate in generating an SCFT.
First, we only need to consider curves of genus zero, which are topologically $\mathbb{CP}^{1}$'s. Further, these curves must have
self-intersection $-n$, with $n>0$. So, the local geometry of each
$\mathbb{CP}^{1}$ is $\mathcal{O}(-n)\rightarrow\mathbb{CP}^{1}$.

This condition extends to multiple curves. Indeed, returning to the metric of
the moduli space (\ref{metric}), we require that our
metric is positive definite, so the intersection pairing matrix is
negative definite:%
\begin{equation}
A<0\text{.}%
\end{equation}
For further discussion on why this condition is necessary and sufficient to establish contractibility,
see references \cite{MR0153682, MR0137127, MR0146182}, as well as the discussion in
Appendix B of reference \cite{Heckman:2013pva}.

This negative definiteness has some additional conditions, as imposed by the
geometry of our curves. Each such curve can only intersect a distinct curve at
one point. Otherwise, there would be a closed loop and one would generate an
associated length scale from objects wound around it.

This can also be seen in more mathematical terms by analyzing the order of vanishing of $f$ and $g$ in
the associated Weierstrass model: If we have a closed loop of curves in a configuration, we find that
the condition of contractibility $A < 0$ leads to an order of vanishing for $f$ and $g$ which is incompatible with the
existence of a Weierstrass model \cite{Heckman:2013pva}. As a special case, this also means that
no three curves can ever meet at a single point in such an F-theory model, since we can deform this ``triple'' to form a closed loop.
It also means we cannot have any tangential intersections of curves because again, we could deform this to form a closed loop.

In the resolved phase, this leads to a very helpful rule of thumb for what sorts of configurations
of curves can intersect. First of all, all curves must intersect transversely (namely, no tangencies). Second of
all, a $-1$ curve can intersect at most two other curves. The reason is that if it were to intersect three curves, we could collapse
the $-1$ curve to zero size, reaching a configuration of curves forming a triple intersection in the base, a contradiction
with the statements of the previous paragraph.

The operating assumption is that after choosing a base $\mathcal{B}$
with some configuration curves satisfying the above properties, we can also
find a consistent elliptic fibration which generates a Calabi-Yau
threefold.\ Note that in most cases, a non-trivial elliptic fibration may be
required, as per our discussion of $-n$ curves in subsection \ref{ssec:6Dvac}.

Building up a consistent base for a 6D\ SCFT amounts to taking the
$-1$ curves, $-2$ curves and NHCs discussed in subsection
\ref{ssec:NHC} and joining them together in such a way that we have a
consistent elliptic fibration and a negative definite intersection pairing $A$.

In general, we can see that the basic building blocks consist of the NHCs, as
well as configurations of $-2$ curves which can simultaneously contract to zero
size. The latter building blocks are classified by the Dynkin diagrams of the
simply laced algebras:%
\begin{align}
A_{N}  &  :2,...,2 \label{eq:Andiagram}\\
D_{N}  &  :2,\overset{2}{2}...,2\\
E_{6}  &  :2,2,\overset{2}{2},2,2\\
E_{7}  &  :2,2,\overset{2}{2},2,2,2\\
E_{8}  &  :2,2,\overset{2}{2},2,2,2,2,
\label{eq:E8diagram}
\end{align}
so the basic \textquotedblleft atoms\textquotedblright\ of any 6D\ SCFT
consist of the following possibilities:%
\begin{equation}
-1\text{ curve; \ \ }ADE\text{ with }-2\text{ curves; \ \ NHCs.}%
\end{equation}
We build larger configurations of curves in the base by a gluing construction, as in (\ref{gluing}). In the special context of 6D\ SCFTs, we note
that a $-1$ curve can only touch a curve of self-intersection $-n$ for $n>1$.
The reason is that the intersection of two $-1$ curves does not yield a
negative definite intersection pairing. The general configuration of curves in
a base therefore takes the form of a tree-like structure consisting of ADE
configurations of $-2$ curves and NHCs, glued together by $-1$ curves.

In addition to our conditions on the base, we also need to ensure that an
elliptic fibration exists so that the total space is a Calabi-Yau threefold.
Here, we focus on the physical aspects of these constraints. The most
stringent condition is one that we can already see in the low energy effective
field theory: 6D gauge anomaly cancellation. This severely
restricts the matter content and neighboring gauge groups. In geometric terms,
this translates to the condition that the elliptic fibers over each curve
remain in Kodaira-Tate form.

Another constraint comes from how we use our $-1$ curves to \textquotedblleft
glue together\textquotedblright\ neighboring configurations of curves. For
example, when there is no gauge algebra over the $-1$ curve, we know that in
the limit where this $\mathbb{CP}^{1}$ collapses to zero size, we get an
E-string theory with flavor symmetry $E_{8}$. We can consistently gauge a
subalgebra of this $E_{8}$, and from the perspective of the tensor branch,
this amounts to the replacement:%
\begin{equation}
1,[E_{8}]\rightarrow\overset{\mathfrak{g}_{L}}{a},1,\overset{\mathfrak{g}%
_{R}}{b},
\end{equation}
for gauge algebras $\mathfrak{g}_{L}\times\mathfrak{g}_{R}\subset\mathfrak{e}_{8}$. We
already encountered this condition in the context of building a consistent
base, but we see that it naturally generalizes to the case where we allow
non-minimal fiber types as well. Note, however, that this condition only
applies when the $-1$ curve is not paired with any gauge algebra.  If it is
paired with a gauge algebra, the conditions on neighboring gauge algebras
$\mf{g}_L$ and $\mf{g}_R$ and charged matter follow from the Katz-Vafa collision rules
introduced earlier, or in field theory terms, from cancellation of gauge and mixed gauge-gauge anomalies.

As we have already mentioned, the description in terms of an
elliptically fibered Calabi-Yau threefold over a smooth base $\mathcal{B}$
implicitly makes reference to the tensor branch of a putative 6D\ SCFT. We
need to take this model and collapse all of the available curves to actually
reach a 6D SCFT. Doing this provides an alternate characterization of the same 6D\ SCFTs, in
terms of F-theory on a singular base $\mathcal{B}_{\text{sing}}$. A helpful
intermediate case is to consider what happens as we collapse all of the $-1$
curves to zero size. This is blowdown of a curve, and the homology class of
neighboring curves which intersect the $-1$ curve will be affected by such a
blowdown. To illustrate, suppose we have a pair of curves of self-intersection
$-n$ and $-m$ which intersect at a point. Blowing up that point amounts to introducing this additional $-1$
curve. In doing so, the homology class for neighboring curves changes as:%
\begin{equation}
\lbrack\Sigma_{\text{old}}]\rightarrow\lbrack\Sigma_{\text{old}}%
]-E=[\Sigma_{\text{new}}],
\end{equation}
where $E$ refers to the \textquotedblleft exceptional
divisor\textquotedblright\ of self-intersection $-1$, and by definition,
\begin{equation}
[\Sigma_{\text{new}}]\cdot E=+1.
\end{equation}
The self-intersection of the new homology class is:
\begin{equation}
\lbrack\Sigma_{\text{new}}]\cdot\lbrack\Sigma_{\text{new}}]=-(n+1).
\end{equation}
So, in configurations of curves which touch at a point, we have the blowup and
blowdown operations:
\begin{align}
\text{Blowup}  &  \text{: }m,n\overset{up}{\rightarrow}%
(m+1),1,(n+1)\label{blowagain}\\
\text{Blowdown}  &  \text{: }(m+1),1,(n+1)\overset{dn}{\rightarrow}m,n.
\label{blowdowneq}
\end{align}
Proceeding in this way, we can express the actual base as a resolution of a more singular space, $\mathcal{B}_{\text{sing}}$. Note that as we
continue to blowdown from $\mathcal{B}$ to $\mathcal{B}_{\text{sing}}$,
additional $-1$ curves may appear, as can be seen from line (\ref{blowdowneq}), setting either $m$ or $n$ equal to 1.
Continuing to blowdown all curves until all have self-intersection $-n$ for
$n>1$, we reach an \textquotedblleft endpoint configuration\textquotedblright%
\ of curves. Collapsing this set of curves takes us to the singular base
$\mathcal{B}_{\text{sing}}$.

Starting from an endpoint configuration of curves, we can also reverse the
procedure, blowing up until we reach a genuine smooth base, which describes the full
tensor branch of a candidate SCFT. Some such blowups are actually required to
reach an elliptic fibration with all fibers in Kodaira-Tate form. To see why,
recall that for any curve of self-intersection $-n$ with $n>2$, we know that
there is a singular elliptic fiber over this curve. If a $-n$ and $-m$ curve
with $n,m>2$ intersect at a point, then we already know that a blowup is
required because the configuration $n,m$ is not an NHC. Blowing up once, we
get%
\begin{equation}
m,n\overset{up}{\rightarrow}(m+1),1,(n+1).
\end{equation}
Note that the putative elliptic fibers over the curves to the left and right of
the $-1$ curve now have a lower self-intersection, so the minimal candidate
elliptic fiber has become more singular. Proceeding in this way until we reach
a base with all fibers in Kodaira-Tate form, we may need to blowup
additional times.

Let us also note that in some cases, the $-1$ curve will necessarily carry a
non-trivial fiber type in these situations. A configuration which appears repeatedly in many examples is the sub-collection
of curves:
\begin{equation}
5,1,3,2,2
\end{equation}
where there is an $\mathfrak{f}_4$ algebra on the $-5$ curve and a $\mathfrak{g}_2$ algebra on the $-3$ curve.
On the $-1$ curve there is also a type $II$ fiber.

\subsubsection{Examples of Blowups}

Let us illustrate with a few examples. We focus
on the minimal resolutions of a base, though we emphasize that additional blowups of the base can sometimes be included, yielding new consistent bases with minimal singularity type for the elliptic fiber.

As a first example, consider the configuration of curves $3,3$. Since we have
two $-3$ curves, we know that the elliptic fiber is at least as singular as a
type IV fiber, though it could be worse. The point of intersection of these two curves is then too singular to be compatible with F-theory, as can be seen immediately from the fact that $3,3$ is not an NHC. We
thus need to blowup the intersection point between the two curves, which yields:%
\begin{equation}
3,3\overset{up}{\rightarrow}4,1,4.
\end{equation}
Now, the minimal fiber on each $-4$ curve yields an $\mathfrak{so}(8)$
algebra, so since $\mathfrak{so}(8)\times\mathfrak{so}(8)$ is a subalgebra of
the flavor symmetry $\mathfrak{e}_{8}$ of the E-string theory, we have reached
a base compatible with our gluing rules.

As another example, consider the configuration of curves $4,4$. Following the
same logic as before, we know that the elliptic fiber on each curve is at
least as singular as a type $I_{0}^{\ast}$ fiber, though it could be more
singular. Indeed, blowing up once, the self-intersection shifts as:%
\begin{equation}
4,4\overset{up}{\rightarrow}5,1,5.
\end{equation}
In this case, the minimal fiber on each $-5$ curve is $\mf{f}_{4}$, and
$\mathfrak{f}_{4}\times\mathfrak{f}_{4}$ is not a subalgebra of $\mathfrak{e}%
_{8}$, so we must blowup further. Blowing up the intersection point of the
leftmost $-5$ curve and the $-1$ curve yields:%
\begin{equation}
5,1,5\overset{up}{\rightarrow}6,1,2,5
\end{equation}
so again, we do not have a consistent base since $2,5$ is not an NHC. Blowing
up the intersection point of the $-2$ curve and the $-5$ curve yields:%
\begin{equation}
6,1,2,5\overset{up}{\rightarrow}6,1,3,1,6.
\end{equation}
The minimal fiber on each $-6$ curve yields an $\mathfrak{e}_{6}$ algebra and
the minimal fiber on each $-3$ curve yields an $\mathfrak{su}(3)$ algebra.
Observe that we can consistently gauge the $\mathfrak{e}_{6}\times
\mathfrak{su}(3)$ subalgebra of the $\mathfrak{e}_{8}$ flavor symmetry for the
leftmost $-1$ curve, and we can do the same for the $\mathfrak{su}%
(3)\times\mathfrak{e}_{6}$ subalgebra of the $\mathfrak{e}_{8}$ flavor
symmetry for the rightmost $-1$ curve, so we have indeed realized a consistent base.

With the rules above stated, we present some additional clarifying
examples.\ For a pair of intersecting $-5$ curves, we have:%
\begin{equation}
5,5\overset{up}{\rightarrow}6,1,6\overset{up}{\rightarrow}%
7,1,3,1,7\overset{up}{\rightarrow}8,1,2,3,2,1,8.
\end{equation}
We can stop at this stage because the $\mathfrak{e}_{7}$ algebra localized on
the $-8$ curve pairs with the $\mathfrak{su}(2)$ algebra localized on the $-2$
curve to form an $\mathfrak{e}_{7}\times\mathfrak{su}(2)$ subalgebra of
$\mathfrak{e}_{8}$.

For a pair of intersecting $-6$ curves we similarly have the minimal blowups:
\begin{align}
&  6,6\overset{up}{\rightarrow}7,1,7\overset{up}{\rightarrow}%
8,1,3,1,8\overset{up}{\rightarrow}9,1,2,3,2,1,9\overset{up}{\rightarrow
}(10),1,2,2,3,2,2,1,10\\
&  \overset{up}{\rightarrow}(11),1,2,2,2,3,2,2,2,1,(11).  \overset{up}{\rightarrow}(11),1,2,2,3,1,5,1,3,2,2,1,(11).
\end{align}
In principle, we also need to blowup a point on each $-11$ curve to resolve
the small instantons, but we shall leave this implicit in much of what
follows. With this caveat stated, note that we can stop blowing up in the
middle because all gauged flavor symmetries of the $-1$ curve theory are
subalgebras of $\mathfrak{e}_{8}$ algebras. For example, on each $-3$ curve we
have a $\mathfrak{g}_{2}$ algebra and on the $-5$ curve we have an
$\mathfrak{f}_{4}$ gauge algebra, and $\mathfrak{f}_{4}\times\mathfrak{g}_{2}$
is indeed a subalgebra of $\mathfrak{e}_{8}$. Again,
we note that the $-1$ curve between each such pair
has a type $II$ fiber on it.

Finally, we can consider a pair of intersecting $-7$ curves, where we have:%
\begin{align}
&  7,7\overset{up}{\rightarrow}8,1,8\overset{up}{\rightarrow}%
9,1,3,1,9\overset{up}{\rightarrow}(10),1,2,3,2,1,(10)\overset{up}{\rightarrow
}(11),1,2,2,3,2,2,1,(11)\\
&  \overset{up}{\rightarrow}(11),1,2,3,1,5,1,3,2,1,(11)  \overset{up}{\rightarrow}(12),1,2,2,3,1,5,1,3,2,2,1,(12).
\end{align}

\subsubsection{6D\ Conformal Matter}

It is helpful to take the leftmost and rightmost
curves of the examples just considered and decompactify them. For example, in
the case of the $7,7$ configuration which supports an $\mathfrak{e}_{8}$ gauge symmetry
on each $-7$ curve (or a $-12$ curve after fully resolving the base),
decompactifying makes this into a non-compact divisor, and we can present the
same resolved geometry as:%
\begin{equation}
\lbrack E_{8}],1,2,2,3,1,5,1,3,2,2,1,[E_{8}]. \label{decompactify}%
\end{equation}
In the unresolved phase where we have two colliding $E_{8}$ seven-branes, the
Weierstrass model is:%
\begin{equation}
y^{2}=x^{3}+u^{5}v^{5},
\end{equation}
where we have introduced local coordinates $u$ and $v$ for $\mathbb{C}^{2}$.
Note that along the locus $u=v$, the order of vanishing for the Weierstrass
model is clearly too large to be described by an elliptic fiber in Kodaira-Tate form. It
nevertheless defines a consistent F-theory background because we can resolve
all collisions, as in line (\ref{decompactify}).

Similar considerations hold for the other examples of minimal resolutions we
have encountered. Let us list the associated Weierstrass model and resolved
base for each such case:%
\begin{align}
y^{2}  =x^{3}+u^{5}v^{5}\text{: \ \ } & [E_{8}],1,2,2,3,1,5,1,3,2,2,1,[E_{8}]\\
y^{2}  =x^{3}+u^{3}v^{3}x\text{: \ \ } & [E_{7}],1,2,3,2,1,[E_{7}]\\
y^{2}  =x^{3}+u^{4}v^{4}\text{: \ \ } & [E_{6}],1,3,1,[E_{6}]\\
y^{2}  =x^{3}+\alpha u^{2}v^{2}x+u^{3}v^{3}\text{: \ \ } & [D_{4}],1,[D_{4}],
\label{eq:D4confmat}
\end{align}
in the obvious notation. We can also consider \textquotedblleft hybrid
situations\textquotedblright\ where the left and right seven-brane gauge
symmetries are different. Again, there is an algorithmic procedure for working
out the resolution of the base such that all fibers are in Kodaira-Tate form.
See Appendix \ref{app:BLOW} for examples of this blowup procedure.

The physical interpretation of such collisions was first analyzed in \cite{Bershadsky:1996nu, Aspinwall:1997ye} and subsequently in references
\cite{Morrison:2012np, Heckman:2013pva, DelZotto:2014hpa, Heckman:2014qba}. In \cite{DelZotto:2014hpa}, it was given the name \textquotedblleft6D\ conformal matter\textquotedblright\
since it appears at a localized point of an intersecting seven-brane configuration, much
as an ordinary hypermultiplet.

The same structures appear repeatedly in more elaborate bases. For example,
given a pair of conformal matter theories, we can gauge a diagonal subgroup and introduce an additional tensor multiplet to build the tensor branch of
another 6D\ SCFT. As an example, consider the case of a single $E_{8}$ gauge
group:%
\begin{equation}
\text{\ }[E_{8}],1,2,2,3,1,5,1,3,2,2,1,(12),1,2,2,3,1,5,1,3,2,2,1,[E_{8}].
\end{equation}
A more concise way to encode the same data involves blowing down to the
endpoint configuration. This is given by a $-2$ curve supporting an $\mathfrak{e}_{8}$
gauge algebra which collides with two $E_{8}$ seven-branes:\footnote{Here and in what follows, we will typically omit the commas between curves in the base when the gauge algebras are included. This notational choice has no deeper significance.}
\begin{equation}
\text{\ }[E_{8}] \,\,\overset{\mathfrak{e}_{8}}{2} \,\, [E_{8}].
\end{equation}
The same structure can be repeated over and over again with $N$ curves of
self-intersection $-2$:%
\begin{equation}
\text{\ }[E_{8}]\,\,\underset{N}{\underbrace{\overset{\mathfrak{e}_{8}%
}{2} \,\,\overset{\mathfrak{e}_{8}}{2} \,\, ... \,\, \overset{\mathfrak{e}_{8}}{2}\,\,
\overset{\mathfrak{e}_{8}}{2}}}\,\,[E_{8}].
\end{equation}
The same sort of structure also persists for the seven-branes supporting each of
the ADE\ algebras, which we denote by a gauge algebra $\mathfrak{g}_{ADE}$:%
\begin{equation}
\text{\ }[G_{ADE}] \,\, \underset{N}{\underbrace{\overset{\mathfrak{g}_{ADE}%
}{2} \,\,\overset{\mathfrak{g}_{ADE}}{2} \,\,...\,\,\overset{\mathfrak{g}_{ADE}\,\,
}{2} \,\,\overset{\mathfrak{g}_{ADE}}{2}}}\,\,[G_{ADE}]. \label{confmattADE}%
\end{equation}
Here we have also included the A-type case, which is realized by $I_{n}$ fibers.

\subsection{Decorating with Singular Fibers}\label{ssec:DecoratingFibers}

In the previous sections we primarily focused on the structure of the
base in the full-resolved phase of the geometry. However, even on a fully-resolved base, we can have additional decoration of the fibers. Recall that the singular fibers of a 6D SCFT at complex codimension 1 and 2
tell us about the spectrum of gauge algebras and charged hypermultiplets,
respectively. We saw in Section \ref{ssec:anomalies} that these are heavily
constrained by anomalies. Indeed, the geometric constraints on singular fibers are more or less identical to the anomaly cancellation constraints, so it suffices for most purposes to consider the latter.
Suppose we have a curve $\Sigma$ of
self-intersection $\Sigma\cdot\Sigma=-n$ with a singular fiber, yielding a
gauge algebra $\mathfrak{g}$:
\begin{equation}
\overset{\mathfrak{g}}{n}%
\end{equation}
Equation (\ref{eq:gaugeanom1a}) gives one constraint on the spectrum of
charged hypermultiplets,
\begin{equation}
x_{\text{adj}}-\sum_{\rho}n_{\rho}x_{\rho}=0, \label{eq:gaugeanom1}%
\end{equation}
with $x_{\rho}$ a group theory constant and $n_{\rho}$ the number of
hypermultiplets charged under $\mathfrak{g}$ in the representation $\rho$. Two
additional constraints come from imposing the vanishing of gauge anomalies and
gauge-gravitational anomalies:
\begin{align}
y_{\text{adj}}-\sum_{\rho}n_{\rho}y_{\rho}  &  =3n\label{eq:gaugeanom2}\\
\text{Ind}_{\text{adj}}-\sum_{\rho}n_{\rho}\text{Ind}_{\rho}  &  =-12+6n.
\label{eq:gaugeanom3}%
\end{align}
The left-hand side of these equations encodes the 1-loop contribution to the
anomaly polynomial, whereas the right-hand side encodes the Green-Schwarz
term, which is related in a very simple way to the self-intersection number
$-n$. These conditions are actually \emph{very} constraining. For a given $n$
and $\mathfrak{g}$, the charged matter is uniquely fixed in all but one case:
for $n=1$ and $\mathfrak{g}=\mathfrak{su}(6)$, one has two possible choices
for the charged matter.

Most 6D SCFT F-theory bases have more than one curve, and new anomaly
cancellation constraints arise in this case. If our curve $\Sigma$ of
self-intersection $-n$ intersects another curve $\Sigma^{\prime}$ of
self-intersection $-n^{\prime}$ carrying gauge algebra $\mathfrak{g}^{\prime}%
$, there is an additional constraint coming from mixed gauge-gauge anomalies,
\begin{equation}
\sum_{\rho}n_{\rho, \rho^{\prime}} \text{Ind}_{\rho} \text{Ind}_{\rho^{\prime
}} = \Sigma\cdot\Sigma^{\prime}= 1, \label{eq:mixed}%
\end{equation}
where $\rho$ runs over the representations of $\mathfrak{g}$, $\rho^{\prime}$
runs over the representations of $\mathfrak{g}^{\prime}$, and $n_{\rho,
\rho^{\prime}}$ is the number of mixed representations $(\rho, \rho^{\prime})$
charged under $\mathfrak{g }\oplus\mathfrak{g}^{\prime}$. For instance,
consider the NHC
\begin{equation}
\overset{\mathfrak{su}(2)}{2 }\,\, \overset{\mathfrak{g}_{2}}{3}%
\end{equation}
The gauge anomaly constraints for the respective gauge groups tell us that
there must be four fundamentals (or, equivalently, eight half-fundamentals)
charged under $\mathfrak{su}(2)$ and a single fundamental $\mathbf{7}$ charged
under $\mathfrak{g}_{2}$. Mixed anomaly constraints tell us that there must be
a half-bifundamental in the mixed representation $(\mathbf{2}, \mathbf{7})$.
This means that the required fundamental of $\mathfrak{g}_{2}$ must transform
as a half-doublet under $\mathfrak{su}(2)$, and seven of the eight
half-fundamentals of $\mathfrak{su}(2)$ transform as a singlet under
$\mathfrak{g}_{2}$. This leaves precisely one half-fundamental of
$\mathfrak{su}(2)$, which transforms as a singlet under $\mathfrak{g}_{2}$.
So, the full spectrum of charged hypermultiplets in this theory is simply
\begin{equation}
\frac{1}{2} (\mathbf{2} , \mathbf{7}) \oplus\frac{1}{2} (\mathbf{2},
\mathbf{1}).
\end{equation}

\subsection{Early Examples Revisited}\label{ssec:revisited}

Having introduced the basic building blocks of 6D\ SCFTs in F-theory, it is
helpful to see how to realize the theories already encountered in subsections
\ref{ssec:OLDIES} and \ref{ssec:EXAMPLES}. At a general level, we should note
that the very fact that we expect a 6D\ SCFT to emerge at long distances means
that the actual UV\ realization in string theory is not so important. From
this perspective, we just need to ensure that we can correctly identify how
the ingredients of one construction map to their F-theory counterparts.

With this in mind, let us first reproduce all of the $\mathcal{N}=(2,0)$ theories. These
are engineered by taking F-theory on a \textquotedblleft Calabi-Yau
threefold\textquotedblright\ where the elliptic fibration is actually
constant:%
\begin{equation}
\text{\ }X=\mathbb{C}^{2}/\Gamma\times T^{2},
\end{equation}
where $\Gamma$ is a discrete subgroup of $SU(2)$. In this case, the base of
the F-theory model is just the orbifold singularity $\mathbb{C}^{2}/\Gamma$.
Resolving the orbifold singularity, we obtain a configuration of $-2$ curves
which intersect according to an ADE\ Dynkin diagram. The reason this
generates the $(2,0)$ theories is that we are free to tune the modulus of the
elliptic fiber to weak coupling, in which case we arrive at type IIB\ string
theory on an ADE\ singularity. As expected, then, we recover all of the
$(2,0)$ theories in one fell swoop.

Recall that M-theory provides an alternate way to realize the A-type
$(2,0)$ theory in terms of a stack of $N$ M5-branes. From the perspective of
F-theory, these are realized by taking $N-1$ curves arranged in an A-type
Dynkin diagram:%
\begin{equation}
\text{\ }A_{N-1}\text{ Theory: \ \ }\underset{N-1}{\underbrace{2,...,2}}.
\label{AtypeAgain}%
\end{equation}
It is also instructive to compare the tensor branch of the two theories in
$\mathcal{N}=(1,0)$ language. To illustrate, consider the case of two
M5-branes, namely the $A_{1}$ theory. In this case, we have a single $-2$
curve in the F-theory description. The relative separation between the
M5-branes maps to the volume of this curve.

Similar considerations also lead us to the rank $N$ E-string theories of
heterotic M-theory, namely $N$ M5-branes near an $E_{8}$ wall. Indeed, we know
that the configuration of curves $1,2,...,2$ enjoys an $E_{8}$ flavor symmetry
in the limit where all curves collapse to zero size. The F-theory
description for the tensor branch of the E-string theory is therefore \cite{Morrison:1996pp}:
\begin{equation}
\text{\ E-string Theory: \ \ }[E_{8}],\underset{N}{\underbrace{1,2,...,2}}.
\label{eq:Estringtheory}
\end{equation}
Where the relative distance between the
$E_{8}$ wall and the closest M5-brane $M5_{1}$ is the volume of the $-1$
curve, the intersecting $-2$ curve is the relative separation between this
M5-brane and the next closest M5-brane $M5_{2}$, and so on. In other words, we
can write, in the obvious notation:%
\begin{equation}
\text{\ Vol}\left(  \Sigma_{1}\right)  =\text{dist}(E_{8},M5_{1})\text{,
\ \ Vol}\left(  \Sigma_{i}\right)  =\text{dist(}M5_{i},M5_{i+1})\text{ \ \ for
\ \ }1\leq i\leq N-1.
\end{equation}
Moving all M5-branes to the $E_{8}$ wall then takes us to the conformal fixed point.

The F-theory realization of the SCFT obtained from $N$ M5-branes probing an
ADE\ singularity is similarly obtained by combining the construction of
(\ref{AtypeAgain}) with a non-trivial elliptic fibration, yielding the theories
of (\ref{confmattADE}). On a partial tensor branch, these are realized by the
F-theory model \cite{DelZotto:2014hpa}:
\begin{equation}
\text{\ }[G_{ADE}] \,\, \underset{N}{\underbrace{\overset{\mathfrak{g}_{ADE}%
}{2} \,\, \overset{\mathfrak{g}_{ADE}}{2} \,\, ...\overset{\mathfrak{g}_{ADE}%
}{2} \,\, \overset{\mathfrak{g}_{ADE}}{2}}} \,\, [G_{ADE}].
\end{equation}
In this case, the relative separation of neighboring M5-branes along a noncompact transverse direction $\mathbb{R}_\perp$
is captured by the volume of each $-2$
curve. A new feature of this example is that in between each M5-brane
factor we have a 7D\ super Yang-Mills theory compactified on a finite
interval. Additionally, there is a bulk 7D topological field theory of a
three-form (and its superpartners) with a Chern-Simons-like action
\cite{Heckman:2017uxe} (see also \cite{Witten:1998wy}).

To see how the fiber decoration works in more detail, let us first consider the case of $\mathfrak{g}_{ADE}=\mathfrak{su}(n)$:
\begin{equation}
[SU(n)] \,\, \overset{\mathfrak{su}(n)}{2 }\,\,
\overset{\mathfrak{su}(n)}{2 }\,\,...\,\, \overset{\mathfrak{su}(n)}{2 }\,\,
[SU(n)] \label{eq:Anquiver}%
\end{equation}
To satisfy the gauge anomaly cancellation conditions in (\ref{eq:gaugeanom1}%
)-(\ref{eq:gaugeanom3}), one needs $2n$ fundamentals charged under each gauge
algebra. To satisfy the mixed anomaly cancellation constraint (\ref{eq:mixed}), one needs
a bifundamental hypermultiplet $(\mathbf{{n},{n})}$ between each pair of
adjacent gauge algebras. For every gauge node in the interior of the quiver,
all $2n$ fundamental hypermultiplets are part of bifundamentals, and there are
no fundamentals leftover. For the first and last gauge groups in the quiver,
on the other hand, there are $n$ leftover fundamentals, which transform as a
fundamental under an $\mathfrak{su}(n)$ flavor symmetry. The resulting theory
is precisely that of $N+1$ M5-branes probing a $\mathbb{C}^2/\mathbb{Z}_{n}$
orbifold singularity that we saw earlier.

Next, we consider the case of $\mathfrak{g}_{ADE}=\mathfrak{so}(2n)$. In this case,
the anomaly cancellation conditions cannot be satisfied by bifundamentals: whereas the fundamental of $\mathfrak{su}(n)$ has
index $1$, the fundamental of $\mathfrak{so}(2n)$ has index $2$, so a
bifundamental of $\mathfrak{so}(2n)$-$\mathfrak{so}(2n)$ does not satisfy
(\ref{eq:mixed}). In geometric terms, the point of intersection between
adjacent $-2$ curves becomes too singular to be described in F-theory and must
be blown up. After blowing up, we are left with a theory of the form
\begin{equation}
[SO(2n)] \,\, \overset{\mathfrak{sp}(n-4)}{1 }\,\,
\overset{\mathfrak{so}(2n)}{4 }\,\,...\,\, \overset{\mathfrak{so}(2n)}{4 }\,\,
\overset{\mathfrak{sp}(n-4)}{1 }\,\, [SO(2n)] \label{eq:Dnquiver}%
\end{equation}
Now, there is a half hypermultiplet in the bifundamental representation of
$\mathfrak{so}(2n) \times \mathfrak{sp}%
(n-4)$ between every pair of adjacent nodes, and gauge anomalies and mixed
anomalies both cancel provided one also includes such matter fields between the leftmost and rightmost
$\mathfrak{sp}(n-4)$ gauge algebra factors and the respective leftmost and rightmost $\mathfrak{so}(2n)$
flavor symmetries. The $-1$ curve carrying
$\mathfrak{sp}(n-4)$ gauge group is $SO(2n)$-$SO(2n)$ conformal
matter, generalizing the case of $n=4$ in (\ref{eq:D4confmat}). The resulting theory in (\ref{eq:Dnquiver}) is the theory
of $N+1$ M5-branes probing a $\mathbb{C}^2/\Gamma_{D_{n}}$ orbifold singularity.

For $N$ M5-branes probing a $\Gamma_{E_{6}}$, $\Gamma_{E_{7}}$, and
$\Gamma_{E_{8}}$ singularity, we have the following tensor branch
descriptions of the respective 6D SCFTs:
\begin{equation}
[E_6] \,\, 1 \,\, \overset{\mathfrak{su}(3)}{3 }\,\, 1 \,\,
\overset{\mathfrak{e}_{6}}{6 }\,\, 1\,\, \overset{\mathfrak{su}(3)}{3 }\,\, 1
\,\, ... \,\, \overset{\mathfrak{e}_{6}}{6}\,\, 1 \,\, \overset{\mathfrak{su}%
(3)}{3 }\,\, 1 \,\, [E_6] \label{eq:E6quiver}%
\end{equation}
\begin{equation}
[E_7] \,\, 1 \,\, \overset{\mathfrak{su}(2)}{2 }\,\,
\overset{\mathfrak{so}(7)}{3 }\,\,\overset{\mathfrak{su}(2)}{2 }\,\, 1 \,\,
\overset{\mathfrak{e}_{7}}{8 }\,\, 1 \,\, \overset{\mathfrak{su}(2)}{2 }\,\,
\overset{\mathfrak{so}(7)}{3 }\,\,\overset{\mathfrak{su}(2)}{2 }\,\, 1 \,\,
... \,\, \overset{\mathfrak{e}_{7}}{8 }\,\, 1 \,\, \overset{\mathfrak{su}%
(2)}{2 }\,\, \overset{\mathfrak{so}(7)}{3 }\,\,\overset{\mathfrak{su}(2)}{2
}\,\, 1 \,\, [E_7] \label{eq:E7quiver}%
\end{equation}
\begin{equation}
[E_8] \,\, 1 \,\, 2 \,\, \overset{\mathfrak{su}(2)}{2 }\,\,
\overset{\mathfrak{g}_{2}}{3 }\,\, 1 \,\, \overset{\mathfrak{f}_{4}}{5 }\,\, 1
\,\, \overset{\mathfrak{g}_{2}}{3 }\,\,\overset{\mathfrak{su}(2)}{2 }\,\, 1
\,\, \overset{\mathfrak{e}_{8}}{(12)} \,\, \,\, 1 \,\, 2 \,\,
\overset{\mathfrak{su}(2)}{2 }\,\, ... \,\, \overset{\mathfrak{e}_{8}%
}{(12)}\,\, 1 \,\, 2 \,\, \overset{\mathfrak{su}(2)}{2 }\,\,
\overset{\mathfrak{g}_{2}}{3 }\,\, 1 \,\, \overset{\mathfrak{f}_{4}}{5 }\,\, 1
\,\, \overset{\mathfrak{g}_{2}}{3 }\,\,\overset{\mathfrak{su}(2)}{2 }\,\, 2 \,\, 1
\,\, [E_8] \label{eq:E8quiver}%
\end{equation}
Here, we have introduced ``$E_{6}$-$E_{6}$ conformal matter,"
\begin{equation}
[E_6] \,\,1 \,\, \overset{\mathfrak{su}(3)}{3 }\,\, 1 \,\, [E_6]
\label{eq:E6E6confmat}
\end{equation}
``$E_{7}$-$E_{7}$ conformal matter,"
\begin{equation}
[E_7] \,\,1 \,\, \overset{\mathfrak{su}(2)}{2 }\,\, \overset{\mathfrak{so}(7)}{3
}\,\,\overset{\mathfrak{su}(2)}{2 }\,\, 1 \,\, [E_7]
\label{eq:E7confmat}
\end{equation}
and ``$E_{8}$-$E_{8}$ conformal matter,"
\begin{equation}
[E_8] \,\,1 \,\, 2 \,\, \overset{\mathfrak{su}(2)}{2 }\,\, \overset{\mathfrak{g}_{2}}{3
}\,\, 1 \,\, \overset{\mathfrak{f}_{4}}{5 }\,\, 1 \,\, \overset{\mathfrak{g}%
_{2}}{3 }\,\,\overset{\mathfrak{su}(2)}{2 }\,\, 2\,\,1 \,\, [E_8]
\label{eq:E8confmat}
\end{equation}

On the partial tensor branch where all M5-branes are separated, we see that for $G$ a D- or E-type
singularity, there is actually already an interacting SCFT present, even for a
single M5-brane \cite{DelZotto:2014hpa}. This can also be interpreted as a theory of chiral
edge mode states localized on the brane.

For another class of examples, consider the theory of $N$ M5-branes near an $E_{8}$
nine-brane wall which fills the 10D geometry $\mathbb{R}^{5,1}\times
\mathbb{C}^{2}/\Gamma$, where the second factor is an
ADE\ singularity.\footnote{As stated, we have not actually fully specified the
heterotic string construction. The reason is we need to also specify
non-trivial boundary conditions for the $E_{8}$ connection \textquotedblleft
at infinity.\textquotedblright\ Here, we are simply positing trivial boundary
conditions. We will revisit this point in great detail in Section \ref{ssec:HOM}.} The
F-theory description of this model is \cite{Aspinwall:1997ye, DelZotto:2014hpa}:
\begin{equation}
\text{\ }[E_{8}] \,\, \underset{N}{\underbrace{\overset{\mathfrak{g}_{ADE}%
}{1} \,\, \overset{\mathfrak{g}_{ADE}}{2} \,\, ... \,\, \overset{\mathfrak{g}_{ADE}%
}{2} \,\, \overset{\mathfrak{g}_{ADE}}{2}}} \,\, [G_{ADE}],
\end{equation}
in the obvious notation. Here, we pick up an $E_{8}$ flavor
symmetry from the nine-brane, and a $G_{ADE}$ flavor symmetry from the
presence of the ADE\ singularity.

Let us consider the fully-resolved geometry for these theories. For $\mathfrak{g}_{ADE} = \mathfrak{su}(n)$, we have
\begin{equation}
[E_8] \,\, 1 \,\, \overset{\mathfrak{su}(1)}{2}\,\,
\overset{\mathfrak{su}(2)}{2 }\,\, ...\,\, \overset{\mathfrak{su}(n-1)}{2}\,\,
\underset{N}{\underbrace{\underset{[N_{f}=1]}{\overset{\mathfrak{su}(n)}{2}}
\,\, \overset{\mathfrak{su}(n)}{2 }\,\,...\,\, \overset{\mathfrak{su}(n)}{2}}}
\,\, [SU(n)]
\end{equation}
Note the presence of a ``ramp" of $\mathfrak{su}$ gauge algebras starting from
$\mathfrak{su}(1)$ and increasing up to $\mathfrak{su}(n)$. Here, the
left-most $-2$ curve in the ramp does not carry any gauge algebra, but it does
support a type $I_{1}$ fiber (indicating the presence of a single D7-brane
wrapping the curve), and we have written ${\mathfrak{su}(1)}$ to remind
ourselves of this fact. As discussed, an empty $-1$ curve comes with a
$\mathfrak{e}_{8}$ global symmetry, which explains the global symmetry on the
left-hand side of the quiver. This ramp of gauge algebras can therefore be
thought of as $E_{8}$-$SU(n)$ conformal matter, since it is the minimal
collection of curves connecting a $\mathfrak{e}_{8}$ symmetry algebra to a
$\mathfrak{su}(n)$ symmetry algebra. Finally, one can check that gauge and
mixed anomalies cancel in this theory if we add a bifundamental between each
pair of adjacent nodes along with a single fundamental of the left-most
$\mathfrak{su}(n)$ and $n$ fundamentals of the right-most $\mathfrak{su}(n)$,
transforming as a fundamental under an $\mathfrak{su}(n)$ flavor symmetry.

For $N$ M5-branes probing a $\mathbb{C}^2/\Gamma_{D_{n}}$ singularity, we have
\begin{equation}
[E_8] \,\, 1 \,\, 2 \,\, \overset{\mathfrak{su}(2)}{2 }\,\,
\overset{\mathfrak{g}_{2}}{3 }\,\, 1 \,\, \overset{\mathfrak{so}(9)}{4 }\,\,
\overset{\mathfrak{sp}(1)}{1 }\,\, ... \,\, \overset{\mathfrak{sp}(n-5)}{1
}\,\, \overset{\mathfrak{so}(2n-1)}{4 }\,\,
\underset{2N-1}{\underbrace{\overset{\mathfrak{sp}(n-4)}{1 }\,\,
\overset{\mathfrak{so}(2n)}{4 }\,\,...\,\, \overset{\mathfrak{so}(2n)}{4 }\,\,
\overset{\mathfrak{sp}(n-4)}{1}}} \,\, [SO(2n)]
\label{eq:E8Dnquiver}%
\end{equation}
In this case, we have a ramp of $\mathfrak{so}$ and $\mathfrak{sp}$ gauge
algebras increasing up to $\mathfrak{so}(2n)$ and $\mathfrak{sp}(n-4)$.
Anomalies cancel, and the $E_{8}$ gauging condition for the $-1$ curve to the
left of the $-3$ curve is satisfied because $\mathfrak{g}_{2} \oplus
\mathfrak{so}(9) \subset\mathfrak{e}_{8}$. For $N=1$, we get a collection of
curves starting from the $-1$ curve on the far left up and including to the
first $-1$ curve with $\mathfrak{sp}(n-4)$ gauge algebra, which is called
$E_{8}$-$SO(2n)$ conformal matter:
\begin{equation}
[E_8] \,\, 1 \,\, 2 \,\, \overset{\mathfrak{su}(2)}{2 }\,\,
\overset{\mathfrak{g}_{2}}{3 }\,\, 1 \,\, \overset{\mathfrak{so}(9)}{4 }\,\,
\overset{\mathfrak{sp}(1)}{1 }\,\, ... \,\, \overset{\mathfrak{sp}(n-5)}{1
}\,\, \overset{\mathfrak{so}(2n-1)}{4 }\,\, \overset{\mathfrak{sp}(n-4)}{1
}\,\, [SO(2n)]
\end{equation}

For the case of a $\mathbb{C}^2/\Gamma_{E_{6}}$ orbifold singularity, we get
\begin{equation}
[E_8] \,\, 1 \,\, 2 \,\, \overset{\mathfrak{su}(2)}{2 }\,\,
\overset{\mathfrak{g}_{2}}{3 }\,\, 1 \,\, \overset{\mathfrak{f}_{4}}{5 }\,\, 1
\,\, \overset{\mathfrak{su}(3)}{3 }\,\, 1 \,\, \underset{4(N-1)}{\underbrace{
\overset{\mathfrak{e}_{6}}{6 }\,\, 1\,\, \overset{\mathfrak{su}(3)}{3 }\,\, 1
\,\, ... \,\, \overset{\mathfrak{e}_{6}}{6}\,\, 1 \,\, \overset{\mathfrak{su}%
(3)}{3 }\,\, 1}} \,\, [E_6]
\end{equation}

For a $\mathbb{C}^2/\Gamma_{E_{7}}$ orbifold singularity, we get
\begin{equation}
[E_8] \,\, 1 \,\, 2 \,\, \overset{\mathfrak{su}(2)}{2 }\,\,
\overset{\mathfrak{g}_{2}}{3 }\,\, 1 \,\, \overset{\mathfrak{f}_{4}}{5 }\,\, 1
\,\, \overset{\mathfrak{g}_{2}}{3 }\,\,\overset{\mathfrak{su}(2)}{2 }\,\, 1
\,\, \underset{6(N-1)}{\underbrace{\overset{\mathfrak{e}_{7}}{8 }\,\, 1 \,\,
\overset{\mathfrak{su}(2)}{2 }\,\, \overset{\mathfrak{so}(7)}{3 }%
\,\,\overset{\mathfrak{su}(2)}{2 }\,\, 1 \,\, ... \,\, \overset{\mathfrak{e}%
_{7}}{8 }\,\, 1 \,\, \overset{\mathfrak{su}(2)}{2 }\,\, \overset{\mathfrak{so}%
(7)}{3 }\,\,\overset{\mathfrak{su}(2)}{2 }\,\, 1}} \,\, [E_7]
\end{equation}
Here, $N$ is the number of $\mathfrak{e}_{7}$ gauge algebras, and $N=1$
corresponds to $E_{8}$-$E_{7}$ conformal matter.

Finally, for a $\mathbb{C}^2/\Gamma_{E_{8}}$ orbifold singularity, we have
\begin{equation}
[E_8] \,\, 1 \,\, 2 \,\, \overset{\mathfrak{su}(2)}{2 }\,\,
\overset{\mathfrak{g}_{2}}{3 }\,\, 1 \,\, \overset{\mathfrak{f}_{4}}{5 }\,\, 1
\,\, \overset{\mathfrak{g}_{2}}{3 }\,\,\overset{\mathfrak{su}(2)}{2 }\,\, 1
\,\, \overset{\mathfrak{e}_{8}}{(11)} \,\, \,\, 1 \,\, 2 \,\,
\overset{\mathfrak{su}(2)}{2 }\,\, ... \,\, \overset{\mathfrak{e}_{8}%
}{(12)}\,\, 1 \,\, 2 \,\, \overset{\mathfrak{su}(2)}{2 }\,\,
\overset{\mathfrak{g}_{2}}{3 }\,\, 1 \,\, \overset{\mathfrak{f}_{4}}{5 }\,\, 1
\,\, \overset{\mathfrak{g}_{2}}{3 }\,\,\overset{\mathfrak{su}(2)}{2 }\,\, 1
\,\, [E_8]
\end{equation}
Note that we have used the shorthand $\overset{\mathfrak{e}_{8}}{(11)} $ for
$\overset{\mathfrak{e}_{8}}{(12)}$ meeting a single $-1$ curve, which is
sometimes called a ``small instanton."

Clearly, then, F-theory provides a unifying framework for realizing a vast
class of previously understood examples. An important point, however, is that
in these other duality frames certain aspects of the 6D\ SCFT may be more
transparent. One can already see an example of this by considering the R-symmetry of the model. In the description of the A-type $(2,0)$
theories realized by coincident M5-branes, we can identify a
transverse $\mathbb{R}^{5}$, and consequently the isometries $\mathfrak{so}%
(5)$ furnish us with the R-symmetry algebra. This is more challenging to
identify for type IIB\ on an A-type orbifold singularity. Similar
considerations hold for the E-string theories. In M-theory language, we have a
transverse $\mathbb{R}^{4}$ inside the nine-brane wall which means we expect
an $\mathfrak{so}(4)\simeq\mathfrak{su}(2)_{L}\times\mathfrak{su}(2)_{R}$
symmetry. We identify the second factor with the R-symmetry present in all $\mathcal{N} = (1,0)$
SCFTs, and the first factor with a global symmetry specific to the E-string
theory. Note that this symmetry is not realized as an isometry of
the geometry in the F-theory description $1,2,...,2$ at the collapsed point.

Indeed, in field theory terms, the $SU(2)$ R-symmetry acts as a local isometry on the
tangent space to the hypermultiplet moduli space. In F-theory, this moduli space is constructed
from the complex structure moduli and the intermediate Jacobian (which forms a torus fibration over the complex structure moduli).
The action of the R-symmetry is thus geometrically realized on the moduli space of the Calabi-Yau rather
than the target space itself. For further discussion on this, see reference \cite{Anderson:2013rka}.

All this is to say that the various string constructions provide complementary
perspectives on the structure of 6D\ SCFTs. That being said, the fact that all
known SCFTs embed in F-theory motivates the study of determining all possible
F-theory models.

\section{Classification via F-theory \label{sec:CLASSIFY}}

In the previous section we saw that F-theory provides a remarkably flexible
framework for engineering 6D\ SCFTs. It is therefore natural to ask whether
one can classify all possible F-theory backgrounds which can yield a 6D\ SCFT.
Nearly all known string theory examples have a counterpart in a geometric
realization of F-theory, and the ones that do not can be realized by allowing
mild non-geometric behavior known as \textquotedblleft frozen
singularities.\textquotedblright\ Since all known stringy examples can
therefore be embedded in F-theory, this provides evidence for a stronger claim
that this actually classifies \textit{all} 6D\ SCFTs.

Our aim in this section will be to review the classification of geometric
backgrounds of F-theory. This naturally splits into two pieces:\ the
classification of bases which have a configuration of simultaneously
contractible curves, and the classification of elliptic fibrations over a
given base.

\subsection{Classification of Bases (All Your Base Are Belong to Us) \label{ssec:KEEPITCLASSY}}

We now review how to classify F-theory bases for 6D\ SCFTs. We have already
listed the primary conditions necessary to realize a consistent base in
subsection \ref{ssec:CONDITIONS}. The goal here will be to illustrate how
these conditions come about, and to then summarize the general structure of
all bases.

To begin, it is helpful to identify several repeating structures which will
occur many times. We refer to a \textquotedblleft node\textquotedblright\ as a
curve of self-intersection $-n$ which supports a minimal gauge algebra of D-
or E-type. We refer to a ``link" as a configuration of curves which do not
involve any nodes. The nodes are simply a subset of the non-Higgsable clusters
already encountered:%
\begin{equation}%
\begin{tabular}
[c]{|l|l|l|l|l|l|}\hline
Nodes & $n=4$ & $n=6$ & $n=7$ & $n=8$ & $n=12$\\\hline
Minimal Fiber & $I_{0}^{\ast}$ & $IV^{\ast}$ & $III^{\ast}$ & $III^{\ast}$ &
$II^{\ast}$\\\hline
Gauge Algebra & $\mf{so}(8)$ & $\mf{e}_{6}$ & $\mf{e}_{7}+\frac{1}{2}\mathbf{56}$ & $\mf{e}_{7}$ & $\mf{e}_{8}$\\\hline
\end{tabular}
\ \ .
\end{equation}
As for the \textquotedblleft links,\textquotedblright\ we have also
encountered some such objects in the context of conformal matter; they involve
all of the other non-Higgsable clusters, as well as the $-1$ curves and the
ADE\ configurations of $-2$ curves. The full list of links is classified in
in an Appendix of \cite{Heckman:2015bfa}. In addition to the conformal matter links, there are also other
``molecules" which can be built up purely from non-DE-type structures.

Given the full collection of nodes and links, we build up larger bases by
gluing these structures together according to the rules outlined in subsection
\ref{ssec:CONDITIONS}. We denote the type of node by the corresponding minimal
gauge group, and leave implicit the particular choice of link, unless it is
ambiguous. For example, the configuration of $-12$ curves suspended between
conformal matter can be written in a compressed notation as:%
\begin{equation}
\lbrack E_{8}]-E_{8}-...-E_{8}-[E_{8}],
\end{equation}
where each $E_{8}$ not enclosed in square brackets denotes a $-12$ curve, and
each $-$ link denotes the configuration of curves $1,2,2,3,1,5,1,3,2,2,1$.

By definition, we build a large base by connecting two nodes via a link. Let
us note that some \textquotedblleft links\textquotedblright\ actually cannot
attach to any nodes. These were referred to in reference \cite{Heckman:2015bfa} as
\textquotedblleft noble molecules.\textquotedblright\ Additionally, as we
shortly explain, the actual kinds of links which can be suspended between two
nodes are typically of minimal type, namely minimal conformal matter.

To see how these conditions come about, it is perhaps simplest to
present some illustrative examples.

The first surprise is that a node can join to at most two other nodes via
links. To see why, consider for example the case of a $-4$ curve joined to
three other $-4$ curves:%
\begin{equation}
\text{NOT AN SCFT: \ \ }4,1,\overset{\overset{4}{1}}{4},1,4\ \ .
\end{equation}
Blowing down the three $-1$ curves, we encounter the configuration:%
\begin{equation}
\text{NOT AN SCFT: \ \ }4,1,\overset{4}{\overset{1}{4}},1,4\overset{dn}{\rightarrow}3,\overset{3}{1}%
,3
\end{equation}
This violates the condition of normal crossing, namely we have three
curves touching a $-1$ curve.

Similar considerations apply for all other nodes.
For example, for a $-6$ curve joined to three other $-6$ curves, we
have:%
\begin{equation}
\text{NOT AN SCFT: \ \ }6,1,3,1,\overset{6}{\overset{1}{\overset{3}{\overset{1}{6}}}}%
,1,3,1,6\overset{dn}{\rightarrow}5,1,\overset{5}{\overset{1}{3}}%
,1,5\overset{dn}{\rightarrow}4,\overset{4}{0},4\text{,}%
\end{equation}
which clearly violates the condition of having a negative definite
intersection form on the base. A similar condition holds when we try to use a
link attached to three different nodes. The issue boils down to the fact that
while we can join a pair of nodes using conformal matter, the link in between
blows down to the trivial configuration. Adding one more $-1$ curve anywhere
on such a configuration will take us to a base which is not suitable for a
6D\ SCFT. So with this in mind, we see that trivalent links connecting three
nodes cannot exist.\footnote{More precisely, one can have links which exhibit
non-trivial tree-like structure, but the specific condition that this tree can
attach to three nodes turns out to be impossible.} We thus obtain our first
condition on any base which contains nodes:

\begin{itemize}
\item Any node can join at most two other nodes.
\end{itemize}

Consequently, nodes must be arranged in a single spine, joined by links.
Denoting this sequence of nodes by the associated minimal gauge group $G$
(coming from the prescribed elliptic fiber), we narrow down the list of
possible bases to:%
\begin{equation}
-\overset{|}{G}_{1}-\overset{|}{G}_{2}-...-\overset{|}{G}_{k-1}-\overset{|}{G}%
_{k}-, \label{InitialSpine}%
\end{equation}
where the \textquotedblleft$-$\textquotedblright\ is shorthand for possible
links attached to these nodes.

In fact, most nodes can only attach to at most two links. So in a
configuration such as line (\ref{InitialSpine}), most of the vertical lines on
the graph are actually absent. Again, it is helpful to illustrate this by way
of example. Consider a candidate configuration and its blowdowns:%
\begin{equation}
\text{NOT AN SCFT: \ \ }1,\overset{1}{4},1,\overset{1}{4},1\overset{dn}{\rightarrow}%
1,1\overset{dn}{\rightarrow}0,
\end{equation}
so such a base is inconsistent. As a more elaborate example, consider:%
\begin{equation}
\text{NOT AN SCFT: \ \ }1,3,1,\overset{1}{6},1,3,1,\overset{1}{6},1,3,1\overset{dn}{\rightarrow
}1,3,1,3,1\overset{dn}{\rightarrow}1,1\overset{dn}{\rightarrow}0.
\end{equation}
The exception to the rule is that if we are close enough to the left or the
right of our spine of nodes, then we can in principle allow for decoration by
links. We summarize this with the \textquotedblleft generic
rule:\textquotedblright

\begin{itemize}
\item Most nodes only attach to two links.
\end{itemize}

Thus, the structure of nearly all bases is:%
\begin{equation}
-\overset{|}{G}_{1}-\overset{|}{G}_{2}-G_{3}-...-G_{k-2}-\overset{|}{G}%
_{k-1}-\overset{|}{G}_{k}-.
\end{equation}
There are some exceptions to this generic behavior which are summarized in
reference \cite{Heckman:2015bfa}, but this is the essence of the structural classification
of possible bases.

At a broad level, there is one additional condition we can see which is that
the actual minimal gauge algebra on each node satisfies a nested sequence of
containment relations of the form:%
\begin{equation}
G_{1}\subseteq G_{2}\subseteq...\subset G_{m}\supseteq ... \supseteq G_{k-1}\supseteq G_{k}, \label{nested}%
\end{equation}
where $m$ denotes the location of some interior node which serves the local
maximum for the rank. It may be closer to one side or the other of the diagram. These containment relations can be thought of as a generalization of the ``convexity condition" on a linear chain of $\mf{su}(n_i)$ gauge algebras, as shown below in (\ref{eq:Angeneral}). In that case, as discussed below in Section \ref{ssec:HOM}, the convexity condition gives $2 n_i \geq n_{i+1} + n_{i-1}$, which ensures that $n_i$ must increase as one moves into the interior of the quiver from the left or right.

To see why this condition comes about, we again turn to some examples
illustrating what can and cannot happen. First, consider the case of the
configuration $E_{6}-D_{4}-E_{6}$:%
\begin{equation}
\text{NOT AN SCFT: \ \ }6,1,3,1,4,1,3,1,6\overset{dn}{\rightarrow}5,1,2,1,5\rightarrow4,0,4,
\end{equation}
which is inconsistent. On the other hand, for $E_{6}-E_{7}-E_{6}$, there is no
such issue:%
\begin{equation}
6,1,2,3,2,1,8,1,2,3,2,1,6\overset{dn}{\rightarrow}%
5,1,3,1,6,1,3,1,5\overset{dn}{\rightarrow}4,1,4,1,4\overset{dn}{\rightarrow
}3,2,3\text{.}%
\end{equation}

At a mathematical level, we thus recover the condition that the
self-intersection is lowest in the middle of a base (as per line
(\ref{nested})), and becomes smaller in magnitude near the edges of a graph.
We summarize this with the rule:

\begin{itemize}
\item The minimal gauge groups on the nodes exhibit a nested sequence of
containment relations.
\end{itemize}

Physically, we can think of this as the \textquotedblleft
biggest\textquotedblright\ or \textquotedblleft highest
tension\textquotedblright\ seven-branes sitting in the middle of a base, with
the smallest ones on the periphery.

Thus, the suggestive structure of all of these 6D\ SCFTs clearly
resembles a generalized quiver in which the bifundamentals (composed from
links) are themselves 6D\ SCFTs. We summarize the general
classification results by the qualitative statement:

\begin{itemize}
\item On their (partial) tensor branch, 6D\ SCFTs resemble generalized quivers.
\end{itemize}

\subsection{Classification of Endpoints}

In the previous subsection we focused on the structure of the tensor branch,
as realized in F-theory. To reach the conformal fixed point, we must
collapse this configuration of curves to zero size. In this limit, the base
geometry will typically be a singular K\"ahler surface. Quite surprisingly, the
structure of these singularities exhibits a very uniform structure, and is
always of the form:%
\begin{equation}
B_{\text{sing}}=\mathbb{C}^{2}/\Gamma_{U(2)},
\end{equation}
where $\Gamma_{U(2)}$ is a discrete subgroup of $U(2)$. The possible choices
of $\Gamma_{U(2)}$ compatible with the existence of an elliptic fibration were
classified in \cite{Heckman:2013pva}, where it was also found that not all discrete subgroups of
$U(2)$ actually appear. In this subsection we review the procedure used to
obtain this classification result.

This relies on the notion of an ``endpoint'' introduced below line (\ref{blowagain}), namely we take a smooth base and start blowing down
all $-1$ curves. If additional $-1$ curves crop up in this process, we blow these down as well until no $-1$ curves appear. This
final configuration of curves is referred to as the endpoint of a base.

The main idea is to systematically study configurations of collapsed curves
which can be resolved to a smooth base compatible with the existence of an
elliptic fibration. To see the sort of restrictions we can expect to
encounter, consider for example the configuration of curves:%
\begin{equation}
\text{BAD\,END:}\,\,\,3,\overset{3}{3},3.
\end{equation}
Performing the requisite blowups to place all elliptic fibers in Kodaira-Tate
form, we have:%
\begin{equation}
\text{BAD\,END:}\,\,\,3,\overset{3}{3},3\overset{up}{\rightarrow}4,1,\overset{4}{\overset{1}{6}%
},1,4\overset{up}{\rightarrow}%
5,1,3,1,\overset{5}{\overset{1}{\overset{3}{\overset{1}{9}}}}%
,1,3,1,5\overset{up}{\rightarrow}%
5,1,3,2,1,\overset{5}{\overset{1}{\overset{3}{\overset{2}{\overset{1}{12}}}}%
},1,2,3,1,5,
\end{equation}
but we cannot stop here because the $-12$ curve is too close to a $-2$ curve
carrying a gauge algebra. Any further blowups on this $-12$ curve will make
the elliptic fiber too singular. We thus conclude that such a configuration
cannot be realized in F-theory.

In reference \cite{Heckman:2013pva} this strategy was used to classify the available endpoint
configurations, which all turn out to have a remarkably simple form: they take
the form of A-, D- and E-type Dynkin diagrams in which we change the diagonal
entries of the intersection pairing:
\begin{align}
\text{A-type Endpoint}  &  \text{: }n_{1},...,n_{k}\\
\text{D-type Endpoint}  &  \text{: }2\overset{2}{2},...,2,m_{l-1}m_{l}\label{Dtype}\\
E_{6}\, \text{Endpoint}  &  \text{: }2,2\overset{2}{2},2,2\\
E_{7}\, \text{Endpoint}  &  \text{: }2,2\overset{2}{2},2,2,2\\
E_{8}\, \text{Endpoint}  &  \text{: }2,2\overset{2}{2},2,2,2.
\end{align}
In the D-type case, we note that it is possible to have\ a curve of lower
self-intersection than $-2$ in the trivalent node location. For example, the
$D_4$-type endpoint configuration with $m_{3} = 3$, $m_4=2$, namely $2,\overset{2}{3},2$, resolves to the base:%
\begin{equation}
2,\overset{2}{3},2\overset{up}{\rightarrow}3,1,\overset{3}{\overset{1}{6}%
},1,3.
\end{equation}
However, the configuration $2,\overset{2}{4},2$ cannot occur because it
involves too many blowups to retain Kodaira-Tate form for the elliptic fibers:%
\begin{equation}
2,\overset{2}{4},2\overset{up}{\rightarrow}3,1,\overset{3}{\overset{1}{7}%
},1,3\overset{up}{\rightarrow}3,2,1,\overset{3}{\overset{2}{\overset{1}{10}}%
},1,2,3\overset{up}{\rightarrow}%
3,2,2,1,\overset{3}{\overset{2}{\overset{2}{\overset{1}{13}}}},1,2,2,3,
\end{equation}
which is clearly inconsistent due to the presence of a $-13$ curve.

All of this is to say that the general form of the endpoint configurations is
quite constrained. For example, when there are a sufficient number of curves
in an A- or D-type configuration, then the vast majority of curves have self-intersection
$-2$. For example, the interior of an A-type configuration consists of all
$-2$ curves, and only the two rightmost curves of the D-type configuration in
line (\ref{Dtype}) can be different from $-2$.

The next surprise is that each such endpoint configuration actually defines an
orbifold of $\mathbb{C}^{2}$, namely $\mathbb{C}^{2}/\Gamma_{U(2)}$ for
$\Gamma_{U(2)}$ a discrete subgroup of $U(2)$ rather than $SU(2)$. In the
E-type case, we are simply dealing with the E-type discrete subgroups of
$SU(2)$. In the A- and D-type cases, however, the orbifold has non-trivial
curvature in the base. The reason we nevertheless retain supersymmetry has to
do with the presence of a non-trivial elliptic fibration.

In the A-type case, the orbifold group action on the base coordinates of
$\mathbb{C}^{2}$ is given by:%
\begin{equation}
(u,v)\rightarrow(\omega u,\omega^{q}v)\text{ \ \ with \ \ }\omega=\exp(2\pi
i/p),
\end{equation}
where the integers $p$ and $q$ are positive relatively prime numbers defined
by the continued fraction:%
\begin{equation}
\text{A-type: }\frac{p}{q}=n_{1}-\frac{1}{n_{2}-...\frac{1}{n_{k}}}.
\end{equation}
This is a well-known result from the theory of Hirzebruch--Jung resolutions
\cite{jung,MR0062842,Riemen:dvq}. In \cite{Heckman:2017uxe} this rational number was interpreted in terms of a fractional conductivity
in a generalization of the fractional quantum Hall effect to six dimensions.
Additional patterns in the set of orbifold group actions appearing in F-theory models were
noted in \cite{Morrison:2016nrt} (see also \cite{Merkx:2017jey}).

For the D-type series, there is a related, though more involved expression for
the orbifold group action. The associated continued fraction for a
configuration of curves such as:%
\begin{equation}
2\overset{2}{n},m_{1},...,m_{l-1}m_{l}%
\end{equation}
is given by:%
\begin{equation}
\text{D-type: }\frac{p}{q}=(n-1)-\frac{1}{m_{1}-...\frac{1}{m_{l}}},
\end{equation}
where the specific group action on the $\mathbb{C}^{2}$ is specified by the
generators (see \cite{DelZotto:2015isa} and references therein):%
\begin{equation}
D_{p+q,q}=\left\{
\begin{array}
[c]{c}%
\left\langle \psi_{2q},\varphi_{2p},\tau\right\rangle \text{ \ \ for
\ \ }p\text{ odd}\\
\left\langle \psi_{2q},\lambda_{2p}\right\rangle \text{ \ \ \ \ \ for
\ \ }p\text{ even}%
\end{array}
\right\}  ,
\end{equation}
with explicit representatives acting on the $\mathbb{C}^{2}$ coordinates:
\begin{align}
\psi_{k}  &  =\left[
\begin{array}
[c]{cc}%
e^{2\pi i/k} & \\
& e^{-2\pi i/k}%
\end{array}
\right]  \text{, \ \ }\varphi_{k}=\left[
\begin{array}
[c]{cc}%
e^{2\pi i/k} & \\
& e^{2\pi i/k}%
\end{array}
\right]  \text{,}\\
\tau &  =\left[
\begin{array}
[c]{cc}
& i\\
i &
\end{array}
\right]  \text{, \ \ }\lambda_{k}=\varphi_{k}\tau=\left[
\begin{array}
[c]{cc}
& ie^{2\pi i/k}\\
ie^{2\pi i/k} &
\end{array}
\right]  .
\end{align}

\subsection{Classification of Fibers}

Proceeding in our classification of 6D\ SCFTs, we have now reviewed the full
list of possible bases which can appear in an F-theory model. With this in
place, we can ask what elliptic fibers can be fibered over the base so
that the total space is a non-compact elliptically fibered Calabi-Yau threefold.

The possible choices of gauge algebra $\mathfrak{g}$ and the associated
charged matter for each $n$ are listed in Section 6.1 of
\cite{Heckman:2015bfa} and are shown in table \ref{tab:algebralist}. Note that
$\mathfrak{e}_{8}$ can only appear on a curve of self-intersection $-12$.
Whenever $\mathfrak{e}_{8}$ is put on a curve with $n < 12$, that curve must
be blown up at $12-n$ points, resulting in a curve of self-intersection $-12$.

\begin{table}[ptb]%
\begin{tabular}
[c]{|c|c|c|c|c|c|c|c|c|c|}\hline
& \multicolumn{9}{|c|}{$n$}\\\hline
$\mathfrak{g}$ & 1 & 2 & 3 & 4 & 5 & 6 & 7 & 8 & 12\\\hline
$\mathfrak{sp}(N)$ & $n_{f} = 2N+8$ &  &  &  &  &  &  &  & \\
$N \geq1$ &  &  &  &  &  &  &  &  & \\\hline
$\mathfrak{su}(2)$ & $n_{f}=10$ & $n_{f}=4$ &  &  &  &  &  &  & \\\hline
$\mathfrak{su}(3)$ & $n_{f}=12$ & $n_{f}=6$ & NHC &  &  &  &  &  & \\\hline
$\mathfrak{su}(N)$ & $n_{f}=N+8$ & $n_{f}=2N$ &  &  &  &  &  &  & \\
$N \geq4$ & $n_{\Lambda}^{2}=1$ &  &  &  &  &  &  &  & \\\hline
$\mathfrak{su}(6)$ & $n_{f}=15$ &  &  &  &  &  &  &  & \\
& $n_{\Lambda}^{3}=\frac{1}{2}$ &  &  &  &  &  &  &  & \\\hline
$\mathfrak{so}(7)$ & $n_{v}=2$ & $n_{f}=1$ & $n_{s}=2$ &  &  &  &  &  & \\
& $n_{s}=6$ & $n_{s}=4$ &  &  &  &  &  &  & \\\hline
$\mathfrak{so}(8)$ & $n_{v}=3$ & $n_{v}=2$ & $n_{v}=1$ & NHC &  &  &  &  & \\
& $n_{s}=3$ & $n_{s}=2$ & $n_{s}=1$ &  &  &  &  &  & \\
& $n_{c}=3$ & $n_{c}=2$ & $n_{c}=1$ &  &  &  &  &  & \\\hline
$\mathfrak{so}(9)$ & $n_{v}=4$ & $n_{v}=3$ & $n_{v}=2$ & $n_{v}=1$ &  &  &  &
& \\
& $n_{s}=3$ & $n_{s}=2$ & $n_{s}=1$ &  &  &  &  &  & \\\hline
$\mathfrak{so}(10)$ & $n_{v}=5$ & $n_{v}=4$ & $n_{v}=3$ & $n_{v}=2$ &  &  &  &
& \\
& $n_{s}=3$ & $n_{s}=2$ & $n_{s}=1$ &  &  &  &  &  & \\\hline
$\mathfrak{so}(11)$ & $n_{v}=6$ & $n_{v}=5$ & $n_{v}=4$ & $n_{v}=3$ &  &  &  &
& \\
& $n_{s}=\frac{3}{2}$ & $n_{s}=1$ & $n_{s}=\frac{1}{2}$ &  &  &  &  &  &
\\\hline
$\mathfrak{so}(12)$ & $n_{v}=7$ & $n_{v}=6$ & $n_{v}=5$ & $n_{v}=4$ &  &  &  &
& \\
& $n_{s}=\frac{3}{2}$ & $n_{s}=1$ & $n_{s}=\frac{1}{2}$ &  &  &  &  &  &
\\\hline
$\mathfrak{so}(13)$ &  & $n_{v}=7$ &  & $n_{v}=5$ &  &  &  &  & \\
&  & $n_{s}=\frac{1}{2}$ &  &  &  &  &  &  & \\\hline
$\mathfrak{so}(N)$ &  &  &  & $n_{v} = N-8$ &  &  &  &  & \\
$N \geq14$ &  &  &  &  &  &  &  &  & \\\hline
$\mathfrak{g}_{2}$ & $n_{f}=7$ & $n_{f}=4$ & $n_{f}=1$ &  &  &  &  &  &
\\\hline
$\mathfrak{f}_{4}$ & $n_{f}=4$ & $n_{f}=3$ & $n_{f}=2$ & $n_{f}=1$ & NHC &  &
&  & \\\hline
$\mathfrak{e}_{6}$ & $n_{f}=5$ & $n_{f}=4$ & $n_{f}=3$ & $n_{f}=2$ & $n_{f}=1$
& NHC &  &  & \\\hline
$\mathfrak{e}_{7}$ & $n_{f}=\frac{7}{2}$ & $n_{f}=3$ & $n_{f}=\frac{5}{2}$ &
$n_{f}=2$ & $n_{f}=\frac{3}{2}$ & $n_{f}=1$ & $n_{f}=\frac{1}{2}$ & NHC &
\\\hline
$\mathfrak{e}_{8}$ &  &  &  &  &  &  &  &  & NHC\\\hline
\end{tabular}
\caption{List of gauge algebras $\mathfrak{g}$ and representations allowed for
a curve of self-intersection $-n$. Empty boxes mean that this gauge algebra
cannot appear on a curve with this self-intersection number, while ``NHC"
indicates that this gauge algebra appears with no matter.}%
\label{tab:algebralist}%
\end{table}

Note that there are infinite towers of gauge algebras for curves of low $n$:
for $n=1$, we have infinite towers of $\mathfrak{sp}(N), N\geq1$ and
$\mathfrak{su}(N), N\geq2$ gauge algebras; for $n=2$, we have infinite towers
of $\mathfrak{su}(N), N\geq2$ gauge algebras, and for $n=4$, we have an
infinite tower of $\mathfrak{so}(N), N\geq8$ gauge algebras.

For each $n$, the smallest gauge algebra allowed is the ``minimal gauge
algebra" shown in (\ref{eq:NHCalgebras}). Furthermore, in each such case,
there is insufficient matter charged under this gauge algebra for it to be
Higgsed (hence the term ``non-Higgsable cluster"). As the fiber singularity
type is enhanced and the gauge algebra is increased, the amount of charged
matter similarly grows. The process of smoothing out the fiber then
corresponds in field theory language to the process of Higgsing. For instance,
a $-3$ curve with $\mathfrak{so}(7)$ gauge algebra necessarily has two spinors
$\mathbf{8}$ charged under this gauge algebra. Giving one of these
$\mathbf{8}$'s a vev Higgses the gauge group down to $\mathfrak{g}_{2}$, and
there is a single fundamental $\mathbf{7}$ charged under this $\mathfrak{g}%
_{2}$. Giving this $\mathbf{7}$ a vev Higgses the gauge group down to
$\mathfrak{su}(3)$. This $\mathfrak{su}(3)$ is the minimal gauge algebra on a
$-3$ curve, and there is no charged matter:
\begin{equation}
\underset{[N_{s}=2]}{\overset{\mathfrak{so}(7)}{3}} ~~\rightarrow~~
\underset{[N_{f}=1]}{\overset{\mathfrak{g}_{2}}{3}} ~~ \rightarrow~~
{\overset{\mathfrak{su}(3)}{3}}%
\end{equation}
Note that for $n=9, 10$, and $11$, the minimal gauge algebra is $\mathfrak{e}%
_{8}$ with some number of ``small instantons." This is just another way of
saying that these curves must be blown up by adding $-1$ curves until one gets
a $-12$ curve, as per our discussion near lines (\ref{ninecurve}) - (\ref{elevencurve}).

To understand theories with gauge algebras supported on intersecting curves,
we must also classify the full set of mixed representations that can
satisfy (\ref{eq:mixed}). The fundamental of $\mathfrak{sp}(n)$ is pseudo-real
and has index 1, which means that we can consider half-fundamentals of index
$1/2$. The fundamental representation of $\mathfrak{su}(n)$ has index $1$, and
these are the only representations of simple Lie algebras with index less than
or equal to $1$. This means that any mixed representation necessarily involves
a fundamental of either $\mathfrak{sp}(n)$ or $\mathfrak{su}(n)$. The only
representations of index $2$ are the fundamental of $\mathfrak{g}_{2}$, the
fundamental of $\mathfrak{so}(n)$, the antisymmetric $\Lambda^{2}$ of
$\mathfrak{su}(4)$, and the spinors of $\mathfrak{so}(7)$ and $\mathfrak{so}%
(8)$. Thus, the list of possible mixed representations $\rho$ satisfying
(\ref{eq:mixed}) are as follows:

\begin{itemize}
\item $\mathfrak{g}_{a} = \mathfrak{su}(N_{a})$, $\mathfrak{g}_{b} =
\mathfrak{su}(N_{b})$, $\rho= (\mathbf{{N}_{a},{N}_{b})}$

\item $\mathfrak{g}_{a} = \mathfrak{su}(N_{a})$, $\mathfrak{g}_{b} =
\mathfrak{sp}(N_{b})$, $\rho= (\mathbf{{N}_{a},{2N}_{b})}$

\item $\mathfrak{g}_{a} = \mathfrak{sp}(N_{a})$, $\mathfrak{g}_{b} =
\mathfrak{so}(N_{b})$, $\rho= \frac{1}{2}(\mathbf{{2N}_{a},{N}_{b})}$

\item $\mathfrak{g}_{a} = \mathfrak{sp}(N_{a})$, $\mathfrak{g}_{b} =
\mathfrak{so}(N_{b})$, $N_{b} = 7, 8$, $\rho= \frac{1}{2}(\mathbf{{2N}_{a}%
,{8}_{s,c})}$

\item $\mathfrak{g}_{a} = \mathfrak{sp}(N_{a})$, $\mathfrak{g}_{b}=
\mathfrak{g}_{2}$, $\rho= \frac{1}{2} (\mathbf{{2N}_{a},{7})}$

\item $\mathfrak{g}_{a} = \mathfrak{sp}(N_{a})$, $\mathfrak{g}_{b} =
\mathfrak{sp}(N_{b})$, $\rho= (\mathbf{{2N}_{a},{2N}_{b})}$

\item $\mathfrak{g}_{a} = \mathfrak{sp}(N_{a})$, $\mathfrak{g}_{b} =
\mathfrak{su}(4)$, $\rho= \frac{1}{2}(\mathbf{{2N}_{a},{6})}$
\end{itemize}

In practice, the last two items on this list do not arise in any 6D SCFT,
because the representations involved only appear on curves of
self-intersection $-1$, and two such curves can never intersect without
violating negative-definiteness of the base. Thus, any mixed representation
appearing in a 6D SCFT must be among the first five entries of this
list (ignoring the case of frozen singularities).

Finally, there are additional constraints, not related to anomalies, on curves
of self-intersection $-1$ or $-2$ that do not carry gauge groups (so-called
``unpaired" curves). In particular, a $-1$ curve must still satisfy the
$E_{8}$ gauging condition of (\ref{eq:E8condition}). Thus, for instance, the
following theories are not allowed, because $\mathfrak{so}(10) \oplus
\mathfrak{so}(8) \not \subset \mathfrak{e}_{8}$ and $\mathfrak{e}_{6}
\oplus\mathfrak{g}_{2} \not \subset \mathfrak{e}_{8}$:
\begin{align}
&  \text{NOT ALLOWED: } ~~~~~~~ \overset{\mathfrak{so}(10)}{4 }\,\, 1 \,\,
\overset{\mathfrak{so}(8)}{4}\\
&  \text{NOT ALLOWED: } ~~~~~~~~~~\overset{\mathfrak{e}_{6}}{6 }\,\, 1 \,\,
\overset{\mathfrak{g}_{2}}{3}%
\end{align}
The conditions on unpaired $-2$ curves are somewhat more complicated \cite{Morrison:2016djb}. They
obey an $\mathfrak{su}(2)$ gauging condition, which means that an unpaired
$-2$ curve can only meet a single curve carrying a gauge algebra, and that
gauge algebra must be $\mathfrak{su}(2)$. However, this condition is not quite
strong enough: an unpaired $-2$ curve that meets a curve carrying
$\mathfrak{su}(2)$ gauge algebra cannot also meet another unpaired $-2$ curve.
This can be shown (see p. 16 of \cite{Morrison:2016djb}) by looking at the
residual orders of vanishing of $f$, $g$, and $\Delta$ at the point of intersection.
Any choice of fiber types ($I_0$, $I_1$, or $II$) for the adjacent unpaired $-2$ curves
will lead to a residual vanishing that is too large on at least one of the curves. Morally, one could think of this adjacent $-2$ curve as carrying an $\mf{su}(1)$ gauge symmetry, in which case we violate the $\mf{su}(2)$ gauge condition because $\mf{su}(2) \times \mf{su}(1) \not\subset \mf{su}(2)$. \footnote{One must be careful, however, because unpaired $-2$ curves are not always associated with $\mf{su}(1)$ gauge algebras. In particular, an unpaired $-1$ curve \emph{can} meet both an unpaired $-2$ curve and a $-12$ curve carrying $\mf{e}_8$ gauge algebra, even though $\mf{e}_8 \times \mf{su}(1) \not\subset \mf{e}_8$.}
Thus, for example, the following theories are not allowed:
\begin{align}
&  \text{NOT ALLOWED: } ~~~~~~~ \overset{\mathfrak{su}(2)}{2 }\,\, 2 \,\, 2\\
&  \text{NOT ALLOWED: } ~~~~~~~~~~ \overset{\mathfrak{sp}_{3}}{1 }\,\, 2
\end{align}

\subsection{A Consistency Check: 6D SCFTs and Homomorphisms}\label{ssec:HOM}

A theory of $N$ M5-branes probing an $E_{8}$ wall and an orbifold
$\mathbb{C}^2/\Gamma_{\text{ADE}}$ allows for a flat $E_{8}$ connection to be
turned on at the ``infinity" of the orbifold $\simeq S^{3}/\Gamma_{\text{ADE}%
}$. Such connections are in 1-1 correspondence with homomorphisms
$\Gamma_{\text{ADE}} \rightarrow E_{8}$. In terms of the worldvolume 6D SCFT,
one expects a flat $E_{8}$ connection at infinity to show up as a (Higgs
branch) deformation of the theory. We will now see that this expectation is
borne out, focusing on the particular example of $\Gamma_{\text{ADE}%
}=\mathbb{Z}_{3}$.

Homomorphisms $\mathbb{Z}_{k} \rightarrow E_{8}$ were classified by Kac in
\cite{MR739850}, and they can be labeled in a simple way by appropriately
deleting the nodes of the affine $\widehat{E}_{8}$ Dynkin diagram. To begin, one
assigns numbers to each of the nodes as follows:
\begin{equation}
\includegraphics{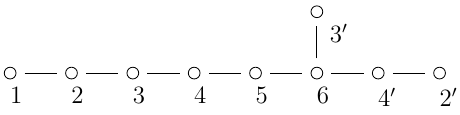}
\end{equation}
Next, one considers all combinations of nodes whose numbers sum to $k$, where
any given node may be used multiple times. For instance, for $k=3$, we have
the following choices of nodes:
\begin{equation}
1+1+1,~~1+2,~~1+2^{\prime},~~3,~~3^{\prime}.
\end{equation}
Next, one deletes these nodes from the affine $\widehat{E}_{8}$ Dynkin diagram:
what remains is the Dynkin diagram of the commutant of the homomorphism, which
is also the global symmetry (on the left-hand side) of the associated 6D SCFT.
For instance, for the homomorphisms $\mathbb{Z}_{3} \rightarrow E_{8}$, we
have 6D SCFTs with tensor branch descriptions:

\[
1+1+1 ~~\leftrightarrow~~ \includegraphics{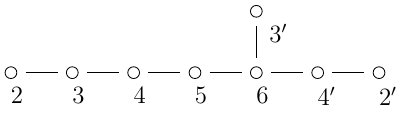}~~
\leftrightarrow~~ [E_8] \,\, 1 \,\,2 \,\, \overset{\mathfrak{su}%
(2)}{2 }\,\, \overset{\mathfrak{su}(3)}{2 }\,\, {\overset{\mathfrak{su}%
(3)}{2}} \,\,... [SU(3)]
\]
\[
1+2 ~~\leftrightarrow~~\includegraphics{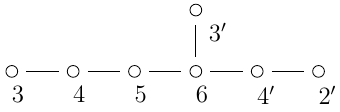}
~~\leftrightarrow~~ [E_7] \,\, 1 \,\, \underset{[N_{f}%
=1]}{\overset{\mathfrak{su}(2)}{2}} \,\, \overset{\mathfrak{su}(3)}{2 }\,\,
{\overset{\mathfrak{su}(3)}{2}} \,\,... [SU(3)]
\]
\[
1+2^{\prime}~~\leftrightarrow~~\includegraphics{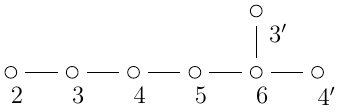}
~~\leftrightarrow~~ [SO(14)] \,\, \overset{\mathfrak{sp}_{1}}{1
}\,\, \underset{[N_{f}=1]}{ \overset{\mathfrak{su}(3)}{2}} \,\,
{\overset{\mathfrak{su}(3)}{2}} \,\,... [SU(3)]
\]
\[
3 ~~\leftrightarrow~~\includegraphics{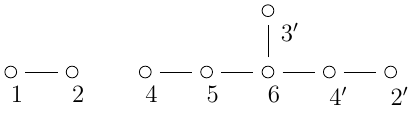}~~\leftrightarrow~~
[E_6] \,\, 1 \,\, \underset{[\mathfrak{su}(3)
]}{\overset{\mathfrak{su}(3)}{2 }} \,\, {\overset{\mathfrak{su}(3)}{2}}
\,\,... [SU(3)]
\]
\[
3^{\prime}~~\leftrightarrow~~ \includegraphics{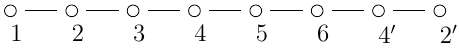}
~~\leftrightarrow~~ [SU(9)] \,\, {\overset{\mathfrak{su}(3)}{1 }}
\,\, {\overset{\mathfrak{su}(3)}{2 }} \,\, {\overset{\mathfrak{su}(3)}{2}}
\,\,... [SU(3)]
\]

There is a slightly different class of 6D SCFTs that is in 1-1 correspondence
with a slightly different class of homomorphisms. Namely, consider 6D SCFTs of
the form
\begin{equation}
[SU(n_0)] \,\, \overset{\mathfrak{su}(n_{1})}{2}\,\,
\overset{\mathfrak{su}(n_{2})}{2}\,\, \overset{\mathfrak{su}(n_{3})}{2 }\,\,
...\,\, \overset{\mathfrak{su}(n)}{2 }\,\,...\,\, \overset{\mathfrak{su}(n)}{2
}\,\, [\SU(n)]
\label{eq:Angeneral}
\end{equation}
with $n_{0} \leq n_{1} \leq n_{2} ... \leq n$. Such a theory is a deformation
of the theory of M5-branes probing a $\mathbb{Z}_{n}$ singularity that we saw
in (\ref{eq:Anquiver}), and it can be associated with a partition of $n$.
Namely, we define our partition $\mu$ by setting
\begin{equation}
\mu_{1}^{T} = n_{1}\,,~~~ \mu_{2}^{T}= n_{2} - n_{1} \,,~~~... ~~~ \mu_{k}%
^{T}= n_{k} - n_{k-1},
\end{equation}
where $\mu_{1}^{T}$ is the number of boxes in the first row of $\mu$. Anomaly
cancellation imposes the convexity condition $n_{i}-n_{i-1} \geq n_{i+1} -
n_{i}$, so indeed we always have $\mu_{i}^{T} \geq\mu_{i+1}^{T}$, as
necessary. Such partitions of $n$ are also in 1-1 correspondence with
homomorphisms $\mathfrak{su}(2)\rightarrow\mathfrak{su}(n)$, so we once again
find a 1-1 correspondence between a class of 6D SCFTs and a class of
homomorphisms. This correspondence might seem accidental so far, but in fact
it carries over to the $D_{n}$ and $E_{n}$ cases as well: deformations of the
left-hand side of the quiver in (\ref{eq:Dnquiver}) are in 1-1
correspondence with homomorphisms $\mathfrak{su}(2) \rightarrow\mathfrak{so}%
(2n)$, and deformations of the quivers in (\ref{eq:E6quiver}%
)-(\ref{eq:E8quiver}) are in 1-1 correspondence with homomorphisms
$\mathfrak{su}(2) \rightarrow\mathfrak{e}_n$. These correspondences were
worked out in \cite{Heckman:2016ssk} and further explored in
\cite{Mekareeya:2017sqh,Mekareeya:2016yal}.

The connection between homomorphisms and 6D SCFTs is important for several
reasons. First, it provides a new and surprising method for classifying
homomorphisms $\Gamma_{\text{ADE}} \rightarrow E_{8}$. This method has been
checked against the classification of homomorphisms $\mathbb{Z}_{k}
\rightarrow E_{8}$ for all $k$ and reproduces
the classification of homomorphisms $\Gamma_{E_{8}} \rightarrow E_{8}$ that
was carried out by Frey in \cite{FREY}.\footnote{As pointed out more recently in
\cite{Mekareeya:2017jgc}, there is a subtlety here for $\mathbb{Z}_{k}
\rightarrow E_{8}$ with $k \geq8$ in that a
quiver with an $\mathfrak{sp}(k-4)$ gauge algebra adjacent to an
$\mathfrak{su}(2k)$ gauge algebra actually counts as two distinct 6D SCFTs, as
there are two distinct ways to embed $\mathfrak{su}(2k)$ in the $\mathfrak{so}%
(4k)$ flavor symmetry associated with the $2k$ hypermultiplets of the
$\mathfrak{sp}(k-4)$ gauge algebra. Once this subtlety is taken into account,
the match with homomorphisms works correctly in these cases.} Further, it allows for a
classification of homomorphisms $\Gamma_{D_{n}} \rightarrow E_{8}$,
$\Gamma_{E_{6}} \rightarrow E_{8}$, and $\Gamma_{E_{7}} \rightarrow E_{8}$ \cite{FreyRudelius}--a novel mathematical result.

From the perspective of 6D SCFTs, this correspondence is important because it
offers a strong consistency check on our claimed classification of 6D SCFTs.
The fact that we do not find a huge shortage of 6D SCFTs relative to the
number of homomorphisms $\mathbb{C}^2/\mathbb{Z}_{k}$ is strong evidence that we
have largely understood the landscape of 6D SCFTs.

\subsection{Aside: F-theory on a Singular Base}

In much of our discussion so far, we have focused on the F-theory realization
of the tensor branch for a 6D\ SCFT. Of course, we are also interested in the
singular limit where the curves of the base have collapsed to zero size. Given
the fact that each base is actually just a blowup of an orbifold singularity
$\mathbb{C}^{2}/\Gamma$, it is natural to ask whether there is a more
intrinsic way to present this data. This can indeed be done, and simply
involves interpreting the Weierstrass model coefficients $f$ and $g$ as
appropriate sections of bundles on the singular base.

Starting from an elliptically fibered model with base $\mathcal{B}%
_{\text{trivial}}=\mathbb{C}^{2}$ and Weierstrass form%
\begin{equation}
y^{2}=x^{3}+f(u,v)x+g(u,v),
\end{equation}
for $\gamma\in\Gamma_{U(2)}$, the variables of the Weierstrass model must
transform as (see \cite{Morrison:2016nrt, DelZotto:2017pti}):%
\begin{align}
(x,y)  &  \mapsto(\det\left(  \gamma\right)  ^{2}x,\det\left(  \gamma\right)
^{3}y)\\
f(u,v)  &  \mapsto\det\left(  \gamma\right)  ^{4}f(u,v)\\
g(u,v)  &  \mapsto\det\left(  \gamma\right)  ^{6}g(u,v).
\end{align}
By tuning the coefficients of the Weierstrass model, we can reach a more
singular model, which in turn may require additional blowups to reach a smooth
base. So in other words, all the data of which 6D\ SCFT we have chosen is
really encoded in just the choice of a discrete group $\Gamma_{U(2)}$ and a
specific choice of tuning for $f$ and $g$.

With this in mind, there are actually two complementary perspectives we can
adopt. On the one hand, we can study F-theory on a smooth base with a
configuration of contractible curves.\ On the other hand, we can simply work
with a singular base, and specify the structure of the Weierstrass model
coefficients in this limit \cite{Morrison:2016nrt, DelZotto:2017pti, DelZotto:2014fia}.
Conceptually, the latter approach is somewhat more direct, but
from the point of view of actually classifying 6D\ SCFTs, it turns out to be
far simpler to work with F-theory on a smooth base.  The issue with classifying 6D\ SCFTs via F-theory on a singular base is that a
priori, we have no direct way of determining the maximal amount of singular
tuning allowed for $f$ and $g$ without eventually blowing up the base. For this
reason, it seems to be simpler to classify F-theory models by working on a
smooth base with contractible configurations of curves.

\subsection{Completeness and a New Swampland Conjecture}

Taking stock of the above, it is tempting to conclude that the F-theory classification is \textit{complete},
namely that we have actually classified \textit{all} 6D SCFTs. This is clearly a rather provocative claim but
there is significant evidence that it is correct.

At the very least, the F-theory approach provides a
systematic way to construct a vast collection of 6D SCFTs.
Moreover, we have also seen that nearly every 6D SCFT which
has ever been constructed in string theory has a realization in a geometric phase of F-theory.

Of course, there are still some outstanding issues: we have not directly classified 6D SCFTs in this
section, but rather tensor branches of 6D SCFTs. If some 6D SCFT does not
possess a tensor branch, it would be missed by our analysis. Indeed,
\cite{Arras:2016evy} found examples of ``$\mathbb{Q}$-factorial terminal
singularities" in 6D F-theory models, which cannot be resolved by moving out
on a tensor branch, but so far the only examples involve free hypermultiplets.
Some field theoretic considerations indicate that there could be 6D SCFTs without a tensor branch
\cite{Shimizu:2017kzs, Chang:2017xmr} based
on the theory of a small instanton in a higher-dimensional gauge theory,
though as usual, the absence of a stringy realization makes it hard to assess whether these will
really provide genuine examples of 6D SCFTs.
It will be interesting to see if examples of interacting 6D SCFTs can be
constructed in this way in the near future, or if a no-go theorem could rule
out such theories.

Another caveat to these considerations is the appearance of frozen singularities \cite{Witten:1997bs, Tachikawa:2015wka}, and
their relation to $O7^{+}$-planes mentioned in subsection \ref{ssec:FROZEN}. These theories still take the form
of generalized quivers, and in all known examples, we can actually view these as a non-geometric phase of an F-theory model,
directly obtained by quotienting a geometric F-theory background \cite{Bhardwaj:2015oru, Apruzzi:2017iqe}. Moreover,
the number of additional theories constructed in this manner is quite limited. Given this,
we conclude that at present, all known stringy realized SCFTs appear to lift to an F-theory avatar, be it geometric, or in some
frozen phase.

The correspondence between 6D SCFTs and homomorphisms is
important in that it helps us understand global symmetries and renormalization
group flows of 6D SCFTs \cite{Heckman:2016ssk,Mekareeya:2016yal}, as
well as their compactifications to lower dimensions \cite{Mekareeya:2017jgc}.
For the particular classes of 6D SCFTs considered in
this subsection, the global symmetry of the 6D SCFT is related to the
commutant of the corresponding homomorphism, which is known for all
homomorphisms $\mathfrak{su}(2) \rightarrow\mathfrak{g}$, as well as
homomorphisms $\mathbb{Z}_{k} \rightarrow E_{8}$ and $\Gamma_{E_{8}}
\rightarrow E_{8}$. This allows us to determine the global symmetry of
any of these 6D SCFTs, which in some cases is difficult to read from the
F-theory perspective \cite{Bertolini:2015bwa, Merkx:2017jey}. Further, the web of renormalization group flows between
6D SCFTs corresponding to homomorphisms $\mathfrak{su}(2) \rightarrow
\mathfrak{g}$ has an analogous structural interpretation from the perspective
of these homomorphisms, as we will discuss further in Section \ref{sec:RGFLOWS}.
Finally, the toroidal compactification of a theory of M5-branes probing
a $\mathbb{C}^2/\mathbb{Z}_k$ singularity and an $E_8$ wall can be cleanly
expressed in terms of the homomorphisms of
$\mathbb{Z}_{k} \rightarrow E_{8}$ \cite{Mekareeya:2017jgc}.

So all told, the mismatch between bottom up and top down considerations is sufficiently mild that one can hope that F-theory is simply
revealing additional field theoretic structures. Examples of this kind are the $-n$ curve theories for $n > 2$, which always
couple to a 6D vector multiplet. Indeed, modulo this subtlety, there is a nearly identical set of conditions imposed by anomaly cancellation
and consistency of an elliptic fibration. This provides reason for optimism that the two approaches will converge.

At a broader level, we can couch these issues in terms of the ``swampland conjectures'' \cite{Vafa:2005ui, Ooguri:2006in, Brennan:2017rbf}.
This is usually stated in terms of the lore that although there could be apparently consistent low energy effective field theories,
coupling to gravity renders these theories inconsistent. This question is particularly sharp in the context of 6D supergravity theories,
and has been vigorously investigated in the context of F-theory compactification starting with reference \cite{Kumar:2009ac}. For a
review of some aspects of this in the context of explicit string compactifications, see reference \cite{Taylor:2011wt}.

Here, the considerations of the present paper appear to be at odds with this philosophy. After all, we have explicitly chosen to
\textit{decouple} 6D supergravity. Note, however, that always lurking in the background is the higher-dimensional
string construction. Indeed, even in the decoupling limit, there is a higher-dimensional gravity theory.

With this in mind, we can put forward a variant of the original swampland conjectures:

\bigskip
\textbf{Conjecture: All self-consistent QFTs can be coupled to a possibly higher-dimensional theory of quantum gravity}.
\bigskip

By construction, all stringy realized QFTs fit in this category, though there are clearly many examples
of QFTs (non-supersymmetric ones being prime examples) which at first have no clear lift to string theory. This
seems more a question about implementation than a genuine counterexample, though. It is tempting
to extend this conjecture to a related statement about continuous global symmetries, though
here some care is required in dealing with symmetries which are emergent in the infrared.

Though it is clearly beyond the scope of the present work to prove this conjecture, we note that it would,
as a byproduct, immediately address whether every 6D SCFT embeds in string theory. We leave this as an exercise for the reader.


\section{Anomaly Polynomials \label{sec:ANOMPOLY}}

Now that we have constructed an infinite class of 6D SCFTs in F-theory,
we would like to study their properties in more detail.  Due to our relatively poor
understanding of 6D SCFTs at the conformal fixed point, however, we are restricted to properties
of the 6D SCFTs that are preserved on the tensor branch. One such property is
the anomaly polynomial $I_8$. Recall that the anomaly polynomial is a sum of two pieces: a 1-loop contribution and
a Green-Schwarz contribution,
\begin{equation}
I_{\text{tot}} = I_{\text{1-loop}}+ I_{\text{GS}} .
\label{eq:Itot}
\end{equation}
The full anomaly polynomial of a 6D SCFT takes the form
\begin{align}
I_{8}  &  = \alpha c_{2}(R)^{2} + \beta c_{2}(R) p_{1}(T) + \gamma
p_{1}(T)^{2} + \delta p_{2}(T)\nonumber\\
&  + \sum_{i} \left[  \mu_{i} \, \mathrm{Tr} F_{i}^{4}
+ \, \mathrm{Tr} F_{i}^{2} \left(  \rho_{i}
p_{1}(T) + \sigma_{i} c_{2}(R) + \sum_{j} \eta_{ij} \, \mathrm{Tr} F_{j}^{2}
\right)  \right]  . \label{eq:anomalypoly}%
\end{align}
Here, $c_{2}(R)$ is the second Chern class of the $SU(2)_{R}$ symmetry,
$p_{1}(T)$ is the first Pontryagin class of the tangent bundle, $p_{2}(T)$ is
the second Pontryagin class of the tangent bundle, and $F_{i}$ is the field
strength of the $i$th symmetry, where $i$ and $j$ run over both the gauge and
global symmetries of the theory.

The 1-loop contribution for a 6D SCFT is simply a sum of contributions from
free tensor multiplets, vector multiplets, and hypermultiplets. These
contributions are, respectively \cite{Ohmori:2014kda}:
\begin{align}
I_{\text{tensor}} = \frac{c_{2}(R)^{2}}{24} + \frac{c_{2}(R)p_{1}(T)}{48}  &
+ \frac{23 p_{1}(T)^{2} -116 p_{2}(T)}{5760},\label{eq:tensor}\\
I_{\text{vector}} = -\frac{\tr_{\text{adj}} F^{4} + 6 c_{2}(R) \tr_{\text{adj}%
} F^{2} + d_{G} c_{2}(R)^{2}}{24}  &  - \frac{\tr_{\text{adj}}F^{2}+d_{G}
c_{2}(R)p_{1}(T)}{48}\nonumber\\
&  - d_{G} \frac{7 p_{1}(T)^{2} - 4 p_{2}(T)}{5760},\label{eq:vector}\\
I_{\text{hyper}} = \frac{\tr_{\rho} F^{4} }{24} + \frac{\tr_{\rho}F^{2}
p_{1}(T)}{48}  &  + d_{\rho}\frac{7 p_{1}(T)^{2} - 4 p_{2}(T)}{5760}.
\label{eq:hyper}%
\end{align}
Here, $\tr_{\rho}$ is the trace in the representation $\rho$, $d_{\rho}$ is
the dimension of the representation $\rho$, and $d_{G}$ is the dimension of
the group $G$. In computing the anomaly polynomial, one should express the
traces over arbitrary representations in terms of the trace in a defining
representation. One may write
\begin{align}
\tr_{\rho}F^{4} = x_{\rho}\, \mathrm{Tr} F^{4} + y_{\rho}(\, \mathrm{Tr}
F^{2})^{2}\\
\tr_{\rho}F^{2} = \text{Ind}_{\rho}\, \mathrm{Tr} F^{2} ,
\end{align}
with $x_{\rho}$, $y_{\rho}$, and $\text{Ind}_{\rho}$ well-known constants in
group theory, listed in Appendix \ref{sec:GROUPTHEORY}. For the adjoint representation, $\text{Ind}_{\rho}$ is also known as
the dual Coxeter number, $h_{G}^{\vee}$. Additionally, the groups $SU(2)$,
$SU(3)$, $G_{2}$, $F_{4}$, $E_{6}$, $E_{7}$, and $E_{8}$ are special in that
they do not have an independent quartic casimir $\, \mathrm{Tr} F^{4}$.

The Green-Schwarz contribution takes the form
\begin{equation}
I_{\text{GS}} = \frac{1}{2} A^{ij} I_{i} I_{j},
\end{equation}
where $A^{ij}$ is the inverse of the Dirac pairing on the string charge
lattice, or equivalently the inverse of the intersection pairing on the
F-theory base $\mathcal{B}$. The term $I_i$ can be written as
\begin{equation}
I_{i} = a_{i} c_{2}(R) + b_{i} p_{1}(T) + \sum_{j} c_{ij} \, \mathrm{Tr}
F_{j}^{2}.
\label{eq:Ii}
\end{equation}

Using the prescription of \cite{Ohmori:2014kda} (see also \cite{Cordova:2018cvg, CordovaMixer},
one can actually compute the anomaly polynomial $I_8$ for any 6D SCFT in our classification.
We now describe the procedure and illustrate with a handful of examples.

\subsection{$(2,0)$ and E-string Theories}

The anomaly polynomial for the $(2,0)$ theory living on a stack of $Q$ M5-branes was first computed in \cite{Harvey:1998bx}.  The total anomaly computed in the 11D M-theory, which is forced to vanish, is a sum of three terms:
\begin{equation}
I_{\text{tot}} = I^{Q \text{ M5s}} + I_{\text{inf}} +  I_{\text{CGG}} = 0.
\end{equation}
Here, $I^{Q \text{ M5s}}$ is the anomaly polynomial for the 6D SCFT of interest, $I_{\text{inf}}$ is the anomaly inflow arising from the coupling
\begin{equation}
\int_{M_{11}} C_3 \wedge I_8^{\text{inf}},
\end{equation}
and $ I_{\text{CGG}} $ the $C\wedge G \wedge G$ Chern-Simons term of 11d supergravity.  Building on the previous work of \cite{Freed:1998tg, Witten:1996hc}, the authors of \cite{Harvey:1998bx} determined $I_{\text{inf}}$ and used this to compute $I^{Q \text{ M5s}}$.

An alternative method of computation was later developed in \cite{Ohmori:2014kda}.  The authors considered the compactification of a 6D SCFT to 5D, under which the 5D theory acquires a Chern-Simons term of the form
\begin{equation}
\frac{1}{2 \pi} S_{\text{CS}} = A^{ij} b_i I_j,
\label{eq:CSterm}
\end{equation}
with $I_j$ the 4-form defined in (\ref{eq:Ii}), and $b_i$ a gauge field that descends from the 6d $2$-form $B^{(i)}$.  In 5D, this Chern-Simons term arises from integrating out massive fermions charged under the gauge field $A_i$, which correspond in 6D to strings charged under the 2-form $B^{(i)}$.  Since we know the spectrum of charged strings in the 6D theory, we also know the spectrum of fermions in 5D, which allows us to compute the Chern-Simons term $S_{\text{CS}} $ and, by extension, the $I_j$ and 6d Green-Schwarz term $A^{ij} I_i I_j$.  To this, we simply add the 1-loop contribution from $Q$ free (2,0) tensor multiplets.  The final result for the anomaly polynomial of the worldvolume theory of a stack of $Q$ M5-branes is
\begin{equation}
I^{Q \text{ M5s}} = \frac{Q^3}{24} (c_2(R)-c_2(L))^2 + \frac{Q}{192} \left( ( 4 c_2(L)+p_1(T) )(4 c_2(R)+p_1(T))-4 p_2(T)\right).
\label{eq:QM5s}
\end{equation}
To get the anomaly polynomial for the $A_{Q-1}$ (2,0) SCFT, one must subtract from this the contribution from the ``center-of-mass" (2,0) tensor multiplet, which is a combination of a (1,0) tensor multiplet, with anomaly polynomial given in (\ref{eq:tensor}), and a free hypermultiplet charged as a doublet under $SU(2)_L$, with
\begin{equation}
I_{\text{free}} = \frac{c_2(L)^2}{24} + \frac{c_2(L) p_1(T)}{48} + \frac{7 p_1(T)^2 }{ 5760 } - \frac{  p_2(T)}{ 1440 }.
\end{equation}
Thus,
\begin{align}
I_{A_{Q-1}} &= I^{Q \text{ M5s}} - I_{\text{tensor}} - I_{\text{free}} \nonumber \\
&=\frac{c_2(L)^2}{24} \left(Q^3-1\right)- \frac{c_2(L) c_2(R)}{12} \left(Q^3-Q\right)
+\frac{c_2(R)^2}{24} \left(Q^3-1\right)+\frac{p_1(T) c_2(L) }{48}\left(Q-1\right) \nonumber \\
&+\frac{p_1(T) c_2(R) }{48}\left(Q-1\right)+\frac{p_1(T)^2}{192}  (Q-1)-\frac{p_2(T)}{48} \left(Q-1\right).
\end{align}
This may be written in a way that generalizes to an arbitrary ADE-type $\mathcal{N} = (2,0)$ SCFT,
\begin{equation}
I_G^{(2,0)} = \frac{h_G^\vee d_G}{24} (c_2(L)-c_2(R))^2 + r_G (I_{\text{tensor}} +I_{\text{free}}).
\end{equation}
Here, $h_G^\vee$, $d_G$, and $r_G$ are the dual Coxeter number, dimension, and rank of $G$, respectively.

Next, we consider the anomaly polynomial for the rank $Q$ E-string theory, first computed in \cite{Ohmori:2014pca} using anomaly inflow in M-theory.  Alternatively, the anomaly polynomial can be calculated by compactifying the 6D SCFT on a circle and computing the Chern-Simons terms, then lifting them to 6D.  The simplest way to do this is to compute the Chern-Simons terms on the special point of moduli space at which all the M5-branes are coincident, but separated from the wall.  At this point in moduli space, the worldvolume theory is simply that of a stack of $Q$ M5-branes considered above, but there is an additional Chern-Simons term associated with the multiplet whose scalar parametrizes the distance between the M5-brane stack and the $E_8$ wall,
\begin{equation}
\frac{1}{2 \pi} S_{\text{CS}} = Q A \left( \frac{Q}{2} c_2(L)-c_2(R) + \frac{1}{4} (\Tr F_{\mf{e}_8}^2 + p_1(T) -2 c_2(L)-2 c_2(R) ) \right)
\end{equation}
From this, we read off the 6D Green-Schwarz contribution $I_{\text{GS}}$ using (\ref{eq:CSterm}), and add this to the contribution from the stack of $Q$ M5-branes given in (\ref{eq:QM5s}).  The final result is
\begin{align}
I^{\text{rank } Q \text{ E-string}} &= I^{Q \text{ M5s}} + \frac{Q}{2} \left( \frac{Q}{2} c_2(L)-c_2(R) + \frac{1}{4} (\Tr F_{\mf{e}_8}^2 + p_1(T) -2 c_2(L)-2 c_2(R) ) \right)
^2 \nonumber \\
&=\frac{Q^3}{6} (c_2(L)-c_2(R))^2 +\frac{Q^2}{8} (c_2(L)-c_2(R)) I_4 + Q(\frac{1}{2} I_4^2 - I_8 ),
\end{align}
with
\begin{equation}
I_4 = \frac{1}{4}(  \Tr F_{\mf{e}_8}^2 + p_1(T) -2 c_2(L) -2 c_2(R)) \,,~~ I_8=\frac{1}{192}[4 p_2(T) - (4 c_2(L)+p_1(T))(4 c_2(R)+p_1(T))].
\end{equation}
As we will see in the next subsection, whenever the rank $Q$ E-string appears as part of a quiver in a 6D SCFT, one must subtract the contribution of a single free hypermultiplet $I_{\text{free}}$.  We therefore define
\begin{equation}
I^1 = I^{\text{rank } 1 \text{ E-string}} - I_{\text{free}} \,,~~I^{12} = I^{\text{rank } 2 \text{ E-string}} - I_{\text{free}} \,, ...~~I^{12...2} = I^{\text{rank } Q \text{ E-string}} - I_{\text{free}}.
\label{eq:Estringdef}
\end{equation}

\subsection{General 6D SCFTs}
We now have all of the ingredients we need to write down the anomaly polynomial for any of our 6D SCFTs.  The procedure consists of two steps: first, we write down the 1-loop contributions to the anomaly polynomial, including those from any attached E-string or $(2,0)$ theories.  Second, we determine the Green-Schwarz contribution to the anomaly polynomial by completing the square with respect to the gauge anomalies, $\Tr F_i^2$.  The most subtle part of this procedure is correctly accounting for the E-strings and $(2,0)$ theories coupled to the quiver.  We show how to handle these in explicit examples.

To begin, let us consider the $E_6$-$E_6$ conformal matter theory of (\ref{eq:E6E6confmat}):
\begin{equation}
[E_6^L] \,\,1 \,\, \overset{\mathfrak{su}(3)}{3 }\,\, 1 \,\, [E_6^R]
\end{equation}
The 1-loop contributions to the anomaly polynomial come from the tensor multiplet associated to the $-3$ curve, the vector multiplets associated with the $\mf{su}(3)$, and the pair of rank 1 E-strings (the curves of self-intersection $-1$).  Note that the $E_8$ flavor symmetry associated with each rank 1 E-string is broken to a subgroup $SU(3) \times E_6^{L,R}$.  We account for this by replacing $\Tr F_{\mf{e}_8}^2 \rightarrow \Tr F_{\mf{su}(3)}^2 + \Tr F_{L,R}^2$.  Thus, using the definitions in (\ref{eq:Itot}), (\ref{eq:tensor}), (\ref{eq:vector}), and (\ref{eq:Estringdef}), we have
\begin{align}
I_{\text{1-loop}} &= I_{\text{tensor}}+ I_{\text{vector}}(F_{\mf{su}(3)}^2)+ I^1(F_{\mf{su}(3)}^2+F_L^2)+ I^1(F_{\mf{su}(3)}^2+F_R^2) \nonumber \\
 &=\frac{1}{5760} \Big[ 4560 c_2(R)^2-180 \Big(\Tr F_R^2 (8 c_2(R)-2 \Tr F_{\mf{su}(3)}^2-2 p_1(T)) \nonumber \\
 &-2 \Tr F_{\mf{su}(3)}^2 (-20 c_2(R)+\Tr F_L^2+p_1(T))- \Tr F_L^2 (-8 c_2(R)+\Tr F_L^2+2 p_1(T))\nonumber \\
 & -(\Tr F_R^2)^2+ (\Tr F_{\mf{su}(3)}^2)^2\Big)-3480 c_2(R) p_1(T)+373 p_1(T)^2-316 p_2(T) \Big]
\end{align}
The total anomaly polynomial must be free of any gauge anomalies, mixed gauge-global anomalies, or mixed gauge-gravitational anomalies, which means that the Green-Schwarz term must cancel these terms from the 1-loop contribution.  Looking carefully at the 1-loop contribution, we see that it is a quadratic polynomial in $\Tr F_{\mf{su}(3)}^2$.  Thus, we can compute the Green-Schwarz term simply by completing the square in $\Tr F_{\mf{su}(3)}^2$, arriving finally at
\begin{align}
I_8 &= I_{\text{1-loop}} + I_{\text{GS}} \nonumber \\&=  \frac{319 }{24}c_2(R)^2-\frac{3}{2} c_2(R) \Tr F_R^2-\frac{3}{2} c_2(R) \Tr F_L^2-\frac{89 }{48}c_2(R) p_1(T)+\frac{1}{8} p_1(T) \Tr F_R^2 \nonumber \\
&+\frac{1}{8} p_1(T) \Tr F_L^2+\frac{1}{16} (\Tr F_R^2)^2+\frac{1}{16} (\Tr F_L^2)^2+\frac{1}{16} \Tr F_R^2 \Tr F_L^2+\frac{553}{5760} p_1(T)^2-\frac{79 }{1440}p_2(T).
\end{align}
Next, we consider the $E_8$-$E_8$ conformal matter from (\ref{eq:E8confmat}),
\begin{equation}
[E_8^L] \,\, 1 \,\, 2 \,\, \overset{\mathfrak{su}(2)}{2 }\,\, \overset{\mathfrak{g}_{2}}{3
}\,\, 1 \,\, \overset{\mathfrak{f}_{4}}{5 }\,\, 1 \,\, \overset{\mathfrak{g}%
_{2}}{3 }\,\,\overset{\mathfrak{su}(2)}{2 }\,\, 2 \,\, 1 \,\, [E_8^R]
\end{equation}
Here, we have rank 2 E-strings on either end of the quiver.  Unlike the rank 1 E-strings we saw in the previous examples, the anomaly polynomial $I^{12}$ of the rank 2 E-string has terms with $c_2(L)$.   In this case, the $SU(2)_L$ symmetry is gauged by the $\mf{su}(2)$ gauge symmetry, so we take $c_2(L) \rightarrow \frac{1}{4} \Tr F_{\mf{su}(2)}^2$.  Furthermore, the rank 2 E-string gobbles up one of the half-hypermultiplets charged under the $\mf{su}(2)$ gauge algebra, so we should only write down the contributions for the seven half-hypermultiplets charged under the $\mf{su}(2)$ gauge algebra.  Thus we have
\begin{align}
I_{\text{1-loop}} &= I^{12}(F_L^2, c_2(L) \rightarrow \frac{1}{4} F_{\mf{su}(2)_L}^2)+ 5 I_{\text{tensor}}+ I_{\text{vector}}(F_{\mf{su}(2)_L}^2)+\frac{1}{2}I_{\text{mixed}}(F_{\mf{su}(2)_L}^2,F_{\mf{g}_{2,L}}^2) \nonumber \\
&+ I_{\text{vector}}(F_{\mf{g}_{2,L}}^2) + I^{1}(F_{\mf{g}_{2,L}}^2+F_{\mf{f}_{4}}^2)+ I_{\text{vector}}(F_{\mf{f}_{4}}^2)
+ I^{1}(F_{\mf{g}_{2,R}}^2+F_{\mf{f}_{4}}^2)+ I_{\text{vector}}(F_{\mf{g}_{2,R}}^2)\nonumber \\&+\frac{1}{2}I_{\text{mixed}}(F_{\mf{su}(2)_R}^2,F_{\mf{g}_{2,R}}^2)+I_{\text{vector}}(F_{\mf{su}(2)_R}^2)+I^{12}(F_R^2, c_2(L) \rightarrow \frac{1}{4} F_{\mf{su}(2)_R}^2).
\end{align}
Here, $\frac{1}{2}I_{\text{mixed}}$ represents the contribution from the half-bifundamental $\frac{1}{2}(\bf{2},\bf{7})$ charged under $\mf{su}(2) \oplus \mf{g}_2$.  The resulting expression is a quadratic polynomial in the gauge field strengths $\Tr F_{\mf{su}(2)_R}^2$, $\Tr F_{\mf{su}(2)_L}^2$, $\Tr F_{\mf{g}_{2,R}}^2$, $\Tr F_{\mf{g}_{2,L}}^2$, and $\Tr F_{\mf{f}_{4}}^2$.  To get the final expression for the anomaly polynomial, we complete the square with respect to each of these field strengths.  This gives
\begin{align}
I_8 &= I_{\text{1-loop}} + I_{\text{GS}} \nonumber \\ &=  \frac{4163 }{8}c_2(R)^2-\frac{45}{4} c_2(R) \Tr F_R^2-\frac{45}{4} c_2(R) \Tr F_L^2-\frac{277 }{16}c_2(R) p_1(T)+\frac{5}{16} p_1(T) \Tr F_R^2 \nonumber \\
&+\frac{5}{16} p_1(T) \Tr F_L^2+\frac{5}{32} (\Tr F_R^2)^2+\frac{5}{32} (\Tr F_L^2)^2+\frac{1}{16} \Tr F_R^2 \Tr F_L^2+\frac{581}{1920} p_1(T)^2-\frac{83 }{480}p_2(T).
\label{eq:E8E8confmatanom}
\end{align}
Anomaly polynomials for other types of conformal matter can be found in Appendix \ref{sec:CONFMAT}.

For a theory with an empty $-2$ curve that is not adjacent to an empty $-1$ curve, we must add the contribution of the $A_1$ $(2,0)$ theory.  As with the rank 2 E-string in the previous example, the $\mf{su}(2)_L$ symmetry is gauged by taking $c_2(L) \rightarrow \frac{1}{4} F_{\mf{su}(2)}^2$, and a single half-hypermultiplet is gobbled up by the empty $-2$ curve.  For instance, for the 6D SCFT with tensor branch:
\begin{equation}
2 \,\, \ov{\mf{su}_2}2 \,\, \ov{\mf{g}_2}3
\end{equation}
we have
\begin{align}
I_{\text{1-loop}} &= I_{A_1}(c_2(L) \rightarrow \frac{1}{4} F_{\mf{su}(2)}^2)+2 I_{\text{tensor}}+ I_{\text{vector}}(F_{\mf{su}(2)}^2)+\frac{1}{2}I_{\text{mixed}}(F_{\mf{su}(2)}^2,F_{\mf{g}_{2}}^2) \nonumber \\
&+ I_{\text{vector}}(F_{\mf{g}_{2}}^2)
\end{align}
Completing the square in $F_{\mf{su}(2)}^2$ and $F_{\mf{g}_2}^2$, we get
\begin{align}
I_8 &= I_{\text{1-loop}} + I_{\text{GS}} = \frac{725 }{84}c_2(R)^2+\frac{53 }{168} c_2(R) p_1(T)+\frac{97 }{6720} p_1(T)^2-\frac{13 }{240}p_2(T).
\end{align}

These rules suffice to compute the anomaly polynomial for any 6D SCFT in our classification.  However, it is also worth noting a few ``tricks" which can simplify the computations of anomaly polynomials.  First off, anomaly polynomials may be ``glued" together by gauging flavor symmetries.  For instance, for the $E_8$-$E_8$ conformal matter, we can gauge the $\mf{e}_8$ flavor symmetries on either side of the quiver:
\begin{equation}
\overset{\mathfrak{e}_8}{(12) } \,\, 1 \,\, 2 \,\, \overset{\mathfrak{su}(2)}{2 }\,\, \overset{\mathfrak{g}_{2}}{3
}\,\, 1 \,\, \overset{\mathfrak{f}_{4}}{5 }\,\, 1 \,\, \overset{\mathfrak{g}%
_{2}}{3 }\,\,\overset{\mathfrak{su}(2)}{2 }\,\, 2 \,\, 1 \,\, \overset{\mathfrak{e}_8}{(12) }
\end{equation}
The anomaly polynomial of this theory can be calculated simply by expanding the $-12$ curves to infinite size, thereby moving to the point on the tensor branch at which the remaining SCFT is simply the $E_8$-$E_8$ conformal matter quiver.  Here, the ``1-loop" contribution to the anomaly polynomial is a sum of the $E_8$-$E_8$ conformal matter anomaly polynomial of (\ref{eq:E8E8confmatanom}) together with the contribution from the pair of free tensors and $\mf{e}_{8,L}$,  $\mf{e}_{8,R}$ vector multiplets:
\begin{equation}
I_{\text{1-loop}} = I_{E_8-E_8 \text{ conf. mat}} + 2 I_{\text{tensor}} + I_{\text{vector}}(F_{L}^2)+ I_{\text{vector}}(F_{R}^2).
\end{equation}
Then, the full anomaly polynomial is obtained simply by completing the square in the field strengths $\Tr F_L^2$, $\Tr F_R^2$ of the $\mf{e}_{8}$ gauge algebras on the left and right of the quiver:
\begin{equation}
I_8 = I_{\text{1-loop}} + I_{\text{GS}} = \frac{34495 }{24}c_2(R)^2+\frac{175 }{48} c_2(R) p_1(T)-\frac{61 }{1920} p_1(T)^2+\frac{21 }{160}p_2(T).
\end{equation}

Analytic continuation is another trick that can be employed in computing anomaly polynomials.  Suppose we have a quiver of the form:
\begin{equation}
[\mathfrak{su}(n_{0})] \,\, \overset{\mathfrak{su}(n_{1})}{2}\,\,
\overset{\mathfrak{su}(n_{2})}{2}\,\, \overset{\mathfrak{su}(n_{3})}{2 }\,\,
...\,\, \overset{\mathfrak{su}(n)}{2 }\,\,...\,\, \overset{\mathfrak{su}(n)}{2
}\,\, [\mathfrak{su}(n)]
\end{equation}
When computing the anomaly polynomial for such a theory, one must first express the traces $\tr_{\rho_i} F_i^2, \tr_{\rho_i} F_i^4$ in terms of the quadratic and quartic casimirs $\Tr F_i^2, \Tr F_i^4$ using the group theory coefficients in Appendix \ref{sec:GROUPTHEORY}.  Note that the group theory coefficients for the gauge algebra $\mf{su}(n_i)$ can be written analytically in terms of $n_i$ for $n_i \geq 4$, while $\mf{su}(2)$ and $\mf{su}(3)$ are different (after all, they do not have independent quartic casimirs).  Thus, one obtains an analytic expression for coefficients of the anomaly polynomial in terms of the $n_i$, which is valid provided $n_i \geq 4$.  Nonetheless, perhaps surprisingly, one may analytically continue the formulae for $n_i \geq 4$ to the special cases of $\mf{su}(2)$ and $\mf{su}(3)$ and obtain the correct result.  Even more surprisingly, one can analytically continue all the way to $\mf{su}(1)$, treating an unpaired $-2$ curve as if it were holding an $\mf{su}(1)$ gauge group.

Similarly, in quivers of the form,
\begin{equation}
 ... \,\, \ov{\mf{so}(n_i)}4 \,\, \ov{\mf{sp}(m_i)}1 \,\, \ov{\mf{so}(n_{i+1})}4 \,\, ...
\end{equation}
one can analytically continue the formulae for the anomaly polynomial coefficients all the way to $m_i = 0$, treating a rank 1 E-string as if it were a $-1$ curve with $\mf{sp}(0)$ gauge group.  In fact, one can even analytic continue to $m_i <0$ and $n_i < 8$ to account for theories with spinors \cite{Mekareeya:2016yal}!  These analytic continuations can be useful in establishing theorems involving the anomaly polynomial coefficients of these classes of theories, as we will see in Section \ref{ssec:atheorem}.

\section{Renormalization Group Flows \label{sec:RGFLOWS}}

The renormalization group (RG) is central to the modern understanding of
quantum field theories. Under a decrease of energy scale, the quantum field
theory describing scattering processes changes, and we can associate this
process with a ``trajectory," or ``flow," through the space of physical
theories. Conformal field theories are the fixed points of such an RG
trajectory, and given two CFTs, it is natural to ask if there is an RG flow
connecting them.

In general, such a flow can be triggered in one of two ways: one can deform
the theory by adding an operator to it, or one can move out onto the moduli
space of the theory by turning on a vev for some operator. In the case of 6D
SCFTs, however, the first of these possibilities cannot arise for a flow
between a UV SCFT and an IR SCFT: as advertised in \cite{Cordova:2015fha} and demonstrated in
\cite{Louis:2015mka, Cordova:2016xhm, Cordova:2016emh}, 6D SCFTs
possess no marginal or relevant supersymmetry-preserving deformations, which
means that any UV SCFT can only be deformed at the expense of breaking
supersymmetry in the IR. This means that we can only trigger an RG flow
between 6D SCFTs by moving out onto the moduli space of the theory, and as we
saw in Section \ref{ssec:anomalies}, this moduli space divides into
branches: a tensor branch and a Higgs branch. Thus, there are two basic types of RG
flows to consider, which are called ``tensor branch flows'' and ``Higgs branch
flows,'' respectively. There are also mixed flows which involve both Higgs and tensor branch flows.

Tensor branch flows are straightforward to understand within the geometric F-theory picture:
one simply picks some subset of curves in the base $\mathcal{B}$ and,
instead of shrinking them to zero size (which would take us to the UV SCFT),
we instead expand them to infinite size and flow to the IR. This has the
effect of ungauging any gauge algebra associated with the expanded curve and
replacing it with a global symmetry. For instance, we may consider the
following tensor branch flow:
\begin{equation}
\lbrack E_6]\,\,1\,\,\overset{\mathfrak{su}(3)}{3}%
\,\,1\,\,\overset{\mathfrak{e}_{6}}{6}\,\,1\,\,\overset{\mathfrak{su}%
(3)}{3}\,\,1\,\,[E_6]~~~~\rightarrow~~~~[E_{6}]\,\,1\,\,\overset{\mathfrak{su}(3)}{3}\,\,1\,\,[E_{6}%
]~~\oplus~~[E_{6}]\,\,1\,\,\overset{\mathfrak{su}(3)}{3}%
\,\,1\,\,[E_{6}] \label{eq:extensorflow}%
\end{equation}
Here, we have expanded the $-6$ curve to infinite volume, thereby ungauging
the associated $\mathfrak{e}_{6}$. Observe that in the infrared, there is
actually an emergent $\mathfrak{e}_{6}$, namely the gauged $\mathfrak{e}_{6}$
of the UV theory sits in the diagonal of an $\mathfrak{e}_{6}\times
\mathfrak{e}_{6}$ algebra.

Both Lorentz invariance and the $\mathfrak{su}(2)_{R}$ R-symmetry are
preserved under a tensor branch flow. If we
consider weakly gauging these symmetries and triggering a tensor branch flow,
we deduce that their anomalies must match between the UV and IR, simply because the coefficients
are rational numbers and no continuous deformation is possible.\footnote{Though it is tempting to
use 't Hooft anomaly matching arguments \cite{tHooft:1979rat}, there
is a subtlety in applying this reasoning since in six dimensions, the anomaly polynomial could
a priori have many linearly independent terms, and weakly coupled spectator fermions may only
generate a subset of such terms. We thank Y. Tachikawa for alerting us to this subtlety.} It is not
hard to check using the prescription of \cite{Intriligator:2014eaa, Ohmori:2014kda} that the anomaly
polynomials of the two theories in (\ref{eq:extensorflow}) do not match.
However, this apparent discrepancy is resolved by the addition of free
multiplets as well as a Green-Schwarz term in the IR
\cite{Intriligator:2014eaa}. In the example of (\ref{eq:extensorflow}), we get
a single free tensor multiplet in the IR coming from expanding a single curve
to infinite size, and we get 78 vector multiplets from ungauging the
$\mathfrak{e}_{6}$ (which has dimension 78). There are no free hypermultiplets
in the IR because there were no hypermultiplets charged under this
$\mathfrak{e}_{6}$ in the UV. Along with the contribution of these free
hypermultiplets, one must also add to the IR anomaly polynomial a
Green-Schwarz term of the form
\begin{equation}
\Delta I_{\text{GS}} = (s c_{2}(R) - t p_{1}(T)+\sum_{i} u_{i} \, \mathrm{Tr}
F_{i}^{2})^{2}, \label{eq:GStensor}%
\end{equation}
for numerical coefficients $s$, $t$, and $u_{i}$, where the sum runs over $i$
runs over global and gauge symmetries of the theory. This contribution to the
anomaly polynomial is universal: whenever a single curve is taken to infinite
size, the difference between the UV and IR anomalies polynomials (after
accounting for free multiplets in the IR) is a perfect square.
Note also that $p_{2}(T)$ cannot show up in this Green-Schwarz term, so the
coefficient $\delta$ of $p_{2}(T)$ must agree in the UV and IR even before the
Green-Schwarz term is added.

One subtlety here is worthy of mention: the IR theory of a single free vector
multiplet in 6D is not actually a conformal field theory, but rather a
``scale-invariant" field theory, or SFT. This means that any IR theory that
results from expanding a curve carrying a non-Abelian gauge algebra to
infinite size will not technically be a CFT, but merely an SFT. Thus, for
instance, the IR theory of (\ref{eq:extensorflow}) is actually just an SFT,
because it has free vector multiplets in the IR. On the other hand, the IR
theory of the flow
\begin{equation}
[E_{6}] \,\, 1 \,\, \overset{\mathfrak{su}(3)}{3 }\,\, 1 \,\,
\overset{\mathfrak{e}_{6}}{6 }\,\, 1\,\, \overset{\mathfrak{su}(3)}{3 }\,\, 1
\,\, [E_{6}] ~~~~\rightarrow~~~~ [E_{6}] \,\, 1 \,\,
\overset{\mathfrak{su}(3)}{3 }\,\, 1 \,\, \overset{\mathfrak{e}_{6}}{6
}~~\oplus~~ \overset{\mathfrak{su}(3)}{3 }\,\, 1 \,\, [E_{6}]
\end{equation}
\emph{is} a CFT because expanding the $-1$ curve to infinite size does not
introduce any free vector multiplets in the IR.

Higgs branch flows, on the other hand, are more difficult to understand
geometrically. Whereas tensor branch flows are ``K\"ahler" deformations of the
base geometry, Higgs branch flows are ``complex structure" deformations
involving the fiber of the geometry. The simplest version of a Higgs branch
flow involves giving a vev to some charged hypermultiplet, which breaks a
gauge algebra down to a subalgebra:
\begin{equation}
\underset{[N_{s}=2]}{\overset{\mathfrak{so}(7)}{3}}~~~~\rightarrow
~~~~\underset{[N_{f}=1]}{\overset{\mathfrak{g}_{2}}{3}}%
\end{equation}
However, more complicated Higgs branch flows arise from giving a vev to ``conformal matter," as studied in \cite{Heckman:2015ola}. As a
particularly important example, we can consider the RG flow that dissolves a
small $E_{8}$ instanton into flux, leaving behind just a collection of 29 free
hypermultiplets:
\begin{equation}
1~~~~\rightarrow~~~~\text{29 Free Hypermultiplets} \label{eq:dissolve}%
\end{equation}

While 6D Lorentz invariance is still preserved under a Higgs branch flow (so that $\Delta \gamma = \Delta \delta =0$), the
$\mathfrak{su}(2)_{R}$ symmetry is broken and re-emerges in the IR. Anomaly matching
is still possible in this case, as the R-symmetry is spontaneously broken. For computations
of anomaly matching on the Higgs branch see e.g. reference \cite{Shimizu:2017kzs}.

Of course, one can also just consider the UV SCFT and the IR SCFT obtained from a
Higgs branch flow, and track the difference in the anomaly, particularly the terms
involving $c_2(R)$:
\begin{equation}
\Delta I_{8} = (\Delta\alpha) c_{2}(R)^{2} + (\Delta\beta) c_{2}(R) p_{1}(T)
+...
\end{equation}
where the ``$...$" includes terms involving broken/emergent global symmetries.
Higgs branch flows do not allow for a Green-Schwarz term in the IR, and they
only introduce free hypermultiplets--not free tensor multiplets or free vector
multiplets. As a corollary, the theory in the IR of a Higgs branch flow is
always a CFT, not merely an SFT, because it does not have any free vector
multiplets. In all known cases, $\alpha$ decreases along an RG flow while
$\beta$ increases, but this has not yet been proven.

\subsection{The $a$-theorem}\label{ssec:atheorem}

It is generally understood that the number of degrees of freedom of any system
should decrease under renormalization group flows. For CFTs, this intuition is
encoded in a weak ``C-function'':\footnote{As opposed to a strong C-function which is monotonic along the entire
RG flow.}
\begin{equation}
\Delta C := C_{UV} - C_{IR} \geq0.
\end{equation}
For even-dimensional CFTs on a curved background, Cardy proposed that the
quantity $a_{D}$ appearing in the trace anomaly (see e.g. \cite{Capper:1974ic}%
):%
\begin{equation}
\left\langle T_{\mu}^{\mu}\right\rangle =-\left(  -\frac{1}{4\pi}\right)
^{D/2}a_{D}E_{D}+... \label{eq:atheorem}%
\end{equation}
is one such C-function \cite{Cardy:1988cwa}. Here, $E_{D}$ is the
$D$-dimensional Euler density constructed with respect to the background
metric. In $D=2$, Cardy's conjecture is called the ``c-theorem," and it was
proven by Zamolodchikov in \cite{Zamolodchikov:1986gt}. In $D=4$, Cardy's
conjecture is called the ``a-theorem," and it was proven by Komargodski and
Schwimmer in \cite{Komargodski:2011vj}. In $D=3$, there is a related F-theorem
\cite{Myers:2010tj, Jafferis:2010un, Casini:2011kv}.

In $D=6$, however, Cardy's conjecture remains unproven. In
\cite{Elvang:2012st}, it was shown that the approach of Komargodski and
Schwimmer that worked in $D=4$ breaks down in $D=6$: there is still a
positivity bound on a coefficient of the effective action describing the
breaking of conformal symmetry along the flow, but unlike in 4D, this
coefficient is not simply the quantity $a_{D}$ appearing in (\ref{eq:atheorem}).

In $D=6$, all known interacting CFTs are supersymmetric. Thus, it is
worthwhile to see if one might be able to prove the $a$-conjecture in 6D under
the assumption of supersymmetry. A first step in this direction was
\cite{Cordova:2015vwa}, which proved the $a$-conjecture for all flows between
$(2,0)$ SCFTs. Shortly thereafter, \cite{Cordova:2015fha} extended this result
to all tensor branch flows between 6D SCFTs (including those with $(1,0)$
supersymmetry). That same work also showed that $a_{6}$ can be written in
terms of the coefficients $\alpha$, $\beta$, $\gamma$, and $\delta$ appearing
in the anomaly polynomial (\ref{eq:anomalypoly}):
\begin{equation}
a_{6} = \frac{8}{3} (\alpha-\beta+ \gamma) + \delta.
\end{equation}
So far, we have been talking specifically about the coefficient $a_{D}$ in
(\ref{eq:atheorem}), but we might wonder more generally about arbitrary weak
C-functions satisfying $C_{UV}\geq C_{IR}$. In 4D, it is known that $a_{4}$
represents the only linear combination of anomaly polynomial coefficients that
can serve as a C-function is precisely \cite{Anselmi:1997ys}. But in 6D, the
situation is not nearly as well-understood. We therefore want to know about a
general linear combination of the coefficients $(\alpha, \beta, \gamma,
\delta):= \vec{\alpha}$,
\begin{equation}
C_{\vec{m}}(\vec{\alpha}) := \vec{m} \cdot\vec{\alpha} := m_{1} \alpha+ m_{2}
\beta+ m_{3} \gamma+ m_{4} \delta.
\end{equation}
for a 4-vector $\vec{m}$. Note that $\vec{m}_{a}$ corresponds to the specific
case of $\vec{m}_{a} = (8/3,-8/3,8/3,1)$. From our discussion in the previous
subsection, we know that $\Delta\delta= 0$ for both tensor branch flows and
Higgs branch flows, so $m_{4}$ is a free parameter that can be ignored. The
overall scale is unimportant, so we can set $| m_{1}|=1$ (in practice $m_{1}
>0$ for all known flows so it suffices to set $m_1$ = 1),
and we are left with just a 2-dimensional parameter space to consider.

For tensor branch flows, we can use the fact that $\Delta I_{8}$ is always a
perfect square to prove a large class of C-theorems under such flows. Namely,
comparing with (\ref{eq:GStensor}), we have
\begin{equation}
\Delta C:=C_{UV}-C_{IR} \propto m_{1} s^{2} + 2 m_{2} s t + m_{3} t^{2} =
(m_{1} + 2 m_{2} \tilde t + m_{3} \tilde t^{2})s^{2},
\end{equation}
where we have defined $\tilde t = t/s$. This is a quadratic polynomial in
$\tilde t$, which will be positive provided $m_{1} > 0$ and the discriminant
is negative, $4 m_{2}^{2}- 4 m_{1} m_{3} <0$. Thus, $\Delta C$ is positive
provided
\begin{align}
m_{1} m_{3} > m_{2}^{2}\,,~~~~m_{1} >0.
\end{align}
Note that $\vec{m}_{a}$ lies right on the boundary of this region, so this
argument is not sufficient to exclude the possibility that $\Delta a = 0$ for some tensor branch flow.
However, this possibility can be excluded by a more sophisticated argument
\cite{Cordova:2015fha}.

Establishing C-theorems for Higgs branch flows is more difficult, and the
$a$-conjecture remains unproven in this case. Nonetheless, there is good
evidence in its favor. First of all, note that the absence of a Green-Schwarz
term together with 't Hooft anomaly matching of gravitational anomalies
implies that $\Delta\gamma$ and $\Delta\delta$ must vanish along any such
flow. Thus, only $m_{1}$ and $m_{2}$ show up in $\Delta C$:
\begin{equation}
\Delta C = m_{1} \Delta\alpha+ m_{2} \Delta\beta.
\end{equation}
Reference \cite{Heckman:2015axa} examined a huge set of Higgs branch flows in 6D SCFTs
and found that in all cases, $\Delta\alpha>0$, $\Delta\beta< 0$. If this holds
true for all Higgs branch flows, it would immediately imply the $a$-theorem,
since $\vec{m}_{a}$ has $m_{1} > 0$, $m_{2} < 0$. Reference \cite{Heckman:2015axa} also
found that the tightest constraint on $m_{2}/m_{1}$ came from the simplest
Higgs branch flow imaginable: that of (\ref{eq:dissolve}). All other Higgs
branch flows studied gave strictly weaker bounds, which gives us strong reason
to think that considering more complicated flows will not yield a
counterexample to the $a$-conjecture (a proof by ``brute force'').
Nevertheless, an analytic proof remains elusive.

Finally, it can be proven that the quantity $a$ is positive for all 6D SCFTs,
as explained in \cite{CordovaMixer} and advertised in \cite{Cordova:2018cvg}.
A free vector multiplet, on the other hand, has $a < 0$. This is one
consequence of the fact that the theory of a free vector multiplet is
scale-invariant but not conformally invariant \cite{Cordova:2015fha}.

\subsection{RG Flows and Homomorphisms}

Finally, we return to the class of theories associated with homomorphisms
$\mathfrak{su}(2) \rightarrow\mathfrak{g}$. For a given $\mathfrak{g}$,
starting from the theory corresponding to the trivial homomorphism, we can
reach all other theories associated with nontrivial homomorphisms by complex
structure deformations of the geometry, which are Higgs branch flows in field
theory language. We will see in this subsection that the correspondence
between theories and homomorphisms can be pushed further: the web of Higgs
branch flows between these theories is related to the ``Hasse diagram" that
connects these homomorphisms.

We saw that for the particular case of $\mathfrak{g} =\mathfrak{su}(k)$,
homomorphisms $\mathfrak{su}(2) \rightarrow\mathfrak{su}(k)$ are in 1-1
correspondence with partitions $\mu$ of $k$. Now, we can define a partial
order on partitions: given two partitions $\mu$ and $\nu$ of $k$, we declare
$\mu\succeq\nu$ if $\sum_{i = 1}^{k} \mu_{i}^{T} \geq\sum_{i = 1}^{k} \nu_{i}^{T}$ for all
$k$, where $\mu^{T}$ is the transpose of $\mu$. For instance, for $k =4$, we
have
\begin{equation}
[1^{4}]={\young(~~~~)} \succeq{\young(~,~~~)} \succeq{\young(~~,~~)}
\succeq{\young(~,~,~~)} \succeq{\young(~,~,~,~)} = [4],
\end{equation}
and the partial order is actually a total order in this case. This order on
partitions has an interpretation in terms of the associated homomorphisms.
Given some homomorphism $\rho: \mathfrak{su}(2) \rightarrow\mathfrak{su}(n)$,
we let $\mathcal{O}$ be the orbit of the nilpotent element of $\mathfrak{su}%
(n)$ specified by the raising operator $J_{+}$ of $\mathfrak{su}(2)$. We then
declare $\rho_{\mu}\succeq\rho_{\nu}$ if the closure of $\mathcal{O}_{\mu}$
contains $\mathcal{O}_{\nu}$, $\overline{{\mathcal{O}}_{\mu}}\supset\mathcal{O}_{\nu}$. This
order on nilpotent orbits agrees with the order on partitions.

This order can also be translated to the language of 6D SCFTs:
$\mu\succeq\nu$ if and only if there is an RG flow from the 6D SCFT associated
with $\mu$ to the 6D SCFT associated with $\nu$. For instance for
$\mu={\young(~,~~~)} $ and $\nu= {\young(~~,~~)}$, we $\mu\succeq\nu$, and
correspondingly there is an RG flow
\begin{equation}
[SU(2)]\,\, {\overset{\mathfrak{su}(3)}{2 }} \,\, \underset{[N_{f}%
=1]}{\overset{\mathfrak{su}(4)}{2 }} \,\, {\overset{\mathfrak{su}(4)}{2}}
\,\,... [SU(4)] ~~~\rightarrow~~~ {\overset{\mathfrak{su}(2)}{2 }}
\,\, \underset{[SU(2)]}{\overset{\mathfrak{su}(4)}{2 }} \,\,
{\overset{\mathfrak{su}(4)}{2}} \,\,... [SU(4)]
\end{equation}

This relationship between the partial order on homomorphisms $\mathfrak{su}(2)
\rightarrow\mathfrak{g}$ and the partial order on theories defined by RG flows
extends to the $\mathfrak{g} = D_{n}$ and $\mathfrak{g} = E_{n}$ cases as
well. On the homomorphism side, the full web of orderings is well known, and
it is encoded in a diagram called the ``Hasse diagram" (see for instance
\cite{Chacaltana:2012zy}). This Hasse diagram therefore encodes the web of RG
flows from the perspective of 6D SCFTs. It is also worth mentioning that for
sufficiently long 6D SCFT quivers, the $a$-conjecture has been proven for
these Higgs branch flows between theories corresponding to homomorphisms
$\mathfrak{su}(2) \rightarrow\mathfrak{g}$.


\section{State Counting \label{sec:INDICES}}

The analysis of RG\ flows presented in the previous section provides a clear
sense in which a 6D\ SCFT can lose degrees of freedom upon flowing from the
UV\ to the IR. A\ natural question in this context is whether it is possible
to count the states in a 6D\ SCFT. One of the aims of calculations involving
the superconformal index of an SCFT is to quantify the operator content of the theory
as a function of various representation theoretic data of the SCFT, such as
the weight of the representation with respect to the superconformal algebra.
At a very broad level, the idea is to place a $D$-dimensional SCFT on the
Euclidean signature background $S^{1}\times M_{D-1}$, with $M_{D-1}$ a
($D-1$)-dimensional manifold. The conditions required for $M_{D-1}$ are that
we can activate fields of the D-dimensional supergravity multiplet to retain
supersymmetry, preserving a Killing spinor for this background. Assuming this
can be done, we can define a supersymmetric index:
\begin{equation}
\mathcal{I}_{S^{1}\times M_{D-1}}=\text{Tr}_{\mathcal{H}}\left(  -1\right)
^{F}\exp\left(  -\beta H+\underset{i}{\sum}\mu_{i}q_{i} \right)  ,
\end{equation}
which is a generalization of the Witten index \cite{Witten:1982df} to more
general curved backgrounds. Here, the $\mu_{i}$ should be viewed as chemical potentials
for the various background charges of the system, such as R-charges and flavor
symmetry charges. Closely related to this is the partition function of the system
$\mathcal{Z}_{S^{1}\times M_{D-1}}$: we place our supersymmetric theory
on the background $S^{1}\times M_{D-1}$ and evaluate the corresponding path integral
in the presence of these fugacities. In what follows we shall gloss over such distinctions.
For further discussion, see e.g. reference \cite{Assel:2015nca}.

In favorable circumstances, it is actually possible to
evaluate the index, though the most well-established computations require a
Lagrangian formulation for the theory and involve localization on a
configuration of fields. For a discussion of these general features, see the
reviews \cite{Pestun:2016zxk} and \cite{Kim:2016usy}.

Although the absence of a Lagrangian description prevents explicit computation of the index in most
cases, it is sometimes possible to calculate the result for suitably chosen
backgrounds, as well as by exploiting the structure of the theory in the
limit where we reduce upon a circle. This can be done by a gauge theoretic
analysis or via methods from topological string theory. Of course, part of the
reason why it is difficult to formulate a Lagrangian formulation of these
theories involves the presence of anti-chiral two-forms (and consequently
anti-chiral strings) on the tensor branch of theory. In this context, two
common choices are the calculation of the partition function on the background
$S^{1}\times S^{5}$, where the metric on $S^{5}$ can either be taken to be
round, or with squashing parameters added.

The other common choice, especially in relation to calculations performed via
topological string theory involves the $\Omega$-background
\cite{Nekrasov:2002qd} $T^{2}\times\mathbb{R}_{\Omega}^{4}$, which has the
metric:%
\begin{align}
ds^{2}  & =dzd\overline{z}+\eta_{\mu\nu}(dx^{\mu}+\Omega^{\mu}dz+\overline
{\Omega}^{\mu}d\overline{z})(dx^{\nu}+\Omega^{\nu}dz+\overline{\Omega}^{\nu
}d\overline{z}),\\
d\Omega & =\varepsilon_{1}dx^{1}\wedge dx^{2}+\varepsilon_{2}dx^{3}\wedge
dx^{4}\text{, \ \ }d\overline{\Omega}=\overline{\varepsilon}_{1}dx^{1}\wedge
dx^{2}+\overline{\varepsilon}_{2}dx^{3}\wedge dx^{4},
\end{align}
with $z$ a local holomorphic coordinate on the $T^{2}$, $x^{\mu}$ real
coordinates on $\mathbb{R}^{4}$ and $\eta_{\mu\nu}$ the flat metric on
$\mathbb{R}^{4}$. The metric depends on two complex parameters $\varepsilon
_{1}$ and $\varepsilon_{2}$. Though it is a non-compact space, it does share some similarities
with Kaluza-Klein reduction because the fugacities in the index suppress contributions
from states with higher quantum numbers.

A second type of index has played an important role in the study of 6D SCFTs:
the index of the theory of the effective strings, which is itself a 2D SCFT.
Aspects of these SCFTs have been analyzed in references \cite{Haghighat:2013gba, Haghighat:2013tka,
Haghighat:2014pva, Haghighat:2014vxa, Gadde:2015tra, Apruzzi:2016iac,
Kim:2016foj, Shimizu:2016lbw, DelZotto:2016pvm, Apruzzi:2016nfr, DelZotto:2018tcj}. In this
case, there are strings which enjoy $\mathcal{N}=(0,4)$ supersymmetry. To get a
non-zero answer for the elliptic genus, it is necessary to insert factors of
the right-mover fermion number. Up to subtleties having to do with having a
non-compact target space for the 2D theory, the index is given by:%
\begin{equation}
Z_{2D}=\text{Tr}\left(  (-1)^{F_{R}}F_{R}^{2}q^{L_{0}}\overline{q}%
^{\widetilde{L}_{0}}\right)  \text{ \ \ with \ \ }q=\exp(2\pi i\tau
)\text{.}\label{elliptico}%
\end{equation}
We remark that for theories with $\mathcal{N}=(0,8)$ supersymmetry, it is
necessary to add a factor of $F_{R}^{4}$ rather than $F_{R}^{2}$ to obtain a
non-zero result.

Our plan in this section will be to review some of the work done on analyzing
the counting of states both for effective strings and 6D\ SCFTs. We begin by
discussing some general properties of effective strings in 6D\ SCFTs and
follow this with a discussion of supersymmetric indices for these theories. We
then turn our attention to indices of 6D\ SCFTs.

\subsection{Effective Strings}

Let us begin with some general qualitative features of effective strings in
6D\ SCFTs. As we have already emphasized, even though such strings are
inevitably present in any 6D\ SCFT with a tensor branch, locality of the
quantum field theory suggests that these degees of freedom are a collective
phenomenon similar to instanton configurations in gauge theory. On the tensor
branch, there are various terms which couple the anti-chiral two-forms to
background curvatures of the field theory. These include metric curvature
terms such as the first Pontryagin class of the tangent bundle $p_1(T)$, the second
Chern class of the R-symmetry bundle $c_2(R)$, and the second Chern class of gauge and
flavor two-form field strengths, $c_2(F)$. In all cases, these background field
configurations induce a chemical potential for effective strings. Such strings
break half of the supersymmetries of the 6D\ SCFT, and as they are chiral,
this leaves us with $\mathcal{N}=(0,4)$ supersymmetry along the worldvolume of
the string. The transverse geometry to the string (and inside the
6D\ SCFT)\ is $\mathbb{R}^{4}$, so there is an $\mathfrak{so}(4)$ global
symmetry which is to be identified with the R-symmetry of the 2D effective
theory. From the perspective of this effective string, the gauge theory
degrees of freedom of the 6D\ SCFT are also global symmetries, so
symmetries preserved by the background field configuration of the 6D\ SCFT
descend to global symmetries of this 2D\ effective theory.

One of the earliest index computations carried out for these theories involves
the calculation of the generalized elliptic genus for the E-string theory. Consider a D3-brane wrapped over the base $\mathbb{CP}^{1}$ of an
F-theory model, and study the 2D worldvolume theory in our 6D spacetime. The
elliptic genus is then given by equation (\ref{elliptico}). In this case, we
can also calculate the elliptic genus by relating it (via F- / M-theory
duality)\ to the computation of an M5-brane wrapped over a $\frac{1}{2}$K3,
i.e. $dP_{9}$ divisor in the local Calabi-Yau geometry defined by the F-theory
model. The partition function is obtained via the partition
function of $\mathcal{N}=4$ super Yang-Mills theory wrapped over this divisor
\cite{Minahan:1998vr}.

More general calculations are now available due to advances in the
understanding of anomaly polynomials for both 6D\ SCFTs as well as the
associated anomaly polynomials for 2D effective strings in such theories.
Using the fact that this string couples to the anti-chiral two-form of the
6D\ SCFT, one can use an anomaly inflow argument to deduce the anomaly
polynomial for this 2D theory \cite{Kim:2016foj, Shimizu:2016lbw}.
The end result is:
\begin{equation}
\mathcal{I}_{4}=\frac{A_{ij}Q^{i}Q^{j}}{2}\left(  c_{2}(L_{\bot}%
)-c_{2}(R_{\bot})\right)  + A_{ij}Q^{i}I^{j},
\end{equation}
where $L_{\bot}\oplus R_{\bot}$ denotes the bundle associated with the $\mathbb{R}^{4}$
directions inside the 6D\ SCFT, and $I^{j}$ denotes the terms appearing in the
Green-Schwarz coupling of the B-field to the curvatures:%
\begin{equation}
A_{ij}I^{j}=\frac{1}{4}\left(  \mu_{i}^{a}\text{Tr}F_{a}^{2}%
-(2- A_{ii})p_{1}(T)\right)  +\mu_{i}^{g}h_{g}^{\vee}c_{2}\left(
\mathcal{R}_{6D}\right)  ,
\end{equation}
where here, $\mu_{i}^{a}$ denote matrices of couplings which pair the B-fields
with the gauge and flavor symmetries, with $a$ running over both sorts of
indices, and $g$ an index running over just the gauge group factors, with
$h_{g}^{\vee}$ the dual Coxeter number for the corresponding gauge algebra.
$\mathcal{R}_{6D}$ denotes the R-symmetry bundle associated with
the 6D\ SCFT, and the quantity $A_{ii}$ has repeated indices, but
they are not to be summed over. This same expression was also obtained for the
case of the single curve NHCs with simple gauge group factor and no matter
fields by direct dimensional reduction of the worldvolume theory of a D3-brane
wrapped over the collapsing $\mathbb{CP}^{1}$ of the 6D\ SCFT
\cite{Apruzzi:2016nfr}.

The worldvolume theory on these effective strings is typically a
non-Lagrangian theory in two dimensions. To see why, consider for example the
case of the single $-12$ curve with gauge algebra $\mathfrak{e}_{8}$. Wrapping
a D3-brane over this curve, the worldvolume theory of the effective string is
described by the 4D theory on the D3-brane near an $E_{8}$ seven-brane, in which
the 4D theory is then placed on the background $\mathbb{R}^{1,1}%
\times\mathbb{CP}^{1}$. In flat space, the 4D worldvolume theory of the
D3-brane is an $\mathcal{N}=2$ superconformal field theory with $E_{8}$ flavor
symmetry known as the $E_{8}$ Minahan-Nemeschansky theory
\cite{Minahan:1996fg, Minahan:1996cj}. This is a theory with no known
Lagrangian description, though many properties of this theory such as its
Seiberg-Witten curve, its Gaiotto curve \cite{Gaiotto:2009we}, and its
anomalies \cite{Aharony:2007dj, Shapere:2008zf} are known. Because of this,
one should not expect a simple weakly coupled description of the resulting 2D
theory, though we can still compute quantities such as the 2D anomaly
polynomial for this theory. Similar considerations hold for all of the single
curve NHC theories.

There are, however, a few examples where more can be said; these
are the situations in which we can return to a perturbative type IIB\ description
of the 6D\ SCFT. In these cases, the microscopic formulation of the D3-brane
worldvolume theory is captured by a quiver gauge theory. This can be carried
out in the case of the pure $-4$ curve theory \cite{Apruzzi:2016nfr}. It can
also be carried out for 6D\ SCFTs composed of $-2$ curves with $I_{n}$-type
fiber decorations over each curve, and for a configuration such as
$1,4,...,4,1$ with alternating $\mathfrak{sp}/\mathfrak{so}$ gauge algebra factors
\cite{Gadde:2015tra}.\ In these cases, the theory of the effective string is
captured by an $\mathcal{N}=(0,4)$ gauged linear sigma model.

\subsection{Indices for Effective Strings}

In favorable circumstances it also possible to extract the elliptic genus for
the effective strings of 6D SCFTs. For the most part, these computations rely
on the method of localization \cite{Witten:1988ze} for the index, i.e. we use
the supersymmetries of the system to construct a cohomology theory such that
the entire action can be written as an exact term plus a tractable classical
contribution. In such situations, the subtleties of path integrals over
functionals are instead replaced by finite-dimensional integrals. These
integrals can often be evaluated through some combination of methods from
matrix models and / or residue integrals. We cannot hope to review all these
methods here, so we refer the interested reader to reference
\cite{Pestun:2016zxk} as well as the literature on two-dimensional
computations, e.g. \cite{Benini:2016qnm}. Our aim here will be to simply
review the results of some of these calculations and to explain in what
situations one can hope to carry out such computations at all.

So far, these computations have primarily been carried out for theories where
the effective strings have a Lagrangian description. This includes, for
example, the effective strings of theories of M5-branes probing a $\mathbb{C}^2/\Gamma$ orbifold, with
$\Gamma$ an A- or D-type discrete subgroup of $SU(2)$. For example, in the
case of $N+1$ M5-branes probing a $\mathbb{C}^{2}/%
\mathbb{Z}
_{k}$ singularity, the effective strings enjoy an $SU(k)^{N+2}$ flavor
symmetry. Two of these factors are inherited from the 6D SCFT. The remaining
$SU(k)$ factors come from the gauge groups of the 6D\ SCFT, which are flavor
symmetries from the perspective of the 2D effective strings. In terms of
occupation numbers $n_{1},...,n_{N}$ for the gauge nodes of the associated
tensor branch theory, the index can be evaluated using a residue integral
(referred to in the literature as a Jeffrey-Kirwan, or JK, residue). The final
expression for even this simple case takes a remarkably intricate form
\cite{Gadde:2015tra}:
\begin{align}
&  \mathcal{Z}_{n_{1},\dots,n_{k}}^{A_{N},k}=\sum_{\mathcal{Y}}\left(
\prod_{i=1}^{N}\prod_{\ell,m=1}^{k}\prod_{\substack{(x_{1},y_{1})\in Y_{\ell
}^{(i)}\\(x_{2},y_{2})\in Y_{m}^{(i)}}}\frac{\theta_{1}(t^{{x_{1}}-{x_{2}}%
}d^{{y_{1}}-{y_{2}}})\theta_{1}(\frac{s_{\ell}^{(i)}}{s_{m}^{(i)}}t^{{x_{1}%
}-{x_{2}}+1}d^{{y_{1}}-{y_{2}}+1})}{\theta_{1}(\frac{s_{\ell}^{(i)}}%
{s_{m}^{(i)}}t^{{x_{1}}-{x_{2}}+1}d^{{y_{1}}-{y_{2}}})\theta_{1}(\frac
{s_{\ell}^{(i)}}{s_{m}^{(i)}}t^{{x_{1}}-{x_{2}}}d^{{y_{1}}-{y_{2}}+1}%
)}\right)  \nonumber\\
&  \times\left(  \prod_{i=1}^{N-1}\prod_{\ell,m=1}^{k}\prod_{\substack{(x_{1}%
,y_{1})\in Y_{\ell}^{(i)}\\(x_{2},y_{2})\in Y_{m}^{(i+1)}}}\frac{\theta
_{1}(c^{-1}\frac{s_{\ell}^{(i)}}{s_{m}^{(i+1)}}t^{{x_{1}}-{x_{2}}+\frac{1}{2}%
}d^{{y_{1}}-{y_{2}}-\frac{1}{2}})}{\theta_{1}(c^{-1}\frac{s_{\ell}^{(i)}%
}{s_{m}^{(i+1)}}t^{{x_{1}}-{x_{2}}-\frac{1}{2}}d^{{y_{1}}-{y_{2}}-\frac{1}{2}%
})}\right)  \nonumber\\
&  \times\left(  \prod_{i=2}^{N}\prod_{\ell,m=1}^{k}\prod_{\substack{(x_{1}%
,y_{1})\in Y_{\ell}^{(i)}\\(x_{2},y_{2})\in Y_{m}^{(i-1)}}}\frac{\theta
_{1}(c\frac{s_{\ell}^{(i)}}{s_{m}^{(i-1)}}t^{{x_{1}}-{x_{2}}+\frac{1}{2}%
}d^{{y_{1}}-{y_{2}}-\frac{1}{2}})}{\theta_{1}(c\frac{s_{\ell}^{(i)}}%
{s_{m}^{(i-1)}}t^{{x_{1}}-{x_{2}}-\frac{1}{2}}d^{{y_{1}}-{y_{2}}-\frac{1}{2}%
})}\right)  \times\nonumber
\end{align}%
\[
\times\left(  \prod_{i=1}^{N}\prod_{\ell,m=1}^{k}\prod_{(x,y)\in Y_{\ell
}^{(i)}}\frac{\theta_{1}(c\frac{s_{\ell}^{(i)}}{s_{m}^{(i-1)}}t^{x+\frac{1}%
{2}}d^{y+\frac{1}{2}})\theta_{1}(c\frac{s_{m}^{(i+1)}}{s_{\ell}^{(i)}%
}t^{-x-\frac{1}{2}}d^{-y-\frac{1}{2}})}{\theta_{1}(\frac{s_{\ell}^{(i)}}%
{s_{m}^{(i)}}t^{x+1}d^{y+1})\theta_{1}(\frac{s_{m}^{(i)}}{s_{\ell}^{(i)}%
}t^{-x}d^{-y})}\right)  .
\]
Here, the sum on $\mathcal{Y=}\left\{  \left\{  Y_{\ell_{i}}^{(i)}\right\}
_{\ell_{i}=1}^{k}\right\}  _{i=1}^{N}$ is over colored Young diagrams where
the $i^{th}$ row has $\ell_{i}$ boxes. The notation $(x,y)\in Y_{\ell}^{(i)}$
indicates a \textquotedblleft coordinate\textquotedblright\ in the diagram, so
the upper lefthand box is at $(0,0)$, moving to the right increases the $x$
coordinate, and moving down increases the $y$ coordinate. The
parameters $s_{m}^{(i)}$ are fugacities associated with chemical potentials
for the Cartan subalgebra of each $SU(k)$ factor. Additionally, the partition
function makes reference to the parameters:%
\begin{equation}
q=e^{2\pi i\tau}\text{, \ \ }t=e^{2\pi i\epsilon_{1}}\text{, \ \ }d=e^{2\pi
i\epsilon_{2}}\text{, \ \ }c=e^{2\pi im},
\end{equation}
where the parameter $\tau$ is the standard parameter appearing in partition
function calculations. The parameters $\epsilon_{1}$,
$\epsilon_{2}$ and $m$ refer to fugacities associated with the Cartan
subalgebra of $SO(4)\times SU(2)_{\mathcal{R}}$. The theta
function $\theta_{1}\left(  z\right)  $ that appears is given by%
\begin{equation}
\theta_{1}\left(  z\right)  =iq^{1/8}(z^{-1/2}-z^{1/2}%
)\underset{i=1}{\overset{\infty}{%
{\displaystyle\prod}
}}(1-q^{i})(1-q^{i}z)(1-q^{i}z^{-1})\text{.}%
\end{equation}

Similar expressions hold for the other effective string theories which possess
a Lagrangian description. An interesting feature of the computations in
reference \cite{Gadde:2015tra} is that it is also possible to (conjecturally)
obtain the elliptic genus for the E-string theories as well. The point is that
although the 6D\ SCFT makes reference to an $E_{8}$ flavor symmetry which is
difficult to engineer in perturbative string theory, it is nevertheless
possible to engineer a suspended brane configuration in type IIA\ string
theory that enjoys an $SO(16)$ flavor symmetry. In the limit where all
suspended branes are coincident, this flows to an SCFT, and the $SO(16)$
flavor symmetry enhances to $E_{8}$. Since the perturbative type IIA
description involves perturbative objects, it is perhaps not surprising that
in this limit, a candidate Lagrangian field theory can be written down for the
effective strings of this 6D\ SCFT and should therefore be directly related to
the elliptic genus of the strings in the E-string theory.

An additional class of theories for which it is now possible to compute the
elliptic genus consists of the single instanton sectors of the rank one NHCs with no
matter, namely the case of a curve with self-intersection $-3$, $-4$, $-5$,
$-6$, $-8$ and $-12$ \cite{DelZotto:2016pvm}, as well as a large class of rank one SCFTs with gauge group of rank $\leq 7$ \cite{DelZotto:2018tcj}. This computation is carried out
by calculating the partition function for a 4D\ $\mathcal{N} = 2$ SCFT
compactified on the background $T^{2}\times S^{2}$.

\subsection{Indices for 6D\ SCFTs}

Let us now turn to a discussion of indices for 6D\ SCFTs. Here, the situation
is intrinsically more challenging because all known interacting fixed points
are non-Lagrangian. Consequently, the application of methods such as
localization will necessarily also be more difficult. Nevertheless, it is
possible to make some progress in the calculation of the partition function by
piecing together consistency conditions present in lower dimensions.

For the most part, attention has focussed on indices associated with either
the background $S^{1}\times S^{5}$ with squashing parameters added, or on the
background $T^{2}\times\mathbb{R}_{\Omega}^{4}$, so we organize our discussion
according to these two special cases.

\subsubsection{The Background $S^{1}\times S^{5}$}

Before turning to specific theories, let us discuss what can be extracted from
general principles alone. One general feature found in \cite{DiPietro:2014bca}
relies on some (mild) assumptions pertaining to supersymmetric backgrounds in
five and six dimensions. In the high temperature limit (a small
thermal circle), general considerations lead to a Cardy-like formula for the
structure of the partition function \cite{DiPietro:2014bca}:
\begin{equation}
\beta\rightarrow0:\log\mathcal{Z}_{S^{1}\times S^{5}}\sim-\frac{\pi^{2}%
}{\omega_{1}\omega_{2}\omega_{3}}\left(  \frac{\kappa_{1}\pi^{2}}{45}%
\frac{R^{2}}{\beta^{3}}+\frac{\left(  \omega_{1}^{2}+\omega_{2}^{2}+\omega
_{3}^{2}\right)  \kappa_{2}+3\left(  \omega_{1}\omega_{2}+\omega_{1}\omega
_{3}+\omega_{2}\omega_{3}\right)  \kappa_{3}}{36}\frac{R}{\beta}\right)
,\label{ZOHARK}%
\end{equation}
where $R$ is the radius of the $S^{5}$ and the $\omega_{i}$ are squashing
parameters related to deformations of the round sphere metric. The
coefficients $\kappa_{i}$ are specified by the anomaly polynomial
coefficients:%
\begin{equation}
\kappa_{1}=-\frac{\delta+5\gamma}{6}\text{, \ \ }\kappa_{2}=\frac
{\delta+2\gamma}{6}+\frac{3}{2}\beta\text{, \ \ }\kappa_{3}=\beta\text{,}%
\end{equation}
where the normalization has been chosen such that $\kappa_{1}=\kappa
_{2}=\kappa_{3}=1$ for a free vector multiplet. One important difference from
the case of the 2D Cardy formula is that in the large $N$ limit, the conformal
anomaly grows as order $N^{3}$ (which happens for theories with a holographic dual \cite{Gubser:1997yh}),
whereas this supersymmetric quantity only grows as order $N$. In this
sense, it is a weaker probe of the underlying microscopic degrees of freedom
of the 6D\ theory.

Turning next to specific theories, we can ask what can be extracted for
6D\ SCFTs. For the most part, the operating principle is to relate the
counting of states in this 6D theory to a state-counting problem in five
dimensions by dimensional reduction on an $S^{1}$. To get back the full
state-counting problem in six dimensions, we can try to retain all of the
Kaluza-Klein modes (which includes both particles and strings) of the parent
6D theory, but packaged as 5D objects. This provides a systematic way to
extract the partition function for these theories provided we have access to
the 5D description.

This sort of strategy has been carried out in the context of 5D gauge theory,
as well as using methods from closed topological strings on the Calabi-Yau
threefold backgrounds provided by the F-theory construction. The choice of
circle on which to dimensionally reduce is actually different in the field
theory computation and the topological string computation, providing strong
evidence in support of the proposed correspondence.

In gauge theory terms, this calculation exploits the fact that the geometry of
$S^{5}$ is actually an $S^{1}$ bundle over $\mathbb{CP}^{2}$. Consequently, it
is natural to expect that the 6D\ index is related to a 5D gauge index
computation on $S^{1}\times\mathbb{CP}^{2}$. The key step here is that
$\mathbb{CP}^{2}$ is also a toric manifold and admits a $\mathbb{C}^{\ast
}\times\mathbb{C}^{\ast}$ action, with three fixed points under the torus
action. In this way, localized field configurations of the index computation
also become localized near these fixed points. Gluing these contributions
together then provides a conjectural relation to the full partition function.
Note that to apply this method, we must know of a 5D gauge theory description for the
theory obtained from dimensional reduction on a circle.

In terms of closed topological strings, one instead exploits the fact that
F-theory on the background $S^{1}\times CY_{3}$ is related, in the limit of a
small $S^{1}$, to M-theory on the same background. It is well known from
\cite{Gopakumar:1998ii, Gopakumar:1998jq} that upon compactification on a
further circle, perturbative closed topological string amplitudes on this
Calabi-Yau threefold can be interpreted as counting BPS particles coming from
M2-branes wrapped along holomorphic curves of the Calabi-Yau threefold, with
momenta along this additional circle. From this perspective, the counting of
states via topological string amplitudes should clearly be related to the
counting of states in the 6D superconformal index. Indeed, in
\cite{Lockhart:2012vp}, it was proposed that the 6D\ SCFT index actually
provides a non-perturbative completion of the topological string amplitude in
appropriate circumstances!

This sort of calculation has been carried out explicitly for several canonical
examples of 6D\ SCFTs, namely the $(2,0)$ theories, the E-string theory, the
rank one non-Higgsable cluster theories, and the worldvolume theory of
M5-branes on a type $A_{n}$ or $D_{n}$ orbifold singularity. In the case of the $(2,0)$ theories, dimensional reduction on a circle leads
to a 5D $\mathcal{N}=2$ super Yang-Mills theory of $ADE$ type in which the
gauge coupling scales as:%
\begin{equation}
\frac{1}{g_{5D}^{2}}\sim\frac{1}{\text{Vol}(S^{1})},
\end{equation}
which is not the relation one would have gotten from dimensional reduction
of a 6D gauge theory. A noteworthy feature of this theory is that it is not
superconformal: the only SCFTs in five dimensions have $\mathcal{N}=1$
supersymmetry. Nonetheless, one may still use this gauge theory to compute the
superconformal index of the associated 6D theory.

The proposed answer for the superconformal index in the case of the theory of
$N$ M5-branes on the background $S^{1}\times S^{5}$ is a function of various
geometric cross ratios:%
\begin{equation}
\mathcal{Z}_{S^{1}\times S^{5}}\left(  -\frac{1}{\omega_{1}},\frac{\omega_{2}%
}{\omega_{1}},\frac{\omega_{3}}{\omega_{1}},\frac{m}{\omega_{1}}\right)  =\int
dt_{i}\text{ }Z^{\text{np,top}},
\end{equation}
where the $t_{i}$ are the scalar vevs parameterizing the tensor branch, and
$Z^{\text{np,top}}$ is a \textquotedblleft non-perturbative\textquotedblright%
\ completion of the topological string amplitude on the Calabi-Yau threefold.
It is so named because of the way it depends on the moduli. In terms of the
perturbative topological string amplitudes, it is given by:%
\begin{equation}
Z^{\text{np,top}}=\frac{Z^{\text{top}}(t_{i},\omega_{1},\omega_{2},\omega
_{3}.m)}{Z^{\prime\text{top}}(\frac{t_{i}}{\omega_{1}},-\frac{1}{\omega_{1}%
},\frac{\omega_{2}}{\omega_{1}},\frac{\omega_{3}}{\omega_{1}},\frac{m}%
{\omega_{1}})\times Z^{\prime\text{top}}(\frac{t_{i}}{\omega_{2}},\frac
{\omega_{1}}{\omega_{2}},-\frac{1}{\omega_{2}},\frac{\omega_{3}}{\omega_{2}%
},\frac{m}{\omega_{2}})},
\end{equation}
where here, $\omega_{i}$ are the squashing parameters of the $S^{5}$, and $m$
is the Cartan of the $SU(2)$ R-symmetry, which upon dimensionally reducing to
five dimensions is interpreted as the mass parameter of an adjoint valued
scalar. The difference between $Z^{\text{top}}$ and its primed version
$Z^{\prime\text{top}}$ is that in the former, we include the \textquotedblleft
center of mass\textquotedblright\ dependence in the theory, whereas in the
latter, we do not. The explicit form of these partition functions are somewhat
involved expressions, so we refer the interested reader to the original
literature for further details. Each factor of $Z^{\text{top}}$ can be
evaluated either using the topological vertex formalism \cite{Aganagic:2003db}%
, or via 5D\ instanton calculations \cite{Nekrasov:2002qd, Nekrasov:2003rj}.

In any case, an important feature of these computations is that the resulting
scaling dependence of the partition function in the low temperature limit
(large $\beta$, in contrast to the small $\beta$ expansion of line
(\ref{ZOHARK})) exhibits an $N^{3}$ scaling dependence, in line with the early
expectation from greybody scattering off of M5-branes carried out in
\cite{Gubser:1997yh}.

\subsubsection{The Background $T^{2}\times\mathbb{R}_{\Omega}^{4}$}

It is also of interest to compute the partition function of the 6D\ SCFT on
the $\Omega$ background $T^{2}\times\mathbb{R}_{\Omega}^{4}$. Here, there are
a number of interlocking conjectures which relate areas of Calabi-Yau geometry
with other quantities of interest. The main point is that BPS\ state counting
in M-theory on Calabi-Yau threefolds is achieved via topological string
amplitudes on such backgrounds \cite{Gopakumar:1998ii, Gopakumar:1998jq}.
Moreover, the BPS\ states counted in this way involve M2-branes wrapped on
two-cycles. In the duality between M-theory on $X$ and F-theory on
$S^{1}\times X$, we can relate the BPS\ particles obtained from M2-branes
wrapped on two-cycles to D3-branes wrapped on a two-cycle of the F-theory base
and the circle factor.

First of all, we can speak of topological string amplitudes $Z^{\text{top}%
}(\varepsilon_{1},\varepsilon_{2},\tau,t_{b},m_{i})$ for the elliptically
fibered Calabi-Yau threefold. Here, the parameters entering this amplitude are
defined as follows: the $\varepsilon_{i}$ are the equivariant parameters of the refined topological string.\footnote{Strictly speaking, this is
only well-defined purely in terms of the Calabi-Yau geometry in the special
case of the unrefined amplitudes $\varepsilon_{1}=-\varepsilon_{2}%
=\varepsilon$. However, once we make reference to the physics of the 6D
spacetime, it is sensible to define the topological string amplitude using the
quantum numbers of the $SU(2)_{L}\times SU(2)_{R}$ spin quantum numbers on
$\mathbb{R}^{4}$.} $\tau$ is the complex structure of the
$T^{2}$ factor appearing in the 6D spacetime. This is joined by the K\"ahler
parameters $t_{b}$ of the base and those of the elliptic fiber, which were
denoted in reference \cite{Haghighat:2014vxa} as $m_{i}$. Turning next to the
structure of 6D SCFTs, we can consider the properties of BPS\ strings of
6D\ SCFTs, as obtained by working on the tensor branch of the theory. A BPS
string can move in a transverse $\mathbb{R}^{4}$, so we can clearly introduce the
corresponding parameters $\varepsilon_{i}$. Additionally, there is $\tau$, the
complex structure of the $T^{2}$ wrapped by the strings. The K\"{a}hler
parameters $t_{b}$ directly appear as background fields from the 6D tensor
multiplets. Additionally, the seven-brane gauge groups appear as flavor symmetries
to BPS\ strings. Correspondingly, we can assign real fugacities $m_{i}$ to
these parameters as well.

Now, precisely because the topological string counts BPS\ states for M-theory
backgrounds, it is natural to conjecture a relation between topological string
amplitudes and the states counted by such effective strings. This leads to the
conjectural relation of reference \cite{Haghighat:2014vxa}:%
\begin{equation}
Z^{\text{top}}(\varepsilon_{1},\varepsilon_{2},\tau,t,m_{i})=Z_{0}%
(\tau,\varepsilon_{1},\varepsilon_{2},m_{i})\left(  1+\underset{k\text{
effective}}{\sum}Q_{k}Z_{k}(\varepsilon_{1},\varepsilon_{2},\tau
,m_{i})\right)  ,
\end{equation}
where here, on the righthand side, we sum over effective classes of the base
which can be wrapped by D3-branes of the model and thus produce BPS\ strings
in the 6D\ spacetime. For each such contribution, we have a corresponding
elliptic genus $Z_{k}(\varepsilon_{1},\varepsilon_{2},\tau,m_{i})$, which is
weighted by $Q_{k}$. This weighting parameter is proportional to $\exp
(-t_{k})$ with $t_{k}$ the value of the corresponding K\"{a}hler volume of the two-cycle.

A further remarkable aspect of this correspondence is that these amplitudes
are in turn (conjecturally) related to the index of the 6D\ SCFT around this
background:%
\begin{equation}
\mathcal{Z}_{T^{2}\times\mathbb{R}_{\Omega}^{4}}\left(  \varepsilon
_{1},\varepsilon_{2},\tau\right)  = Z^{\text{top}}.
\end{equation}
So far, this correspondence has mainly been checked for specific examples,
particularly with a single tensor multiplet and a simply laced gauge group
factor on the tensor branch \cite{Haghighat:2014vxa}. Further computations in
these theories were carried out in \cite{DelZotto:2016pvm, DelZotto:2017mee},
and this remains an active area of investigation.

\subsection{More General Partition Functions}

Even though the bulk of our discussion has focused on the special case of
partition functions on the background $S^{1}\times S^{5}$, it is of course in
principle possible to analyze the structure of the partition function from
working on a general six-manifold background. The condition to preserve
symmetry on this background follows from an analysis of the
off-shell structure of minimal 6D supergravity which would in principle follow
from an extension of the analysis in reference \cite{Festuccia:2011ws} though
as far as we are aware this has only been carried out for some limited
examples \cite{Samtleben:2012ua}. Even though the explicit evaluation of such
partition functions is an open direction for future research, some general
aspects of the resulting structure for such theories can already be deduced.

In general, we should not always expect a partition function. Instead, we should
expect a vector of partition functions \cite{Witten:1998wy} (see also
\cite{Henningson:2010rc, Tachikawa:2013hya, DelZotto:2015isa, Monnier:2017klz}%
. The extra data needed to specify this structure is encoded in the
\textquotedblleft defect group\textquotedblright:%
\begin{equation}
\mathcal{C} = \Lambda^{\ast}/\Lambda,\nonumber
\end{equation}
with $\Lambda$ the lattice of string charges on the tensor branch of a
6D\ SCFT, and $\Lambda^{\ast}$ the dual lattice. The order of this
group is specifed by $|\det A|$, with $A$ the adjacency matrix of a 6D\ SCFT.
For the $(2,0)$ theories it is simply the center of the corresponding simply
laced ADE\ Lie group. It has also been computed for all 6D\ SCFTs with an
F-theory realization in reference \cite{DelZotto:2015isa}. The partition
functions of the theory then transform in a representation of the discrete
three-form fluxes indicated by the Heisenberg group \underline{$H^{3}$}%
$(M_{6},\mathcal{C})$, defined via the short exact sequence:
\begin{equation}
1\rightarrow U(1)\rightarrow\underline{H^{3}}(M_{6},\mathcal{C})\rightarrow
H^{3}(M_{6},\mathcal{C})\rightarrow0,\nonumber
\end{equation}
where $H^{3}(M_{6},\mathcal{C})\simeq H^{3}(M_{6},\mathcal{%
\mathbb{Z}
})\otimes\mathcal{C}$, where we assume there are no torsional elements in the homology of $M_6$.
Note that in the special case of index computations on
$S^{1}\times S^{5}$ this additional data will not make an appearance
because $H^{3}(S^{1}\times S^{5},\mathbb{Z})$ is trivial. If, however, we add
defects to the system (by cutting out two-cycles from the geometry) then it will indeed become important additional data.

If one imposes the more stringent demand that a \textquotedblleft
true\textquotedblright\ quantum field theory possesses a partition function
rather than a vector of partition functions, then one can always add on a
spectator set of free tensor multiplets such that the new defect group is
trivial \cite{DelZotto:2015isa, DelZotto:2014fia}. In the context of F-theory
models coupled to gravity, this is simply the statement that for $\mathcal{B}$
a compact base, the lattice $\Lambda=H_{2}\left(  \mathcal{B},\mathbb{Z}%
\right)$ is self-dual, as follows from Poincar\'{e} duality.

It remains an outstanding open question to calculate the explicit form of
partition functions for general 6D\ SCFTs.

\section{Summary and Future Directions \label{sec:CONC}}

6D\ SCFTs occupy a central role in our modern understanding of quantum field theory. Long
thought not to exist, there is now strong evidence from string theory that not
only do they exist, but that many properties of lower-dimensional quantum
field theories descend from higher-dimensional considerations. In this review
article we have discussed many of the basic elements of these theories,
including the underlying symmetry properties, and how to construct these theories
in a variety of string constructions, particularly in the context of F-theory.
We have also explored some features of these rich physical theories, including
the structure of their tensor branches, RG\ flows between different fixed
points, and superconformal indices related to the degrees of freedom in these
theories. In the remainder of this section we discuss some areas which are
still rapidly developing, and potential avenues for future investigation.

We have only touched lightly on the compactification of 6D SCFTs to
lower dimensions. This is currently an active
area of investigation across many different dimensions, including 5D
\cite{Hayashi:2017jze, DelZotto:2017pti}, 4D \cite{Gaiotto:2015usa,
Ohmori:2015pua, DelZotto:2015rca, Franco:2015jna, Ohmori:2015pia, Coman:2015bqq,
Morrison:2016nrt, Apruzzi:2016nfr, Razamat:2016dpl, Bah:2017gph,
DelZotto:2017pti, Kim:2017toz, Hassler:2017arf, Bourton:2017pee, Kim:2018bpg, Apruzzi:2018oge},
and 2D theories \cite{Gadde:2013sca, Apruzzi:2016nfr, Lawrie:2016axq}.
Quite noticeably, there has been considerably less
analysis of the resulting 3D and 1D theories obtained from compactification on
a three-manifold and five-manifold, respectively. In all of these cases,
this should provide a systematic way to construct and study
lower-dimensional conformal fixed points, as well as strongly coupled quantum
field theories in general.

Along these lines, it is tempting to conjecture that all lower-dimensional
conformal fixed points actually descend from suitable compactifications of
6D\ SCFTs. Indeed, as one proceeds down in dimension, the complexity of the
internal geometry of a string compactification increases, leading to a
corresponding proliferation in possible theories. Developing methods to
systematically classify the resulting theories obtained from compactification
of 6D SCFTs would seem a worthwhile endeavor.

With such results in place, it would also become possible to compute partition
functions for a wide variety of 6D\ SCFTs on many new backgrounds. This would
clearly provide additional insight into the microscopic degrees of freedom in
such theories.

The conjectural classification of 6D\ SCFTs afforded by compactifications of
F-theory provides a natural starting point for a host of additional questions.
As we have already seen, there is now clear support for the 6D\ a-theorem,
which provides a monotonic height function on this restricted class of
elliptic Calabi-Yau threefolds. More ambitiously, it is natural to conjecture
that the hierarchies of RG\ flows can be used to systematically classify RG
flows between 6D\ SCFTs. This remains an outstanding open problem, but it is
now within the realm of possibility given our current advances in understanding 6D\ SCFTs.

A related point is how to connect the present considerations to more generic questions concerning 6D SCFTs
such as the operator content, scaling dimensions and three-point functions in these theories.
For example, using the conformal bootstrap it has become possible
to extract some robust results on certain subsectors of $(2,0)$ theories \cite{Beem:2014kka, Beem:2015aoa}
and some $(1,0)$ theories \cite{Chang:2017xmr}. This is likely to be a fruitful area of investigation
in the future.

Another general question left open by the present discussion is the extent to
which the \textquotedblleft top down\textquotedblright\ nature of our analysis
can be replaced by purely \textquotedblleft bottom up\textquotedblright%
\ considerations. For the most part, there is a tight interplay between
anomaly cancellation considerations on the tensor branch and the sort of
F-theory backgrounds which admit an elliptic fibration. On the other hand,
there also some noticeable places where the field theory analysis seems to be
missing something crucial. An example is the theory of a $-n$ curve for $n>2$.
Whereas the F-theory construction always predicts the existence of a gauge
theory sector, a priori, the field theory analysis would appear to be
compatible with no such sector. It is quite likely that this gap can be
closed, perhaps along the lines of reference \cite{Seiberg:2011dr}.

At a more fundamental level, it is remarkable that in spite of the mysterious
elements surrounding the microscopic origin of these 6D\ SCFTs so much can be
said about them! We leave the issue of how to construct a microscopic
formulation of these theories as an exercise for the interested reader.

\section*{Acknowledgements}

We thank F. Apruzzi, L. Bhardwaj, M. Del Zotto, F. Hassler, N. Mekareeya, I.V.
Melnikov, D.R.\ Morrison, D.S. Park, L. Tizzano, A. Tomasiello and C. Vafa
for a stimulating set of collaborations on which many elements of this work are
based.\ We also thank C. C\'{o}rdova, T.T. Dumitrescu, K. Intriligator, Y. Tachikawa
and K. Yonekura for many helpful discussions on related topics.
We also thank F. Apruzzi, C. C\'{o}rdova, M. Del Zotto, T.T. Dumitrescu,
B. Haghighat, K. Intriligator,
P. Jefferson, L. Lin, D.R. Morrison, T. Rochais, N. Seiberg, Y. Tachikawa, W. Taylor,
K. Yonekura, C. Vafa and E. Witten
for taking the time to read an earlier draft of this article,
for their detailed comments, and for their general encouragement. We thank F. Baume,
V. Saxena, X. Yu, and H.Y. Zhang for alerting us to typos in earlier drafts.
The work of JJH is supported by NSF CAREER grant PHY-1756996.
TR is supported by the Carl P. Feinberg Founders Circle Membership and the NSF Grant PHY-1314311.

\appendix

\section{The Superconformal Algebra \label{app:SCA}}

In this Appendix we present a brief review of the $D=6$ superconformal
algebra in Euclidean signature. A physically convenient basis of generators of the algebra are the
Lorentz generators $M_{\mu\nu}$, the translations $P_{\mu}$, the special
conformal transformations $K_{\mu}$ and the dilatation operator $D$. They
satisfy the commutation relations for the algebra $\mathfrak{so}(D+1,1)$:
\begin{align}
\mathcal{[}M_{\mu\nu},M_{\rho\sigma}]  &  =-i\left(  \delta_{\mu\sigma}%
M_{\nu\rho}+\delta_{\nu\rho}M_{\mu\sigma}-\delta_{\mu\rho}M_{\nu\sigma}-\delta
_{\nu\sigma}M_{\mu\rho}\right) \\
\mathcal{[}M_{\mu\nu},P_{\rho}]  &  =-i\left(  \delta_{\nu\rho}P_{\mu}-\delta
_{\mu\rho}P_{\nu}\right) \\
\mathcal{[}M_{\mu\nu},K_{\rho}]  &  =-i\left(  \delta_{\nu\rho}K_{\mu}-\delta
_{\mu\rho}K_{\nu}\right) \\
\mathcal{[}P_{\mu},K_{\nu}]  &  =-i\left(  2\delta_{\mu\nu}D+2M_{\mu\nu}\right)
\\
\mathcal{[}D,P_{\mu}]  &  =-iP_{\mu}\\
\mathcal{[}D,K_{\mu}]  &  =+iK_{\mu},
\end{align}
with all other commutators vanishing. These generators satisfy the Hermiticity properties:\footnote{In Lorentzian signature, one instead
has the algebra $\mf{so}(D,2)$, and the generators are all Hermitian operators.}
\begin{align}
M_{\mu\nu}^\dagger &= M_{\mu\nu} \\
D^\dagger &= - D \\
P_\mu^\dagger& = K_\mu \\
K_\mu^\dagger &= P_\mu.
\end{align}
It is a non-trivial problem to construct interacting quantum field theories which
enjoy this symmetry. From a physical perspective, states and operators can be
organized according to unitary representations of the conformal algebra.

For superconformal field theories, we enlarge this bosonic algebra to the
superalgebra with $(\mathcal{N},0)$ supersymmetry. Introducing supercharges $Q_{A\alpha}$ and and $S_{A \dot\alpha}$, with $A=1,...,2 \mathcal{N}$, $\alpha = 1,...,4$ , $\dot \alpha=1,...,4$, we have the following commutation relations:
\begin{align}
M_{\mu \nu} Q_{A \alpha} &= \frac{i}{4} [ \Gamma_\mu, \Gamma_\nu]_\alpha^\beta Q_{A \beta} \\
M_{\mu \nu} S_{A \dot\alpha} &= \frac{i}{4} [\tilde\Gamma_\mu, \tilde\Gamma_\nu]_{\dot\alpha}^{\dot\beta} S_{A \dot\beta} \\
[D, Q_{A \alpha}] &= -\frac{i}{2} Q_{A \alpha} \\
[D, S_{A \dot \alpha}] &= \frac{i}{2} S_{A \dot\alpha} \\
[P_\mu ,S_{A \dot \alpha}] &= i (\tilde \Gamma_\mu \tilde \Gamma_6 )_{\dot\alpha}^{\beta} Q_{A \beta} \\
[K_\mu, Q_{A \alpha}] &= i ( \Gamma_\mu  \Gamma_6 )_{\alpha}^{\dot \beta} S_{A \dot\beta}.
\end{align}
Here, $\Gamma_\mu$ are the gamma matrices in six dimensions, and $\Gamma^{1,2,3,4,5}=\tilde \Gamma^{1,2,3,4,5}$, $\Gamma^6 = - \tilde \Gamma^6$.  Definining $\Gamma_7 = - \Gamma^1 \Gamma^2 \Gamma^3 \Gamma^4 \Gamma^5 \Gamma^6$ and the projector $P_+ = \frac{1}{2} (1 + \Gamma_7$), we further have
\begin{align}
\left\{ Q_{A \alpha}, Q_{B\beta} \right\} &= (P_+ \Gamma^\mu P_\mu C)_{\alpha \beta} \Omega_{AB}\\
\left\{ S_{A \dot\alpha}, S_{B \dot\beta} \right\} &= (P_+ \tilde\Gamma^\mu K_\mu C)_{\dot\alpha \dot\beta} \Omega_{AB}.
\end{align}
Here, $C$ is the charge conjugation matrix, $\Omega_{AB} = \epsilon_{AB}$ for $\mathcal{N}=1$, and $\Omega_{AB}$ is the 4D symplectic matrix with nonvanishing elements $\Omega_{14} = - \Omega_{41} = \Omega_{23} = - \Omega_{32} =1 $ for $\mathcal{N}=2$.  Note that $Q$, $S$ satisfy the reality properties,
\begin{align}
Q_{A \alpha} &= i \Omega_{AB} (C \Gamma_6^T)_{\alpha \dot\beta} S^{\dagger B \dot \beta} \\
S_{A \alpha} &= -i \Omega_{AB} (C \tilde\Gamma_6^T)_{\dot \alpha \beta} Q^{\dagger B \beta}.
\end{align}

Finally, for $\mathcal{N}=1$, we have
\begin{align}
\left\{ Q_{A \alpha}, Q^{ \dagger B \gamma } \right\} = \Big[ \delta_{A}^B \frac{i}{2} [(M_{\mu\nu} P_+ \Gamma^\mu \Gamma^\nu)_\alpha^\gamma + 2 (P_+)_\alpha^\gamma D] + 4 (T_a \sigma_a)_{A}^B (P_+)_\alpha^\gamma \Big].
\label{eq:QQdag}
\end{align}
Here, $\sigma_a$ are the Pauli matrices, and $T_a$ are the generators of the $\mf{su}(2)_R$ R-symmetry, satisfying
\begin{align}
[T_a, Q_{A \alpha}] &= -( \sigma_a/2)_A^B Q_{B \alpha} \\
[T_a, S_{A \dot\alpha}] &= -( \sigma_a/2)_A^B S_{B \dot\alpha},
\end{align}

For $\mathcal{N}=2$, we replace the $\mf{su}(2)_R$ generators $T_a$ with $\mf{so}(5)_R$ generators $T_{ab}$, $a,b=1,...,5$, and we exchange $\sigma_a$ for $(-i/8) [\Gamma'_a, \Gamma'_b]$.  Thus we simply replace (\ref{eq:QQdag}) with
\begin{align}
\left\{ Q_{A \alpha}, Q^{ \dagger B \gamma } \right\} = \Big[ \delta_{A}^B \frac{i}{2} [(M_{\mu\nu} P_+ \Gamma^\mu \Gamma^\nu)_\alpha^\gamma + 2 (P_+)_\alpha^\gamma D]  -(i/2) (T_{ab}  [\Gamma'_a, \Gamma'_b])_{A}^B (P_+)_\alpha^\gamma \Big].
\end{align}

\section{Discrete Subgroups of $SU(2)$ and ADE\ Singularities \label{app:ADE}}
In this Appendix we list the discrete subgroups of $SU(2)$ and the
orbifold singularities generated by group actions on $\mathbb{C}^{2}$. Recall
that there is a celebrated ADE\ classification of such discrete groups. The
generators are conveniently listed as:%
\begin{align}
\Gamma_{A_{n-1}}  &  =\left\langle \zeta_{n}\right\rangle \\
\Gamma_{D_{n}}  &  =\left\langle \zeta_{2n-4},\delta\right\rangle \\
\Gamma_{E_{6}}  &  =\left\langle \zeta_{4},\delta,\tau\right\rangle \\
\Gamma_{E_{7}}  &  =\left\langle \zeta_{8},\delta,\tau\right\rangle \\
\Gamma_{E_{8}}  &  =\left\langle -(\zeta_{5})^{3},\iota\right\rangle ,
\end{align}
where we have introduced the generators with matrix representatives
\begin{align}
\zeta_{n}  &  =\left[
\begin{array}
[c]{cc}%
e^{2\pi i/n} & \\
& e^{-2\pi i/n}%
\end{array}
\right]  \text{, \ \ }\delta=\left[
\begin{array}
[c]{cc}
& 1\\
-1 &
\end{array}
\right]  \text{, \ \ }\tau=\frac{1}{\sqrt{2}}\left[
\begin{array}
[c]{cc}%
e^{-2\pi i/8} & e^{-2\pi i/8}\\
e^{10\pi i/8} & e^{2\pi i/8}%
\end{array}
\right] \\
\iota &  =\frac{1}{e^{4\pi i/5}-e^{6\pi i/5}}\left[
\begin{array}
[c]{cc}%
e^{2\pi i/5}+e^{-2\pi i/5} & 1\\
1 & -e^{2\pi i/5}-e^{-2\pi i/5}%
\end{array}
\right]  .
\end{align}
These are the symmetry groups of natural geometric structures. For example,
$\Gamma_{A_{n-1}}$ is the order $n$ cyclic group (discrete rotations),
$\Gamma_{D_{n}}$ is the binary dihedral group (double cover of symmetries of
an $n$-gon), $\Gamma_{E_{6}}$ is the binary tetrahedral group
(symmetries of the tetrahedron platonic solid), $\Gamma_{E_{7}}$ is the binary
octahedral group (symmetries of the octahedron platonic solid), and
$\Gamma_{E_{8}}$ is the binary icosahedral group (symmetries of the
icosahedron platonic solid).

Quotienting $\mathbb{C}^{2}$ by the group action specified by the above
$2\times2$ matrix representatives generates an orbifold singularity,
$\mathbb{C}^{2}/\Gamma$. This singularity can be described as a hypersurface
in $\mathbb{C}^{3}$ by constructing appropriate polynomials in the local
coordinates $(u,v)$ for $\mathbb{C}^{2}$ which are invariant under this group
action. For example, in the case of $\mathbb{C}^{2}/%
\mathbb{Z}
_{n}$, we have the invariants $W=uv$, $U=u^{n}$ and $V=v^{n}$, so we can
introduce coordinates $(U,V,W)$ on $\mathbb{C}^{3}$ subject to the relation:%
\begin{equation}
UV=W^{n}.
\end{equation}
Recall that a singularity is present whenever the zero locus and the vanishing
locus of the first derivatives have a common solution.

Proceeding in this way, we can list all of the Kleinian singularities. To give
a uniform presentation, we adopt coordinates $(X,Y,Z)$ on $\mathbb{C}^{3}$:%
\begin{align}
\mathbb{C}^{2}/\Gamma_{A_{n-1}}  &  :Y^{2}=X^{2}+Z^{n}\label{Atype}\\
\mathbb{C}^{2}/\Gamma_{D_{n}}  &  :Y^{2}=X^{2}Z+Z^{n-1}\\
\mathbb{C}^{2}/\Gamma_{E_{6}}  &  :Y^{2}=X^{3}+Z^{4}\\
\mathbb{C}^{2}/\Gamma_{E_{7}}  &  :Y^{2}=X^{3}+XZ^{3}\\
\mathbb{C}^{2}/\Gamma_{E_{8}}  &  :Y^{2}=X^{3}+Z^{5}.
\end{align}
In algebraic geometry, we resolve these singularities by performing blowups.
This introduces a configuration of $-2$ curves which intersect according to
the associated Dynkin diagram of ADE type.

To illustrate how this works, we now explicitly resolve the $\mathbb{C}%
^{2}/\Gamma_{A_{n-1}}$ singularity. We follow the presentation for physicists
in reference \cite{Hori:1997zj} (see also \cite{ShafI}). To this end, it is helpful
to work in terms of the presentation $UV=W^{n}$ (we can reach the presentation
of line (\ref{Atype}) by the linear change of coordinates $U=Y+X$, $V=Y-X$,
$W=Z$). To resolve the singularity, we introduce two coordinate charts:%
\begin{align}
\text{Chart I coordinates}  &  \text{: }(U,V,\widetilde{W})\\
\text{Chart II coordinates}  &  \text{: }(\widetilde{U},V,W).
\end{align}
These charts are joined together by the presence of an additional compact
$\mathbb{P}^{1}$ (or new curve from the blowup), and we introduce the
relations:%
\begin{equation}
\widetilde{U}\widetilde{W}=1\text{ \ \ and \ \ }U\widetilde{W}=W\widetilde{U}%
\text{.}%
\end{equation}
In terms of the original $\mathbb{C}^{3}$ coordinates we have:%
\begin{equation}
(U,V,W)=(U,V,U\widetilde{W})=(W\widetilde{U},V,W).
\end{equation}
Let us see what has happened to the equation $UV=W^{n}$ in our two charts:%
\begin{align}
\text{Chart I: }  &  UV=U^{n}\widetilde{W}^{n}\\
\text{Chart II}  &  \text{:}\text{ }W\widetilde{U}V=W^{n}.
\end{align}
Eliminating common factors, we see that the candidate singular geometries are:%
\begin{align}
\text{Chart I: }  &  V=U^{n-1}\widetilde{W}^{n}\\
\text{Chart II}  &  \text{:}\text{ }\widetilde{U}V=W^{n-1},
\end{align}
so we see that in chart I, there is no singularity (the derivative with
respect to $V$ is just $1$ which never vanishes), and in Chart II, the
exponent of $W$ has now decreased. Continuing in this way, we eventually reach
a final chart of the form $UV=W$, which is not
singular.\footnote{By abuse of notation we use the same variables to
denote this chart.} This process introduces
precisely $n-1$ compact curves of self-intersection $-2$, which intersect
according to the pattern:%
\begin{equation}
\mathbb{C}^{2}/\mathbb{Z}_{n}\text{ resolution: }\underset{n-1}{\underbrace{2,...,2}%
}\text{.}%
\end{equation}
Similar (albeit more involved) resolutions can be carried out for each of the
D- and E-type singularities. We note that this is the same procedure one needs
to follow to resolve the singular fibers of an elliptic threefold.

As a brief aside, we note that the self-intersection number of the compact
curves follows from the genus formula: $\Sigma\cdot(\Sigma+K)=2g-2$, with $K$
the canonical class of the ambient K\"ahler surface. Since $g=0$ and we are assuming we
have a Calabi-Yau after the resolution, $K$ vanishes, so $\Sigma\cdot
\Sigma=-2$. Note that in the context of the blowups in the base of an
elliptically fibered Calabi-Yau threefold, the exceptional divisors have
self-intersection $-1$, not $-2$.

\section{The Local Geometry $\mathcal{O}(-n)\rightarrow\mathbb{CP}^{1}$ \label{app:LOCO}}

In this Appendix we briefly review some aspects of the local
geometry $\mathcal{O}(-n)\rightarrow\mathbb{CP}^{1}$. It is customary to refer
to the base curve as one of self-intersection $-n$. The terminology is most
straightforward to understand when $-n$ is positive, since in this case we can
locally deform the curve to another location, and the number of points of
self-intersection with the curve and its deformed image is $-n$. When $-n$ is
negative, as will always be the case in our study of 6D\ SCFTs, it is
customary to say that the curve has a negative self-intersection number.

We now provide a few complementary descriptions of the same geometry. To
begin, we provide an intrinsic description of $\mathbb{CP}^{1}$. This can be
viewed as points $[u,v]$ in the complex space $\mathbb{C}^{2}-(0,0)$ which are
identified according to the equivalence relation:%
\begin{equation}
\lbrack u,v]\sim\lbrack\lambda u,\lambda v]\text{ \ \ for \ \ }\lambda
\in\mathbb{C}^{\ast}%
\end{equation}
We refer to $u$ and $v$ as homogeneous coordinates of the projective space. In
a patch where $v$ does not vanish, we can also work in terms of the ratio
$u/v$, and in a patch where $u$ does not vanish we can work in terms of the
ratio $v/u$. One somewhat abstract way to specify a line bundle of degree $-n$
is to work in terms of transition functions between these two patches.
Sections $\lambda$ of the bundle which transform as a power of degree $\lambda^{-n}$
define the line bundle $\mathcal{O}(-n)\rightarrow\mathbb{CP}^{1}$.

A somewhat more familiar construction to physicists involves explictly
specifying the fiber bundle description. To each point of the $\mathbb{CP}%
^{1}$, we attach a copy of the complex line $\mathbb{C}$. This defines a fiber
bundle with total space a non-compact K\"{a}hler surface. Moving from point to
point can then be described by introducing a complexification of a $U(1)$
gauge connection over the $\mathbb{CP}^{1}$. The curvature of this gauge
connection $F$ gives us the Chern class of the bundle, which we can integrate
over the compact $\mathbb{CP}^{1}$:%
\begin{equation}
-n=\underset{\mathbb{CP}^{1}}{\int}\frac{F}{2\pi}.
\end{equation}

Finally, we can also describe this same space using methods from symplectic
geometry, or equivalently in terms of the vacuum moduli space of a
two-dimensional gauged linear sigma model with $\mathcal{N}=(2,2)$
supersymmetry \cite{Witten:1993yc}. Introducing a single $U(1)$ vector multiplet and three
chiral superfields $A,B,C$ of respective charges $+1,+1,-n$, the vacuum D-term
constraint reads:%
\begin{equation}
(|A|^{2}+|B|^{2}-n\left\vert C\right\vert ^{2}=r)/U(1)
\end{equation}
The zero section corresponds to the locus where $C=0$, namely:%
\begin{equation}
C=0\text{ locus: \ }|A|^{2}+|B|^{2}=r.
\end{equation}
This describes a two sphere, as can be seen by decomposing $A$ and $B$ into
real and imaginary parts, and imposing the $U(1)$ gauge constraint.

In the context of engineering 6D\ SCFTs, we are especially interested in the
limit where this compact $\mathbb{CP}^{1}$ collapses to zero size, namely
where $r\rightarrow0$. When this occurs, the variable $C$ is fully determined
by the parameters $A$ and $B$. In particular, we see that $A$ and $B$ sweep
out a copy of $\mathbb{C}^{2}$, but subject to the identification:%
\begin{equation}
\left(  A,B\right)  \sim(\zeta A,\zeta B)\text{ \ \ for \ \ }\zeta
^{n}=1\text{,}%
\end{equation}
that is, we have an orbifold singularity $\mathbb{C}^{2}/%
\mathbb{Z}
_{n}$. Note that this is not a \textquotedblleft supersymmetric
orbifold,\textquotedblright\ because the group action is the same on $A$ and
$B$.

\section{Examples of Blowups for Conformal Matter \label{app:BLOW}}

In this Appendix we give some examples of how the blowup procedure is
implemented for collisions of seven-branes with conformal matter. For specificity,
we suppose that we have local coordinates $(u,v)$ on $\mathbb{C}^{2}$. In
terms of these coordinates, $f$, $g$ and $\Delta$ of the Weierstrass model
will be polynomials which we take to vanish to some order at the orgin
$u=v=0$. We perform a blowup in the base of the local Calabi-Yau threefold by
deleting this point and replacing it with an exceptional divisor which may
still have a singular fiber on it. Continuing in this way, we continue to
blowup until all fibers are in Kodaira-Tate form. To illustrate, we use the
following algorithmic procedure, discussed for example in \cite{Aspinwall:1997ye}.

In the local patch described by the blowup, we can write this as the
birational transform:%
\begin{equation}
u\mapsto u\text{, \ \ }v\mapsto uv\text{,}%
\end{equation}
where here, we assume that the singularity type along $v=0$ is worse than
along $u=0$. Since $f$, $g$ and $\Delta$ transform as sections of
$K_{\mathcal{B}}^{-4}$, $K_{\mathcal{B}}^{-6}$ and $K_{\mathcal{B}}^{-12}$
respectively, under the birational transform they go to:%
\begin{align}
f(u,v)  & \mapsto f(u,uv)/u^{4}\\
g(u,v)  & \mapsto g(u,uv)/u^{6}\\
\Delta(u,v)  & \mapsto\Delta(u,uv)/u^{12}.
\end{align}

Let us illustrate how this works in a few examples. To begin, consider the
case of a single small instanton on an $\mathfrak{e}_{8}$ locus. The
Weierstrass model is:%
\begin{equation}
y^{2}=x^{3}+uv^{5},
\end{equation}
namely $g(u,v)=uv^{5}$. So, under the blowup, we get:%
\begin{equation}
uv^{5}=g(u,v)\mapsto g(u,uv)/u^{6}=v^{5},
\end{equation}
so after the resolution, we have an $\mathfrak{e}_{8}$ meeting along $v=0$ and
it meets a $-1$ curve with no fiber decoration.

Something quite similar happens in the case of a colliding $D_{4}\times D_{4}$
singularity. In this case, the local Weierstrass model is:%
\begin{equation}
y^{2}=x^{3}+\alpha u^{2}v^{2}x+\beta u^{3}v^{3}\text{,}%
\end{equation}
where $\alpha$ and $\beta$ are constants. In this case, after one blowup we
have:%
\begin{align}
\alpha u^{2}v^{2}  & =f(u,v)\mapsto f(u,uv)/u^{4}=\alpha v^{2}\\
\beta u^{3}v^{3}  & =g(u,v)\mapsto g(u,uv)/u^{6}=\beta v^{3},
\end{align}
so again, we have an isolated $I_{0}^{\ast}$ fiber and we realize the
configuration of curves and fiber types:%
\begin{equation}
\lbrack I_{0}^{\ast}],\overset{I_{0}}{1},[I_{0}^{\ast}],
\end{equation}
which we recognize as the tensor branch of $D_{4}\times D_{4}$ conformal
matter:%
\begin{equation}
\lbrack D_{4}],1,[D_{4}],
\end{equation}

Sometimes we must perform additional blowups. An example of this type is the
collision of two $E_{6}$ singularities:%
\begin{equation}
y^{2}=x^{3}+u^{4}v^{4}\text{.}%
\end{equation}
Now after one blowup, we have:%
\begin{equation}
u^{4}v^{4}=g(u,v)\mapsto g(u,uv)/u^{6}=u^{2}v^{4},
\end{equation}
so in this patch the Weierstrass Model is:%
\begin{equation}
y^{2}=x^{3}+u^{2}v^{4},
\end{equation}
which describes the collision of a type $IV$ and $IV^{\ast}$ fiber. This is
still not in Kodaira-Tate form because the order of vanishing along $u=v=0$ is
still too singular. Blowing up this collision again, we have:%
\begin{equation}
u^{2}v^{4}=g(u,v)\mapsto g(u,uv)/u^{6}=v^{4},
\end{equation}
so now we have an isolated $E_{6}$. Now, we are still not done because there
is still the collision of this type $IV$ fiber with the \textquotedblleft
other\textquotedblright\ $E_{6}$ factor. By symmetry, we get the same result
on the two sides, so we get the following configuration of curves and fiber
types:%
\begin{equation}
\lbrack IV^{\ast}],1,\overset{IV}{3},1,[IV^{\ast}],
\end{equation}
which we recognize as the tensor branch of $E_{6}\times E_{6}$ conformal
matter:%
\begin{equation}
\lbrack E_{6}],1,\overset{\mathfrak{su}(3)}{3},1,[E_{6}].
\end{equation}
Similar considerations hold for other collisions of singularities.

\section{F-theory on Hirzebruch Surfaces}\label{sec:Hirzebruch}

In this Appendix, we analyze F-theory on Hirzebruch surfaces. This can be viewed as a technical supplement to the discussion of non-Higgsable clusters in Section \ref{ssec:NHC}.

Starting from the correspondence of (\ref{HETF}), we can fiber each side over a
common manifold to produce additional, lower-dimensional examples of heterotic /
F-theory pairs. To retain supersymmetry, we restrict to the case where the
heterotic string is compactified on an elliptically fibered K3 surface, so
this additional \textquotedblleft common\textquotedblright\ space is a
$\mathbb{CP}^{1}$ which we denote as $\mathbb{CP}_{\text{common}}^{1}$. On the
F-theory side, we have a K3-fibered Calabi-Yau manifold with base a
$\mathbb{CP}_{\text{common}}^{1}$. Now, the base of the F-theory
model is a complex surface, which is itself described by a $\mathbb{CP}^{1}$
bundle over this $\mathbb{CP}_{\text{common}}^{1}$. Such K\"{a}hler surfaces
are known as Hirzebruch surfaces $\mathbb{F}_{n}$, and are classified by a
single integer $n$, the degree of the first Chern-class for the $\mathbb{CP}%
^{1}$ bundle over $\mathbb{CP}_{\text{common}}^{1}$.\footnote{Here we follow
the physicist's conventions to allow $n$ both positive and negative.
Geometrically, the cases of $\pm n$ describe the same K\"ahler surface. The
reason we allow ourselves to range over both choices has to do with possibly
treating the two $E_{8}$ factors of the heterotic dual differently.} On the heterotic side,
this integer $n$ tells us the number of \textit{dissolved} instantons
localized on each $E_{8}$ wall \cite{Morrison:1996pp, Witten:1996qb, Bershadsky:1996nh}:
$(12-n,12+n)$ for the two walls, respectively. As we move from one value of $n$ to the next,
we can interpret the corresponding shift in $n$ as pulling an M5-brane off of
one nine-brane wall and dissolving it in the other.

The Hirzebruch surfaces have $H_2(\mathbb{F}_n , \mathbb{Z})$ which is generated by the base
$\mathbb{CP}_{\text{common}}^{1}$, denoted by $\sigma_{\text{base}}$ and the
fiber class $\sigma_{\text{fiber}}$. These divisor classes satisfy the
intersection pairing:%
\begin{equation}
\sigma_{\text{base}}\cdot\sigma_{\text{base}}=-n\text{, \ \ }\sigma
_{\text{base}}\cdot\sigma_{\text{fiber}}=1\text{, \ \ }\sigma_{\text{fiber}%
}\cdot\sigma_{\text{fiber}}=0. \label{HirzebruchIntersection}%
\end{equation}
The local geometry of the base curve is dictated by its self-intersection,
being given by $\mathcal{O}(-n)\rightarrow\mathbb{CP}^{1}$. Globally, we
projectivize this line bundle to a $\mathbb{CP}^1$.

In terms of the generators $\sigma_{\text{base}}$ and $\sigma_{\text{fiber}}$,
the canonical class of the surface is:%
\begin{equation}
K_{\mathbb{F}_{n}}=-(n+2)\sigma_{\text{fiber}}-2\sigma_{\text{base}}\text{.}%
\end{equation}
To construct an F-theory model with base a Hirzebruch surface, we need $f$
and $g$ of the Weierstrass model to be sections of:%
\begin{equation}
f\sim\mathcal{O}\left(  4(n+2)\sigma_{\text{fiber}}+8\sigma_{\text{base}%
}\right)  \text{ \ \ and \ \ }g\sim\mathcal{O}\left(  6(n+2)\sigma
_{\text{fiber}}+12\sigma_{\text{base}}\right)  . \label{linebundlesexample}%
\end{equation}
In this case, writing out the explicit polynomial forms of these expressions
is a bit more challenging, owing to the fact that we have a non-trivial
intersection pairing structure on the base. To do this, we introduce $[u,v]$
homogeneous coordinates of the base $\mathbb{CP}_{\text{common}}^{1}$, and a
local affine coordinate $w$ to describe the fiber direction so that $w=0$
denotes the divisor class $\sigma_{\text{base}}$. We can express $f$ and $g$
as polynomials:%
\begin{equation}
f=\underset{i}{\sum}w^{i}f_{a(i)}(u,v)\text{, \ \ }g=\underset{j}{\sum}%
w^{j}g_{b(j)}(u,v),
\end{equation}
where $f_{a(i)}$ and $g_{b(j)}$ denote homogeneous polynomials of degree
$a(i)$ and $b(j)$ which we fix according to the Calabi-Yau condition for our
threefold. To determine, this, we note that these polynomials need to be
sections of the following bundles:%
\begin{align}
f_{a(i)}  &  \sim\mathcal{O}\left(  4(n+2)\sigma_{\text{fiber}}+8\sigma
_{\text{base}}-i\sigma_{\text{base}}\right)  |_{\sigma_{\text{base}}}\\
g_{b(j)}  &  \sim\mathcal{O}\left(  6(n+2)\sigma_{\text{fiber}}+12\sigma
_{\text{base}}-j\sigma_{\text{base}}\right)  |_{\sigma_{\text{base}}},
\end{align}
or, in terms of explicit line bundles on the common $\mathbb{CP}^{1}$:%
\begin{align}
f_{a(i)}  &  \sim\mathcal{O}_{\mathbb{CP}^{1}}\left(  4(n+2)\sigma
_{\text{fiber}}\cdot\sigma_{\text{base}}+(8-i)\sigma_{\text{base}}\cdot
\sigma_{\text{base}}\right)  =\mathcal{O}_{\mathbb{CP}^{1}}\left(
8+n(i-4)\right) \\
g_{b(j)}  &  \sim\mathcal{O}_{\mathbb{CP}^{1}}\left(  6(n+2)\sigma
_{\text{fiber}}\cdot\sigma_{\text{base}}+(12-j)\sigma_{\text{base}}\cdot
\sigma_{\text{base}}\right)  =\mathcal{O}_{\mathbb{CP}^{1}}\left(
12+n(j-6)\right)  ,
\end{align}
so we can write the Weierstrass coefficients as:%
\begin{equation}
f=\underset{i}{\sum}w^{i}f_{8+n(i-4)}(u,v)\text{, \ \ }g=\underset{j}{\sum
}w^{j}g_{12+n(j-6)}(u,v), \label{fgexpandobando}%
\end{equation}
in the obvious notation. Here, the sum is restricted to the case where the
degree is actually non-negative.

An important feature of this configuration is the lowest degree terms in the
polynomial $w$, which clearly depends on the value of $n$. First of all, we
observe that in the special case $n=0$, the base is just $\mathbb{CP}%
^{1}\times\mathbb{CP}^{1}$ and $f$ and $g$ are bihomogeneous polynomials of
degree $(8,8)$ and $(12,12)$ respectively.

For $n=1$, we see that $f$ and $g$ are necessarily non-trivial polynomials
with lowest powers $4$ and $6$. That is to say, in the $[u,v]$ directions, we
have a local $dP_{9}$ geometry. In particular, we can coalesce the fibers to
obtain an $E_{8}$ singularity, which is of course nothing but the $E_{8}$
symmetry of the heterotic string. Interestingly, we can also collapse the
common $\mathbb{CP}_{\text{common}}^{1}$ to zero size, and in this limit, all
the fibers are necessarily on top of each other. So in this singular limit,
there is always an emergent $E_{8}$ symmetry in the low energy theory.

In the stable degeneration limit where we focus on just this single $E_{8}$
factor, the heterotic description also simplifies since now we are describing
an M5-brane moving close to the nine-brane. In this picture, the distance between the M5-brane and the
nine-brane maps to the overall volume of the $\mathbb{CP}^{1}$ on the F-theory side.

For $n=2$, we see that it is possible to just keep $f$ and $g$ constant, at
least to leading order. This is compatible with the fact that in this case,
the local geometry of the common $\mathbb{CP}_{\text{common}}^{1}$ is
$\mathcal{O}(-2)\rightarrow\mathbb{CP}^{1}$, which is Calabi-Yau. We can, of
course, perform further tuning in the Weierstrass model to produce higher
order singularities. This corresponds to wrapping seven-branes over the curve.

Starting at $n\geq3$, a singular fiber becomes unavoidable. Indeed, the
condition that we have positive degree homogeneous polynomials in
(\ref{fgexpandobando}) demand:
\begin{equation}
8+n(i-4)\geq0\text{ \ \ and \ \ }12+n(j-6)\geq0\text{,}%
\end{equation}
which for a non-trivial $f$ requires $i\geq 2$, and for a non-trivial $g$
requires $j\geq 2$. In this case, then, a seven-brane is inevitably present,
and cannot be deformed away by variations in the complex structure. We have
actually encountered this situation previously in our discussion of bottom up
constraints. It has to do with the curious feature of six-dimensional gauge
theories:\ In this case, there is no scalar mode in the $\mathcal{N}=(1,0)$
vector multiplet, so there may not always be a way to Higgs the corresponding
gauge theory. Another important feature is that once $n$ becomes too large,
there is no way to rescue the Weierstrass model. The reason is that we also
have an upper bound on $i$ and $j$, as dictated by the $(4,6)$ order of
vanishing of $f$ and $g$. This restricts us to $n\leq12$.

In fact, returning to our discussion of Kodaira fiber types, we can even read
off what this minimal singularity must be in the different cases:%
\begin{equation}%
\begin{tabular}
[c]{|l|l|l|l|l|l|l|l|l|}\hline
 & $n=0,1,2$ & $n=3$ & $n=4$ & $n=5$ & $n=6$ & $n=7$ & $n=8$ &
$n=12$\\\hline
Minimal Fiber  & Trivial  & $IV$ & $I_{0}^{\ast}$ &
$IV^{\ast}$ & $IV^{\ast}$ & $III^{\ast}$ & $III^{\ast}$ & $II^{\ast}$\\\hline
Gauge Algebra   & None & $\mf{su}(3)$ & $\mf{so}(8)$ & $\mf{f}_{4}$ & $\mf{e}_{6}$ &
$\mf{e}_{7}+$matter & $\mf{e}_{7}$ & $\mf{e}_{8}$\\\hline
\end{tabular}
\ .
\end{equation}
Note that in this description we have skipped over the cases $n=9,10,11$.
Returning to our Weierstrass model, we see that here, $g$ takes the form%
\begin{align}
n  &  =9:g=w^{5}g_{3}(u,v)+...\\
n  &  =10:g=w^{5}g_{2}(u,v)+...\\
n  &  =11:g=w^{5}g_{1}(u,v)+...,
\end{align}
so while there is an $\mf{e}_{8}$ singularity present, it actually becomes worse at
specific locations where the $g_{j}$'s vanish. This can of course be dealt
with by performing blowups in the base, one for each location where the
$g_{j}$ vanishes. The local geometry of each additional such curve is
$\mathcal{O}(-1)\rightarrow\mathbb{CP}^{1}$, so it is natural to interpret
this in heterotic terms as the presence of additional \textquotedblleft small
instantons,\textquotedblright\ which are M5-branes near the nine-brane wall which
cannot be dissolved back into flux \cite{Morrison:1996pp, Witten:1996qb}.

The models with $\mathbb{F}_{n}$ base already present us with a very interesting class
of 6D\ SCFTs. Observe that in these models, we have a curve $\mathbb{CP}%
_{\text{common}}^{1}$ of self-intersection $-n$, with local geometry
$\mathcal{O}(-n)\rightarrow\mathbb{CP}^{1}$. Collapsing this $\mathbb{CP}^{1}$
to zero size, the tension vanishes for the effective strings generated from
D3-branes wrapped over this curve, thus providing a set of 6D\ SCFTs. As
discussed in Appendix \ref{app:LOCO}, in this limit of a collapsing $\mathbb{CP}^{1}$,
the local geometry of the base is actually an orbifold singularity
$\mathbb{C}^{2}/\mathbb{Z}_{n}$ in which the local coordinates $u$ and $v$ of
$\mathbb{C}^{2}$ are identified according to the group action:%
\begin{equation}
(u,v)\rightarrow(\zeta u,\zeta v)\text{ \ \ with \ \ }\zeta^{n}=1\text{.}
\label{groupaction}%
\end{equation}
Note that for $n>2$, this is not the same as the \textquotedblleft
supersymmetric orbifold\textquotedblright\ where the group would have embedded
in $SU(2)$. This cannot happen here because the associated linear
transformation has determinant $\zeta^{2}$ rather than 1. The reason that
our geometry is nevertheless supersymmetric is that in addition to the
background profile of the metric, we also have additional sources in the 10D
Einstein field equations associated with localized seven-branes. Indeed, the
actual condition that we realize a supersymmetric 6D theory is that we
compactify F-theory on an elliptically fibered Calabi-Yau threefold,
\textit{not} that the base $\mathcal{B}$ itself be Calabi-Yau.

Another interesting feature of the above local model is that for special
values of $n$ and tuned values of the complex structure moduli, we can present
the entire construction as an orbifold of the form \cite{Witten:1996qb, Heckman:2013pva}:
\begin{equation}
\left(  \mathbb{C}^{2}\times T^{2}\right)  /\Gamma.
\end{equation}
The group action on the $\mathbb{C}^{2}$ coordinates is specified as in line
(\ref{groupaction}). The group action on the $T^{2}$ is constrained by the
condition that we have a Calabi-Yau threefold, so locally, including the
$T^{2}$ coordinate $w$, the full action is:%
\begin{equation}
(u,v,w)\rightarrow(\zeta u,\zeta v,\zeta^{-2}w)\text{ \ \ with \ \ }\zeta
^{n}=1\text{.}%
\end{equation}
Orbifolds of $T^{2}$ only exist for special values of $n$, namely
$n=3,4,6,8,12$, which as we see, covers some, but not all of the local models
encountered here. Finally, note that due to the quotient, the elliptic
fibration is actually non-trivial.

\section{Group Theory Constants \label{sec:GROUPTHEORY}}

\begin{table}[t!]
	\centering
	\begin{tabular}{|c||c|c|c|c|c|c|c|c|c|c|c|}
		\hline
		$G$ &$ SU(2) $&$SU(3)$&$ SU(k\geq 4) $& $SO(8)$ &$ SO(k \neq 8) $&$Sp(k)$&$G_2 $&$F_4 $&$E_6$&$E_7$&$E_8$\\
		\hline\hline
		$r_{G}$&1&2&$k-1$ & 4& $\lfloor k/2 \rfloor$ & $k$ & 2 & 4 & 6 & 7 & 8\\
		\hline
		$d_{G}$&3&8&$k^2-1$ & 28 &$k(k-1)/2$ & $k(2k+1)$ & 14 & 52 & 78 & 133 & 248\\
		\hline
		$h^\vee_{G}$&2&3&$k$ & 6& $k-2$ & $k+1$ & 4 & 9 & 12 & 18 & 30\\
		\hline
		$x_{G}$&0&0&$2k$ & $4$ &$k-8$ & $2k+8$ & 0 & 0 & 0 & 0 & 0\\
		\hline
		$y_{G}$&2&$\frac{9}{4}$&$\frac{3}{2}$ &$(4,4)$&$3$ & $\frac{3}{4}$ & $\frac{5}{2} $&$ \frac{15}{4} $&$ \frac{9}{2} $&$ 6$&$ 9$\\
		\hline
		$d_{\text{fnd}}$&2&3&$k$ &8 & $k$ & $2k$ & 7 & 26 & 27 & 56 & 248\\
		\hline
		$h_{\text{fnd}}$&$\frac{1}{2}$&$\frac{1}{2}$&$\frac{1}{2}$ &1& $1$ & $\frac{1}{2}$ & 1 & 3 & 3 & 6 & 30\\
		\hline
		$x_{\text{fnd}}$&0&0& $1$ & 1 & $1$ & $1$ & 0 & 0 & 0 & 0 & 0 \\
		\hline
		$y_{\text{fnd}}$&$\frac{1}{8}$&$\frac{1}{8}$& 0 & $(0,0)$ & 0 & 0 &  $\frac{1}{4}$ & $\frac{3}{4}$ & $\frac{3}{4}$ & $\frac{3}{2}$ &$9$ \\
		\hline
		$d_{\text{spin}}$&-&-& - & $8_s, 8_c$& $ 2^{\lceil k/2\rceil-1}$ & - & -& - & - & - & - \\
		\hline
		$h_{\text{spin}}$&-&-& - & $1,1$ & $ \frac{1}{8}  d_{\text{spin}}$ & - & -& - & - & - & - \\
		\hline
	$x_{\text{spin}}$&-&-& - & $0,0$& $- \frac{1}{16}  d_{\text{spin}}$& -& - & - & - & - & - \\
		\hline
		$y_{\text{spin}}$&-&- & -  & $(1,0),(0,1)$& $\frac{3}{64}  d_{\text{spin}}$ & - &  - & -  & - & - & - \\\hline
		$d_{\Lambda^2}$ &-&-&  $\frac{k(k-1)}{2}$&- & - & - & -& - & - & - & - \\
		\hline
		$h_{\Lambda^2}$ &-&-&  $\frac{k-2}{2}$ &-& - & - & - & - & - & - & - \\
		\hline
	$x_{\Lambda^2}$ &-&-&  $k-8$ &-& - & - & - & - & - & - & - \\
		\hline
		$y_{\Lambda^2}$ &-&-&  $\frac{3}{4}$ &-& - & - &  - & -  & - & - & -\\\hline
	\end{tabular}
	\caption{Group theory constants for common representations in 6D SCFTs (see also \cite{Grassi:2011hq, Ohmori:2014kda}). 
The special case of $SO(8)$ is further discussed in the text.		\label{constanttab}}
\end{table}

In this Appendix, we present a table of group theory constants, which are especially useful
in computing the anomaly polynomials of 6D SCFTs.  For a given gauge algebra $\mf{g}$, we
express $\tr_{\text{adj}}F^2$, the trace in the adjoint representation,
in terms of the quartic casimir as
\begin{equation}
\tr_{\text{adj}} F^2 = \frac{1}{2} \text{Ind}_{\text{adj}} \Tr F^2 = h_G^\vee \Tr F^2,
\end{equation}
where $h_G^\vee$ is the dual Coxeter number of $\mf{g}$.  We similarly write
\begin{equation}
\tr_{\text{adj}}F^4 = x_G \Tr F^4 + y_G (\Tr F^2)^2.
\end{equation}
Likewise, for a more general representation $\rho$ of $\mf{g}$, we write
\begin{equation}
\tr_{\rho} F^2 = \frac{1}{2}\text{Ind}_\rho \Tr F^2= h_\rho \Tr F^2, ~~\tr_{ \rho }F^4 = x_\rho \Tr F^4 + y_\rho (\Tr F^2)^2.
\end{equation}
The Lie algebras $\mf{su}(2)$, $\mf{su}(3)$, $\mf{g}_2$, $\mf{f}_4$, $\mf{e}_6$, $\mf{e}_7$, and $\mf{e}_8$ are special in that they do not have an independent quartic casimir.  Hence, $x_\rho = x_G=0$ for all representations of these groups.  The algebra $\mf{so}(8)$ is special in that it has three independent quartic casimirs, which rotate into each other under triality.  We may write
\begin{align}
\tr_{\text{adj}} F_{\mf{so}(8)}^2 &= 6 \Tr F_{\mf{so}(8)}^2 \\
\tr_{\text{v}} F_{\mf{so}(8)}^2 = \tr_{\text{s}} F_{\mf{so}(8)}^2 =\tr_{\text{c}} F_{\mf{so}(8)}^2 &= \Tr F_{\mf{so}(8)}^2
\end{align}
for $\rho = 8_v, 8_s$, and $8_c$, respectively.  We may also write
\begin{align}
\tr_{\text{adj}} F_{\mf{so}(8)}^4 &=4 \tr_{v} F_{\mf{so}(8)}^4 +4 \tr_s F_{\mf{so}(8)}^4 +4 \tr_c F_{\mf{so}(8)}^4,
\end{align}
where the basis for the three quartic casimirs is taken to be the trace in the $\bf{8_v}$,  $\bf{8_s}$ and $\bf{8_c}$ representations.

Table \ref{constanttab} depicts the above group theory constants for a variety of representations of each gauge algebra, along with gauge algebra rank $r_G$, gauge algebra dimension $d_G$, and representation dimension $d_\rho$.

\section{Anomaly Polynomials for Conformal Matter}\label{sec:CONFMAT}

In this Appendix, we present the quivers and anomaly polynomials for some common examples of conformal matter. While this list is not exhaustive, it does cover the types of conformal matter that appear most frequently in 6D SCFTs.

In each of these examples, we can express the anomaly polynomial in the form
\begin{align}
I_8 &= \frac{a}{24} c_2(R)^2 - \frac{b}{48} c_2(R) p_1(T) + c \frac{7 p_1(T)^2 - 4 p_2(T)}{5760} \nonumber \\
&+ \left(- \frac{x_L}{8} c_2(R) + \frac{y_L}{96}  p_1(T) \right) (\Tr F_L^2)+ \left(- \frac{x_R}{8} c_2(R) + \frac{y_R}{96}  p_1(T) \right) (\Tr F_R^2) \nonumber  \\
& + \frac{z_L}{32} (\Tr F_L^2)^2 + \frac{z_R}{32}(\Tr F_R^2)^2 + \frac{w}{16} \Tr F_L^2 \Tr F_R^2 +  \frac{t_L}{32} \Tr F_L^4 + \frac{t_R}{32}\Tr F_R^4
\end{align}
For gauge algebras without an independent quartic casimir ($\mf{su}(2)$, $\mf{su}(3)$, $\mf{g}_2$, $\mf{f}_4$, $\mf{e}_6$, $\mf{e}_7$, $\mf{e}_8$), $t_L = t_R = 0$. For $\mf{so}(8)$, there are quartic casimirs associated with each of the 8-dimensional representations, but there is an additional relation:
\begin{equation}
3 (\Tr F^2)^2 = \Tr_{\rm adj.} F^4 = 4 \Tr_{\bf 8_v}F^4 + 4 \Tr_{\bf 8_s}F^4 + 4 \Tr_{\bf 8_c}F^4.
\end{equation}

We first consider the case of $\mf{g}_L = \mf{g}_R:=\mf{g}$. In this case, we have $x_L=x_R:=x$, $y_L=y_R:=y$, $z_L=z_R:=z$, $t_L=t_R:=t$. The anomaly polynomial coefficients in these cases are shown in table \ref{tab:confmat1}.
\begin{table}[t]
	\centering
	\begin{tabular}{|c|c||c|c|c|c|c|c|c|c|}
		\hline
		$\mf{g}$&$a$&$ b$& $c$ & $x$ &$y $&$z$&$w$& $t$  \\
		\hline\hline
		$Sp(n)$& $10 n^2+57 n + 81$ & $2n^2+3n-9$ & $2n^2+n+1$  & $n+3$ & $n+1$ & $1/16$&  $1/4$ & $\frac{n+4}{24}$  \\ \hline
		$SU(n)$& $0$ & $0$ & $n^2$  & $0$ & $n$ & 0&  1&$ \frac{n}{24}$  \\ \hline
		$SO(2k)$  & $10 k^2 -57k +81$ & $2k^2-3k-9$ & $k(2k-1)+1$  & $2k-6$ &$2k-2$ & 1  &1 & $\frac{k-4}{24}$ \\ \hline
		$E_6$  & $319$ & $89$ & $79$  & $12$ &$12$ & 2  & 1 & 0 \\ \hline
		$E_7$  & $1670$ & $250$ & $134$  & $30$ &$18$ & 3  & 1 & 0 \\ \hline
		$E_8$  & $12489$ & $831$ & $249$  & $90$ &$30$ & 5  & 1 & 0 \\ \hline
			\end{tabular}
	\caption{Anomaly polynomial coefficients for conformal matter.	\label{tab:confmat1}}
\end{table}

For these cases with $\mf{g}_L = \mf{g}_R$, we present the explicit form of the conformal matter. For $Sp(n)$-$Sp(n)$ conformal matter, we have a theory of the form:
\begin{equation}
[Sp( n )] \,\, \overset{\mathfrak{so}(2n+8)}{4 }  \,\, [Sp( n )]
\end{equation}
$SU(n)$-$SU(n)$ conformal matter is simply a bifundamental hypermultiplet of $\mf{su}(n) \times \mf{su}(n)$, which we might depict in quiver notation as
\begin{equation}
[SU( n )]  \,\, [SU( n )]
\end{equation}
For $SO(2k)$-$SO(2k)$ conformal matter, we have a theory of the form:
\begin{equation}
[SO( 2k )] \,\, \overset{\mathfrak{sp}(k-4)}{1 }  \,\, [SO( 2k )]
\end{equation}
For the special case of $k=4$, the gauge group $\mf{sp}(0)$ vanishes, and we are left with the rank 1 E-string theory.

$E_{6}$-$E_{6}$ conformal matter was first written down in (\ref{eq:E6E6confmat}):
\begin{equation}
[E_6]  \,\, 1 \,\, \overset{\mathfrak{su}(3)}{3 }\,\, 1 \,\, [E_6]
\end{equation}
$E_{7}$-$E_{7}$ conformal matter was likewise written down in (\ref{eq:E7confmat}):
\begin{equation}
[E_7]  \,\, 1 \,\, \overset{\mathfrak{su}(2)}{2 }\,\, \overset{\mathfrak{so}(7)}{3
}\,\,\overset{\mathfrak{su}(2)}{2 }\,\, 1 \,\, [E_7]
\end{equation}
Finally, $E_8$-$E_8$ conformal matter is given by (\ref{eq:E8confmat}):
\begin{equation}
[E_8] \,\, 1 \,\, 2 \,\, \overset{\mathfrak{su}(2)}{2 }\,\, \overset{\mathfrak{g}_{2}}{3
}\,\, 1 \,\, \overset{\mathfrak{f}_{4}}{5 }\,\, 1 \,\, \overset{\mathfrak{g}%
_{2}}{3 }\,\,\overset{\mathfrak{su}(2)}{2 }\,\, 1 \,\, [E_8]
\end{equation}

Next, we consider some select cases of $\mf{g}_L \neq \mf{g}_R$ conformal matter. We begin with $E_8$-$SU(n)$ conformal matter. This involves a ramp of gauge algebras between $\mf{g}_L = \mf{e}_8$ and $\mf{g}_R=\mf{su}(n)$:
\begin{align}
[E_{8}] \,\, 1 \,\, \overset{\mathfrak{su}(1)}{2}\,\,
\overset{\mathfrak{su}(2)}{2 }\,\, ...\,\, \overset{\mathfrak{su}(n-1)}{2} \,\, [SU(n)]
\end{align}
The anomaly polynomial in this case is given by
\begin{align}
I_8 &= \left( \frac{2 n^5}{5}-3 n^4+\frac{85 n^3}{9}-\frac{383 n^2}{24}+\frac{2713 n}{180} \right) c_2(R)^2 \nonumber \\
&+ \left(-\frac{7 n^3}{36}+\frac{37 n^2}{48}-\frac{79 n}{72} \right) c_2(R) p_1(T)  +\left(  \frac{7 n^2}{5760}+\frac{7 n}{192} \right) p_1(T)^2 + \left( -\frac{n^2}{1440}-\frac{n}{48} \right) p_2(T) \nonumber \\
&+ \left( -\frac{n^3}{6}+\frac{3 n^2}{4}-\frac{4 n}{3} \right) c_2(R) \Tr F_L^2 + \frac{n}{16} p_1(T)  \Tr F_L^2 + \frac{n}{32} (\Tr F_L^2)^2 \nonumber \\
& + \left(-\frac{n^2}{8}+\frac{n}{8}-\frac{1}{4} \right) c_2(R) \Tr F_{R}^2  + \left( \frac{n+5}{96} \right) p_1(T) \Tr F_{R}^2   + \frac{1}{32} (\Tr F_R^2)^2  + \frac{1}{16} \Tr F_L^2 \Tr F_R^2 \nonumber \\
&+\frac{n-1}{24}\Tr F_R^4.
\end{align}

Similarly, we consider the case of $E_8$-$B_k/D_k$ conformal matter:
\begin{align}
[E_{8}] \,\, 1 \,\, 2 \,\, \overset{\mathfrak{su}(2)}{2} \,\, \overset{\mathfrak{g}_2}{3} \,\, 1 \,\, \overset{\mathfrak{so}(9)}{4}\,\,
\overset{\mathfrak{sp}(1)}{1 }\,\, \overset{\mathfrak{so}(11)}{4}\,\,  ...\,\, \overset{\mathfrak{sp}(k-4)}{1} \,\, [SO(2k)/SO(2k+1)]
\end{align}
The anomaly polynomial in the two cases is exactly the same:
\begin{align}
I_8 &= \left(\frac{2 k^5}{5}-3 k^4+\frac{85 k^3}{9}-\frac{383 k^2}{24}+\frac{2713 k}{180} \right) c_2(R)^2 \nonumber \\
&+ \left( -\frac{7 k^3}{36}+\frac{37 k^2}{48}-\frac{79 k}{72}  \right) c_2(R) p_1(T)  +\left( \frac{7 k^2}{5760}+\frac{7 k}{192}  \right) p_1(T)^2 + \left( -\frac{k^2}{1440}-\frac{k}{48} \right) p_2(T) \nonumber \\
&+ \left( -\frac{k^3}{6}+\frac{3 k^2}{4}-\frac{4 k}{3} \right) c_2(R) \Tr F_L^2 + \frac{k}{16} p_1(T)  \Tr F_L^2 + \frac{k}{32} (\Tr F_L^2)^2 \nonumber \\
& + \left( -\frac{k^2}{2}+\frac{7 k}{4}-\frac{3}{2} \right) c_2(R) \Tr F_{R}^2  + \left( \frac{k+2}{48} \right) p_1(T) \Tr F_{R}^2   + \frac{1}{16} (\Tr F_R^2)^2  + \frac{1}{16} \Tr F_L^2 \Tr F_R^2 \nonumber \\
&+ \frac{k-4}{24} \Tr F_R^4.
\end{align}

Next, we have $E_8 $-$E_7$ conformal matter:
\begin{equation}
[E_8] \,\,1 \,\, 2 \,\, \overset{\mathfrak{su}(2)}{2 }\,\, \overset{\mathfrak{g}_{2}}{3
}\,\, 1 \,\, \overset{\mathfrak{f}_{4}}{5 }\,\, 1 \,\, \overset{\mathfrak{g}%
_{2}}{3 }\,\,\overset{\mathfrak{su}(2)}{2 }\,\, 1 \,\, [E_7]
\end{equation}
The anomaly polynomial is:
\begin{align}
I_8 &= \frac{2867 c_2(R)^2}{6}-10 c_2(R) \Tr F_{R}^2-11 c_2(R) \Tr F_{L}^2-\frac{193 c_2(R) p_1(T)}{12}+\frac{1}{16} \Tr F_{R}^2 \Tr F_{L}^2\nonumber \\
&
+\frac{1}{4} p_1(T) \Tr F_{R}^2+\frac{(\Tr F_{R}^2)^2}{8}+\frac{5}{16} p_1(T) \Tr F_{L}^2+\frac{5 (\Tr F_{L}^2)^2}{32}+\frac{77 p_1(T)^2}{288}-\frac{11 p_2(T)}{72}.
\end{align}

For $E_8 $-$E_6$ conformal matter:
\begin{equation}
[E_8] \,\,1 \,\, 2 \,\, \overset{\mathfrak{su}(2)}{2 }\,\, \overset{\mathfrak{g}_{2}}{3
}\,\, 1 \,\, \overset{\mathfrak{f}_{4}}{5 }\,\, 1 \,\, \overset{\mathfrak{su}(3)}{3 }\,\, 1 \,\, [E_6]
\end{equation}
The anomaly polynomial is:
\begin{align}
I_8&=\frac{1124 c_2(R)^2}{3}-\frac{31}{4} c_2(R) \Tr F_R^2-\frac{41}{4} c_2(R) \Tr F_L^2-\frac{41 c_2(R) p_1(T)}{3}+\frac{1}{16} \Tr F_R^2 \Tr F_L^2 \nonumber \\
&+\frac{3}{16} p_1(T) \Tr F_R^2+\frac{3 (\Tr F_R^2)^2}{32}+\frac{5}{16} p_1(T) \Tr F_L^2+\frac{5 (\Tr F_L^2)^2}{32}+\frac{7 p_1(T)^2}{30}-\frac{2 p_2(T)}{15}.
\end{align}

For $E_8 $-$F_4$ conformal matter:
\begin{equation}
[E_8] \,\,1 \,\, 2 \,\, \overset{\mathfrak{su}(2)}{2 }\,\, \overset{\mathfrak{g}_{2}}{3
}\,\, 1  \,\, [F_4]
\end{equation}
The anomaly polynomial is:
\begin{align}
I_8&=51 c_2(R)^2-\frac{5}{2} c_2(R) \Tr F_R^2-4 c_2(R) \Tr F_L^2-\frac{9 c_2(R) p_1(T)}{2} +\frac{1}{16} \Tr F_R^2 \Tr F_L^2\nonumber \\
&+\frac{1}{8} p_1(T) \Tr F_R^2+\frac{(\Tr F_R^2)^2}{16}+\frac{1}{4} p_1(T) \Tr F_L^2+\frac{(\Tr F_L^2)^2}{8}+\frac{119 p_1(T)^2}{720}-\frac{17 p_2(T)}{180}.
\end{align}

For $E_8$-$G_2$ conformal matter:
\begin{equation}
[E_8] \,\,1 \,\, 2 \,\, \overset{\mathfrak{su}(2)}{2 }\,\, [G_2]
\end{equation}
The anomaly polynomial is:
\begin{align}
I_8 &= \frac{21 c_2(R)^2}{2}-\frac{7}{4} c_2(R) \Tr F_L^2-c_2(R) \Tr F_R^2-\frac{7 c_2(R) p_1(T)}{4}+\frac{1}{16} \Tr F_L^2 \Tr F_R^2 \nonumber \\
&+\frac{3}{16} p_1(T) \Tr F_L^2+\frac{3 (\Tr F_L^2)^2}{32}+\frac{1}{12} p_1(T) \Tr F_R^2+\frac{(\Tr F_R^2)^2}{24}+\frac{161 p_1(T)^2}{1440}-\frac{23 p_2(T)}{360}.
\end{align}

$E_7$-$E_6$ conformal matter is given by:
\begin{equation}
[E_7] \,\,1 \,\, \overset{\mathfrak{su}_{2}}{2
} \,\, \overset{\mathfrak{so}(7)}{3 }\,\, \overset{\mathfrak{su}_{2}}{2
}\,\, 1  \,\, [E_6]
\end{equation}
Note that this is exactly the same as $E_7$-$E_7$ conformal matter, so the anomaly polynomial is identical to that one, which can be found in table \ref{tab:confmat1}.

$E_7$-$SO(8)$ conformal matter is given by
\begin{equation}
[E_7] \,\,1 \,\, \overset{\mathfrak{su}_{2}}{2
} \,\, \overset{\mathfrak{g}_2}{3 }\,\, 1  \,\, [SO(8)]
\end{equation}
The corresponding anomaly polynomial is
\begin{align}
I_8 &=\frac{887 c_2(R)^2}{24}-3 c_2(R) \Tr F_L^2-\frac{9}{4} c_2(R) \Tr F_R^2-\frac{169 c_2(R) p_1(T)}{48}+\frac{1}{16} \Tr F_L^2 \Tr F_R^2 \nonumber \\
&+\frac{3}{16} p_1(T) \Tr F_L^2+\frac{3 (\Tr F_L^2)^2}{32}+\frac{1}{8} p_1(T) \Tr F_R^2+\frac{(\Tr F_R^2)^2}{16}+\frac{749 p_1(T)^2}{5760}-\frac{107 p_2(T)}{1440}.
\end{align}

Finally, $E_6$-$SO(8)$ conformal matter is given by
\begin{equation}
[E_6] \,\,1 \,\, \overset{\mathfrak{su}_{3}}{3 }\,\, 1  \,\, [SO(8)]
\end{equation}
The anomaly polynomial is
\begin{align}
I_8 &=\frac{319 c_2(R)^2}{24}-\frac{3}{2} c_2(R) \Tr F_L^2-\frac{3}{2} c_2(R) \Tr F_R^2-\frac{89 c_2(R) p_1(T)}{48}+\frac{1}{16} \Tr F_L^2 \Tr F_R^2 \nonumber \\
&+\frac{1}{8} p_1(T) \Tr F_L^2+\frac{(\Tr F_L^2)^2}{16}+\frac{1}{8} p_1(T) \Tr F_R^2+\frac{(\Tr F_R^2)^2}{16}+\frac{553 p_1(T)^2}{5760}-\frac{79 p_2(T)}{1440}.
\end{align}

\newpage

\bibliographystyle{utphys}
\bibliography{ref}

\end{document}